\documentclass[]{aa}
\usepackage{graphicx,natbib}

\DeclareGraphicsExtensions{.eps,.ps}

\newcommand{\upf}[1]{\raisebox{0.5ex}[0pt]{#1}}

\begin{document}
\title{Structures in surface-brightness profiles of LMC and SMC star clusters: evidence of mergers?}

\author{L. Carvalho\inst{1} \and T.A. Saurin\inst{1} \and E. Bica\inst{1} \and C. Bonatto\inst{1} \and A.A. Schmidt\inst{2}}

\institute{Universidade Federal do Rio Grande do Sul, Departamento de Astronomia, CP\,15051, RS, Porto Alegre 91501-970, Brazil\
\and Universidade Federal de Santa Maria, LANA, RS, Santa Maria 97119-900, Brazil\\
\email{luziane@if.ufrgs.br, tiago.saurin@ufrgs.br, bica@if.ufrgs.br, charles@if.ufrgs.br,\\ alex@lana.ccne.ufsm.br}
\mail{luziane@if.ufrgs.br}}

\date{Received --; accepted --}

\abstract
{The LMC and SMC are rich in binary star clusters, and some mergers are expected. It is important to characterize single clusters, binary clusters and candidates to mergers.}
{We selected a sample of star clusters in each Cloud with this aim. Surface photometry of 25 SMC and 22 LMC star clusters was carried with the ESO Danish 1.54 m telescope. 23 clusters were observed for the first time for these purposes.}
{We fitted Elson, Fall and Freeman (1987, EFF) profiles to the data, deriving structural parameters, luminosities and masses. We also use isophotal maps to constrain candidates to cluster interactions.}
{The structural parameters, luminosities and masses presented good agreement with those in the literature. Three binary clusters in the sample have a double profile. Four clusters (NGC\,376, K\,50, K\,54 and NGC\,1810) do not have companions and present as well important deviations from EFF profiles.}
{The present sample contains blue and red Magellanic clusters. Extended EFF profiles were detected in some blue clusters. We find evidence that important deviations from the body of EFF profiles might be used as a tool to detect cluster mergers.}


\titlerunning{Structures in LMC and SMC}

\authorrunning{L. Carvalho et al.}

\maketitle

%

\section{Introduction}
\label{sec:int}
%

Surface-brightness and number-density profiles can be used to investigate
properties of star clusters in different tidal environments. The standard
description of Globular Clusters (GCs) assumes an isothermal central region and
a tidally truncated outer region (e.g. \citealt{Binney}). However, both
structures evolve with time. Evolved GCs, in particular, can be virtually
considered as dynamically relaxed systems (e.g. \citealt{NoyolaG06}). Since
formation, star clusters are subject to internal and external processes that
affect the spatial distribution of stars and introduce asymmetries in the
luminosity distribution, which in principle, can be detected by 
surface-brightness profiles (SBPs). Among the former are mass loss associated
with stellar evolution, large-scale mass segregation and low-mass star
evaporation. The latter are tidal stress and dynamical friction
(e.g. \citealt{Khal07}; \citealt{Lamers05}; \citealt{Gne97}). These processes
tend to decrease cluster mass, which may accelerate the core collapse phase in
some cases (e.g. \citealt{DjorgM94}). With time, what results is a spatial
distribution of light (or mass) that reflects the combined effect of these
processes associated with physical conditions at the early collapse
(\citealt{BB07} and references therein).

In this context, the present-day internal structure of individual star clusters,
as well as the collective large-scale galactocentric distribution can be used to
probe conditions related to galaxy formation, and to investigate cluster
dynamical evolution (e.g. \citealt{MacVDB05}; \citealt{Bica06}).

SBPs of star clusters, GCs in particular, have been shown to follow analytical
profiles. The most commonly used are the single-mass, modified isothermal sphere
of \citet{King66} that is the basis of the Galactic GC parameters given by
\citet{TKD95} and Harris (1996, and the 2003
update\footnote{\em http://physun.physics.mcmaster.ca/Globular.html}),
the modified isothermal sphere of \citet{Wilson75} that assumes a pre-defined
stellar distribution function which results in more extended envelopes than in
\citet{King66}, and the power-law with a core of EFF (\citealt{EFF87}) that has
been fit to massive young clusters especially in the Magellanic Clouds
(e.g. \citealt{Mack03a,Mack03b}). Each function is characterised by different
parameters that are somehow related to the cluster structure.

Perhaps as interesting as the fact that SBPs can be described by analytical
profiles such as \citet{King66} or EFF, from which structural
parameters can be derived, is the fact that significant deviations have been
detected. They are (i) post-core-collapse (PCC) excesses to power laws in
surface density vs. log radius for Galactic GCs 
(e.g. \citealt{TKD95}) and (ii) in Galactic open clusters
(e.g. \citealt{BB05}; \citealt{BBB06}), and (iii) extensions beyond the tidal
radius in young populous LMC clusters (e.g. \citealt{EFF87}) or R\,136 in
30\,Dor (e.g. \citealt{Mack03a}), which appear to be related to formation
conditions.

Not many SMC clusters have published SBPs. \citet{Mack03b} studied 10 populous
SMC clusters. Using HST, 53 clusters in the LMC were studied by \citet{Mack03a}.
Number-density studies (\citealt{Chryso89}, and references therein) included
SMC and LMC clusters, but inner profiles were limited by the low stellar
resolution in photographic material.

Evidence of cluster binarity and mergers has been reported for both Clouds, which appear to be suitable environments for that. Indeed, the SMC and LMC are very rich in cluster pairs and multiplets
(\citealt{BicaSchM95}; \citealt{Bica99}; \citealt{Dieball}). Studies of radial
variations of parameters related to isophotes and comparisons with N-body
simulations were carried by \citet{deOliv00a,deOliv00b}, where several possible
mergers were discussed. The blue LMC cluster NGC\,2214 has been reported as 
merger, where the secondary component appears to be still conspicuous in images
(\citealt{Bathia88}).

In the present study we explore a sample of 47 SMC and LMC clusters of different
ages, part of them studied in terms of SBP for the first time. The sample selection was mostly based on \citet{vdbergh81}, in view of spanning a wide age range. We argue that
distortions found in the SBP of some clusters may arise from the spatial
evolution of mergers, and that the star clusters NGC\,376, K\,50, K\,54, and
NGC\,1810 have evidence for mergers.

This work is structured as follows. In Sect. \ref{sec:obs} we present the
observations. In Sect. \ref{sec:res} we fit profiles to the data and derive
structural parameters and masses. The structures on the profiles are discussed
in Sect. \ref{sec:dis}, and finally, concluding remarks are given in Sect.
\ref{sec:con}.

\section{Observations and analysis}
\label{sec:obs}

The data were obtained in two observing runs, the first between 1990 november
16 and 20, and the second between 1991 october 25 and november 9 using the ESO
Danish 1.54\,m telescope, in La Silla, Chile. The detectors were RCA CCDs (SID
501 in the first run and 503 in the second run). They have 320$\times$512 and
1024$\times$640\,pixels, with the spatial scales 0.475$''$/pixel and
0.237$''$/pixel (Table \ref{tab:obs}), respectively. V and B Bessel filters were
used. CCD gains were 12.3 and 5.9\,electrons/ADU, and readout noises were 31.6
and 15\,electrons. The mean seeing for the frames was 1.5$''$. In Table
\ref{tab:obs} we summarize the observational data including the exposure time
and seeing for each image together with the most common identifiers
(\citealt{BicaSchM95,Bica99}) for the star clusters of this sample.

For all objects the V band observations were used, except NGC\,2159, for which
the B image had a significantly higher signal-noise ratio.

\begin{table}[!h]
\caption[]{Cluster sample and observational details.}
\label{tab:obs}
\renewcommand{\tabcolsep}{0.4mm}
\renewcommand{\arraystretch}{1.1}
\begin{tabular}{lccrccr}
\hline
\hline
\multicolumn{1}{l}{Cluster name} &&&\multicolumn{1}{c}{Time} &\multicolumn{1}{c}{Seeing} &\multicolumn{1}{c}{CCD scale} &\multicolumn{1}{r}{Date}\\
\multicolumn{1}{l}{} &&&\multicolumn{1}{r}{(s)}\hspace{0.5mm} &\multicolumn{1}{c}{(arcsec)} &\multicolumn{1}{c}{(arcsec/pixel)}\\
\multicolumn{1}{l}{(1)} &&&\multicolumn{1}{r}{(2)}\hspace{0.45mm} &\multicolumn{1}{c}{(3)} &\multicolumn{1}{c}{(4)} &\multicolumn{1}{c}{(5)}\\
\hline
\multicolumn{7}{c}{SMC}\\
\hline
NGC\,121 (K\,2, L\,10)       &&& 360 & 1.6 & 0.475 & 1990\\
NGC\,176 (K\,12, L\,16)      &&& 360 & 1.6 & 0.475 & 1991\\
K\,17 (L\,26)                &&& 360 & 1.2 & 0.475 & 1991\\
NGC\,241+242 (K\,22, L\,29)  &&&  60 & 1.7 & 0.475 & 1991\\
NGC\,290 (L\,42)             &&&  60 & 1.4 & 0.475 & 1991\\
L\,48                        &&& 360 & 1.8 & 0.237 & 1991\\
K\,34 (L\,53)                &&& 240 & 1.4 & 0.475 & 1991\\
NGC\,330 (K\,35, L\,54)      &&& 100 & 1.5 & 0.475 & 1990\\
L\,56                        &&& 120 & 1.6 & 0.475 & 1990\\
NGC\,339 (K\,36, L\,59)      &&& 180 & 1.9 & 0.475 & 1990\\
NGC\,346 (L\,60)             &&&  60 & 1.8 & 0.475 & 1990\\
IC\,1611 (K\,40, L\,61)      &&& 360 & 1.4 & 0.475 & 1991\\
IC\,1612 (K\,41, L\,62)      &&& 360 & 1.4 & 0.475 & 1991\\
L\,66                        &&&  30 & 1.3 & 0.475 & 1991\\
NGC\,361 (K\,46, L\,67)      &&& 120 & 1.9 & 0.475 & 1991\\
K\,47 (L\,70)                &&& 120 & 1.3 & 0.475 & 1991\\
NGC\,376 (K\,49, L\,72)      &&& 240 & 1.5 & 0.237 & 1991\\
K\,50 (L\,74)                &&& 120 & 1.2 & 0.475 & 1991\\
IC\,1624 (K\,52, L\,76)      &&& 180 & 1.6 & 0.475 & 1991\\
K\,54 (L\,79)                &&&  30 & 1.9 & 0.475 & 1991\\
NGC\,411 (K\,60, L\,82)      &&& 180 & 2.1 & 0.475 & 1991\\
NGC\,416 (K\,59, L\,83)      &&& 150 & 1.2 & 0.475 & 1991\\
NGC\,419 (K\,58, L\,85)      &&& 300 & 1.4 & 0.237 & 1991\\
NGC\,458 (K\,69, L\,96)      &&& 240 & 1.6 & 0.475 & 1990\\
L\,114                       &&& 180 & 1.4 & 0.475 & 1990\\
\hline
\multicolumn{7}{c}{LMC}\\
\hline
NGC\,1783 (SL\,148)	     &&& 300 & 1.8 & 0.475 & 1991\\
NGC\,1810 (SL\,194)	     &&& 120 & 1.2 & 0.475 & 1991\\
NGC\,1818 (SL\,201)	     &&&  60 & 1.7 & 0.237 & 1991\\
NGC\,1831 (SL\,227, LW\,133) &&& 300 & 1.6 & 0.237 & 1991\\
NGC\,1847 (SL\,240)	     &&& 120 & 2.2 & 0.475 & 1991\\
NGC\,1856 (SL\,271)	     &&& 120 & 1.7 & 0.475 & 1991\\
NGC\,1866 (SL\,319, LW\,163) &&&  30 & 1.8 & 0.475 & 1990\\
NGC\,1868 (SL\,330, LW\,169) &&& 120 & 1.8 & 0.475 & 1990\\
NGC\,1870 (SL\,317)          &&& 120 & 1.8 & 0.475 & 1991\\
NGC\,1978 (SL\,501)	     &&& 160 & 1.5 & 0.475 & 1990\\
NGC\,2004 (SL\,523)	     &&&  60 & 1.3 & 0.475 & 1991\\
NGC\,2011 (SL\,559)	     &&& 160 & 1.3 & 0.475 & 1990\\
NGC\,2100 (SL\,662)	     &&&  20 & 1.5 & 0.475 & 1991\\
NGC\,2121 (SL\,725, LW\,303) &&& 240 & 1.3 & 0.475 & 1991\\
NGC\,2157 (SL\,794)	     &&& 120 & 2.1 & 0.475 & 1990\\
NGC\,2159 (SL\,799)	     &&&  20 & 1.3 & 0.475 & 1991\\
NGC\,2164 (SL\,808)	     &&& 180 & 2.2 & 0.475 & 1991\\
NGC\,2210 (SL\,858, LW\,423) &&& 180 & 1.4 & 0.475 & 1991\\
NGC\,2213 (SL\,857, LW\,419) &&& 120 & 2.2 & 0.475 & 1990\\
NGC\,2214 (SL\,860, LW\,426) &&&  50 & 1.8 & 0.475 & 1990\\
H\,11 (SL\,868, LW\,437)     &&& 420 & 1.3 & 0.475 & 1991\\
HS\,314                      &&&  60 & 1.7 & 0.475 & 1990\\
\hline
\end{tabular}
\begin{list}{Table Notes.}
\item Col. 1: Cluster identifications in main catalogues. Col. 2:
   CCD exposure time in seconds. Col. 3: Seeing in arcseconds. Col. 4: CCD
   scale in arcseconds per pixel. Col. 5: Year of the observing run.
\end{list}
\end{table}

\begin{figure}[!h]
   \centering
   \fbox{\includegraphics[scale=0.155,viewport=0 0 480 660,clip]{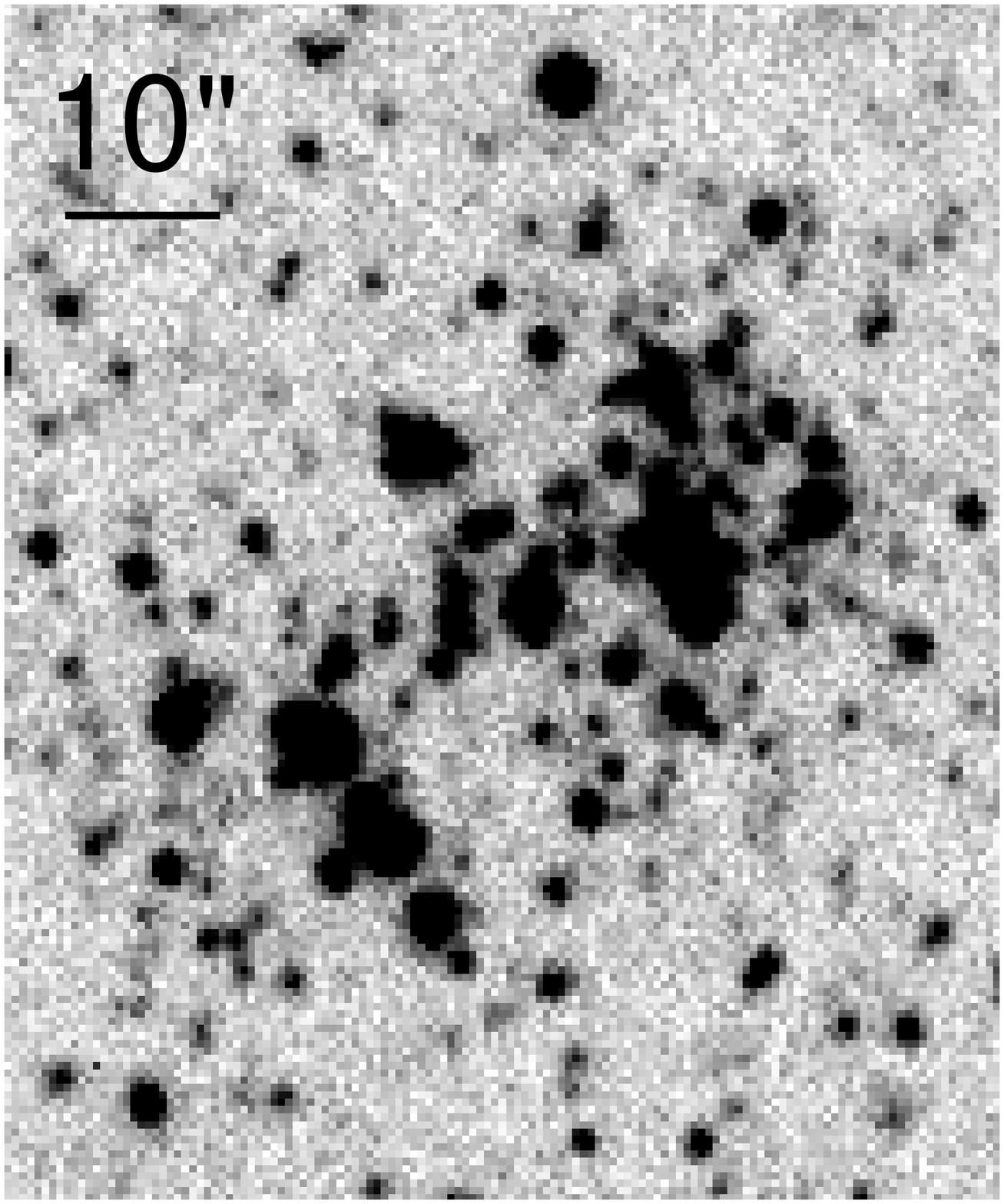}}
   \fbox{\includegraphics[scale=0.155,viewport=0 0 480 660,clip]{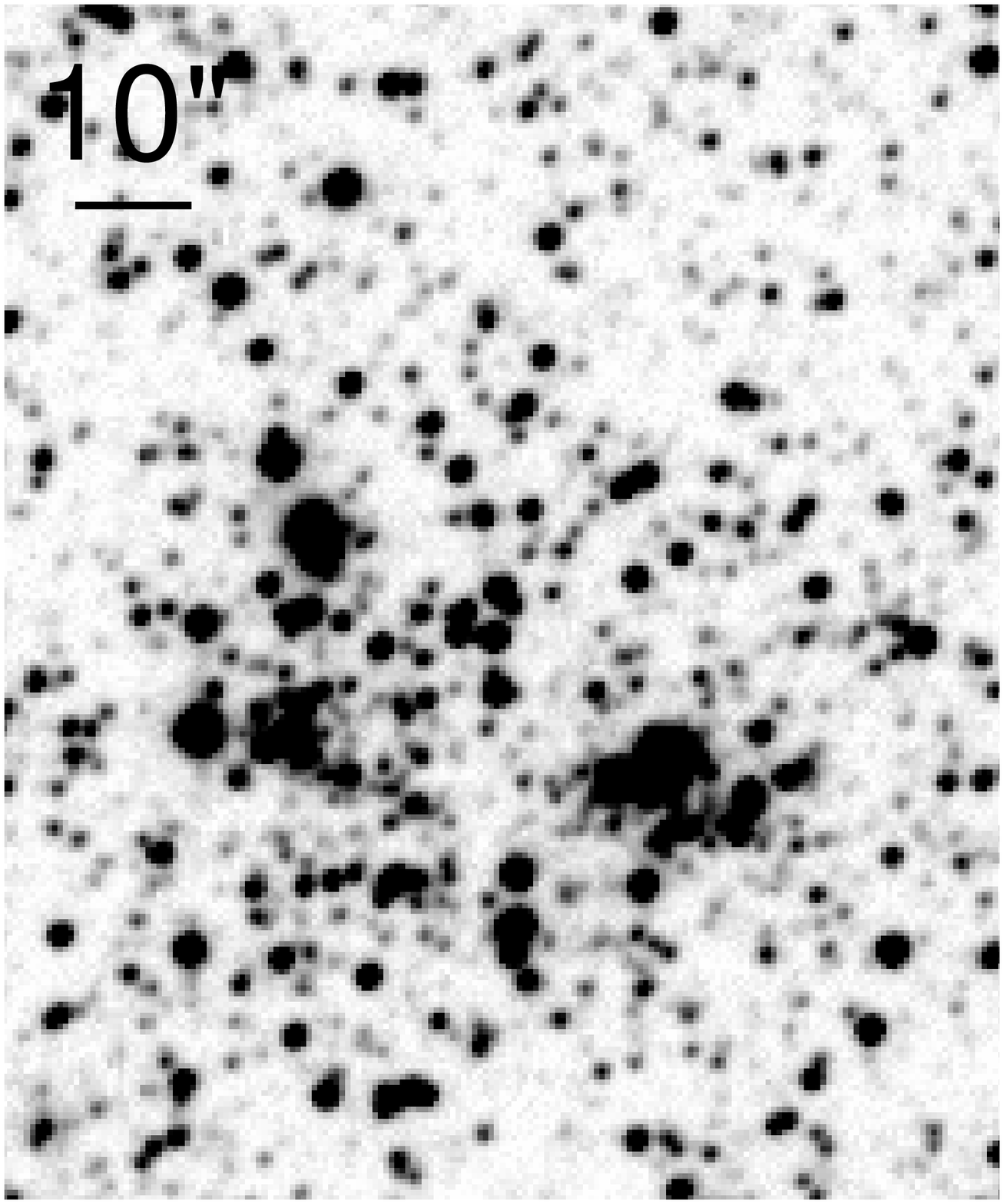}}
   \fbox{\includegraphics[scale=0.155,viewport=0 0 480 660,clip]{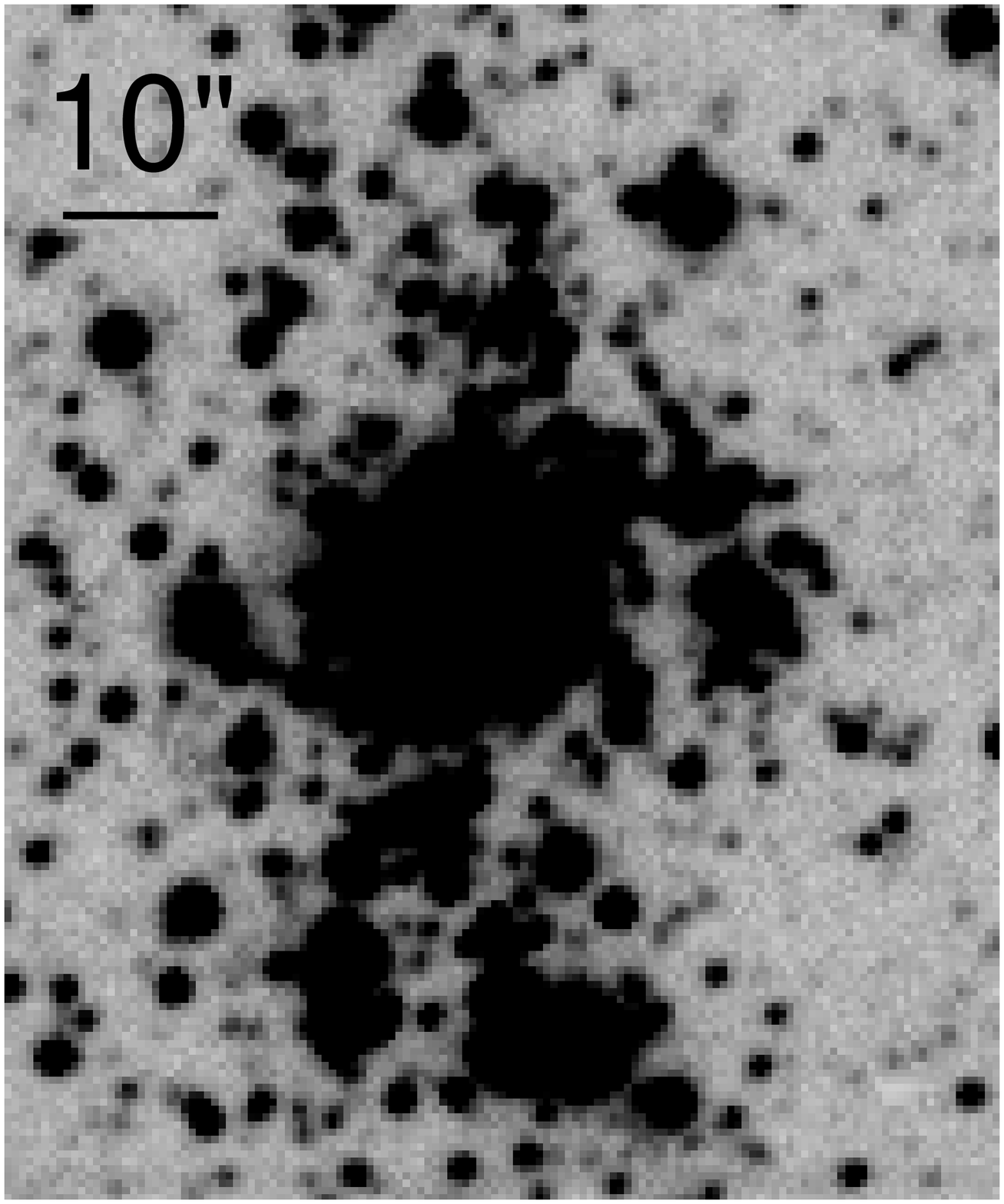}}
   \fbox{\includegraphics[scale=0.155,viewport=0 0 480 660,clip]{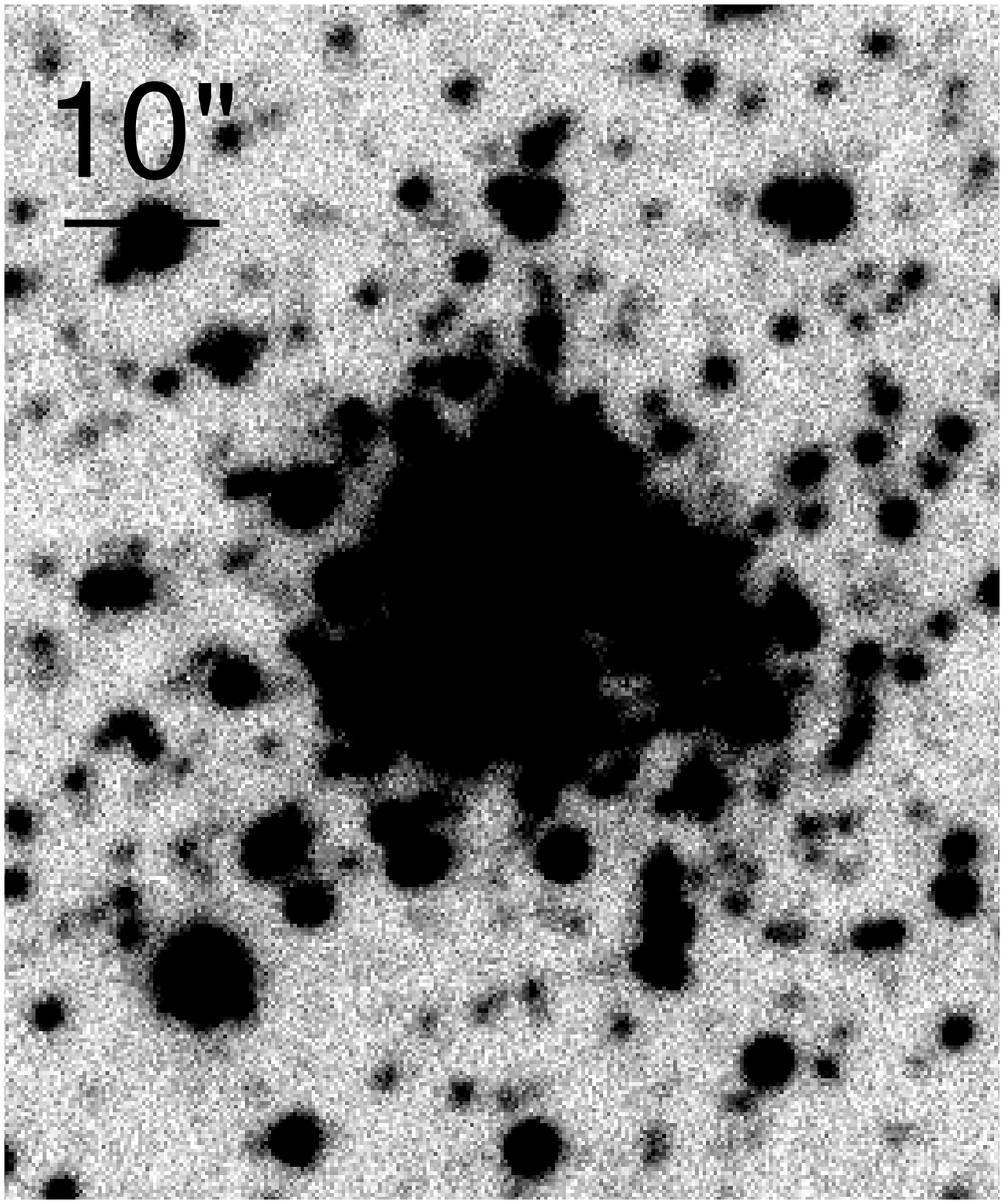}}
   \fbox{\includegraphics[scale=0.155,viewport=0 0 480 660,clip]{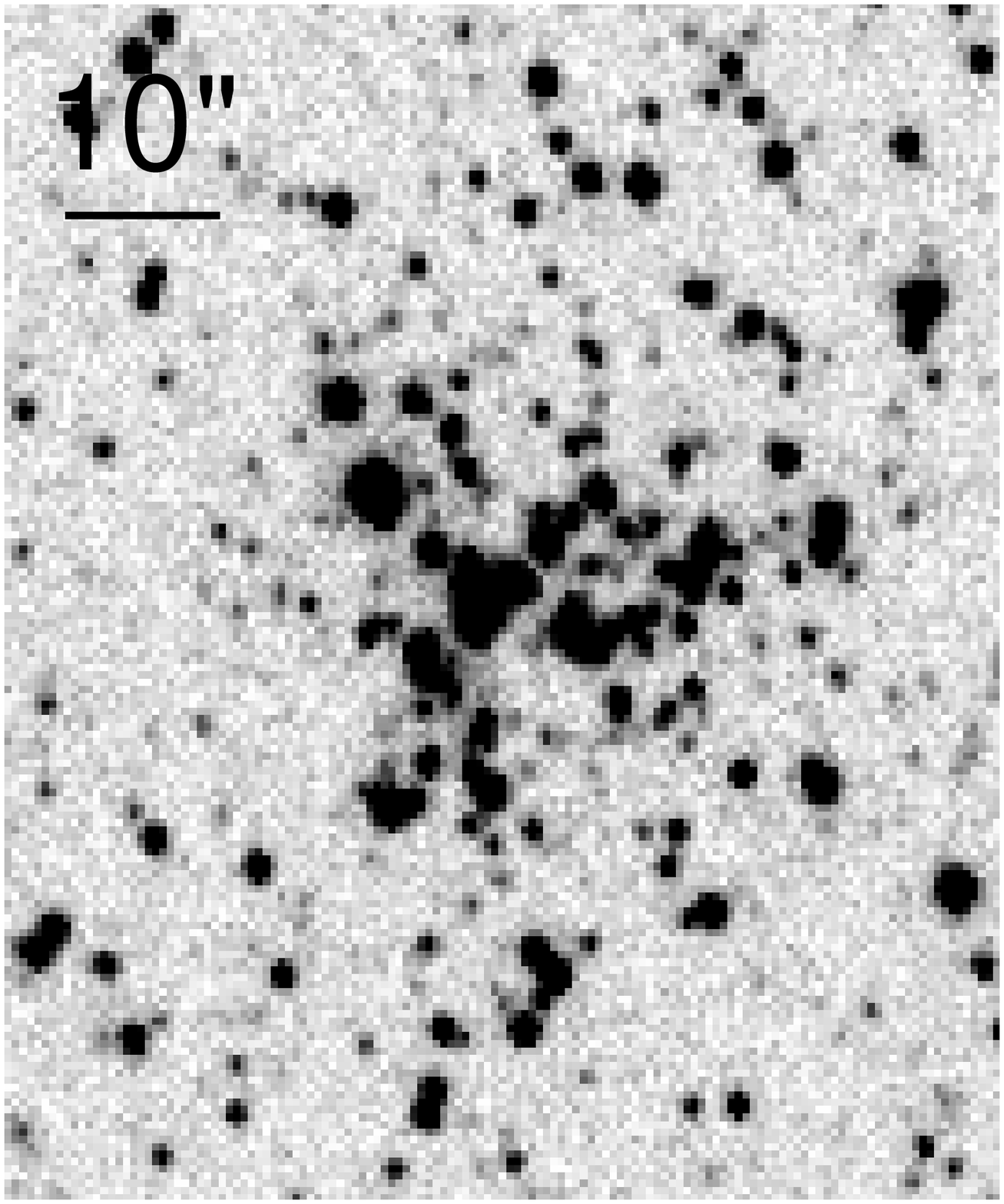}}
   \fbox{\includegraphics[scale=0.155,viewport=0 0 480 660,clip]{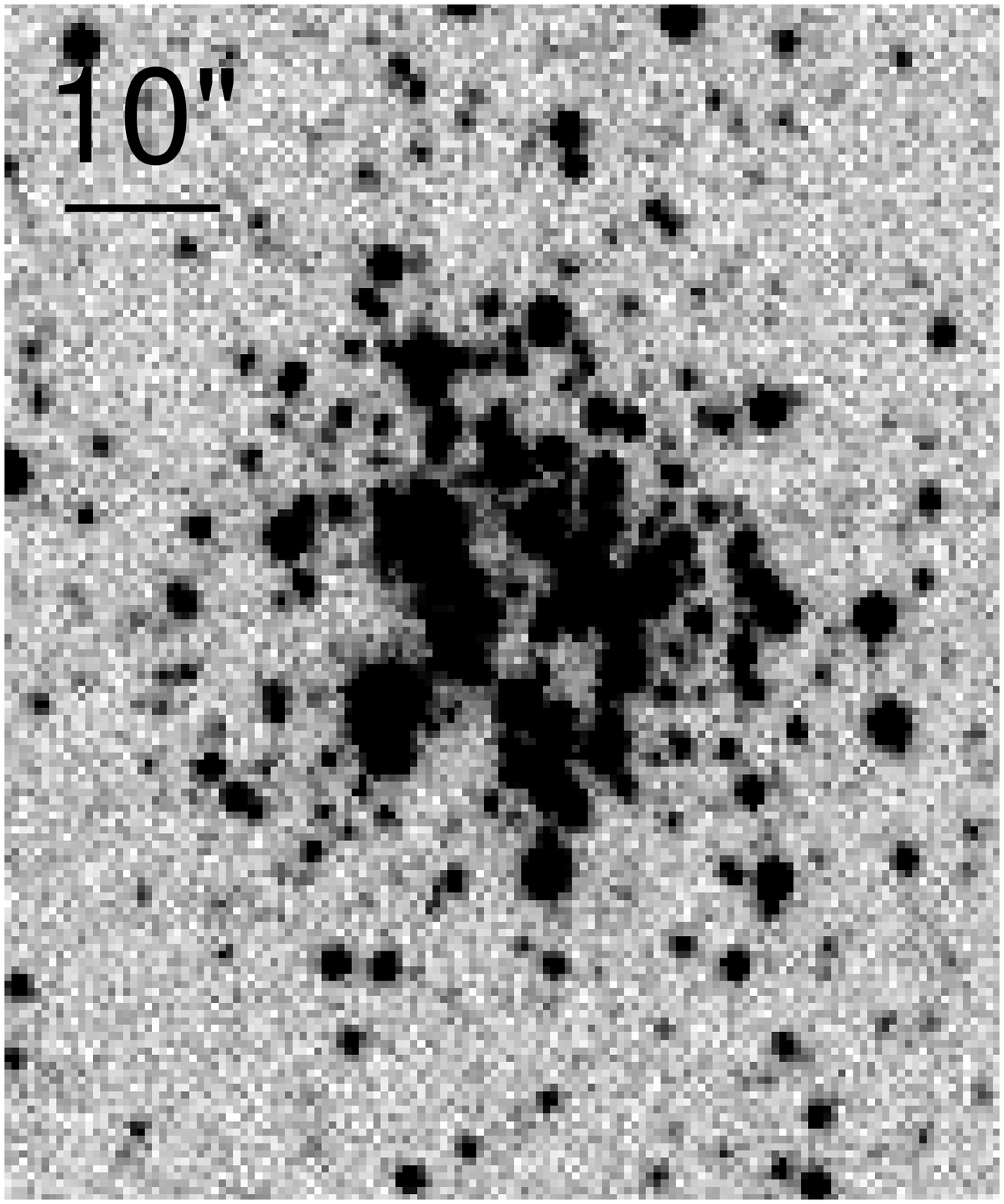}}
   \fbox{\includegraphics[scale=0.155,viewport=0 0 480 660,clip]{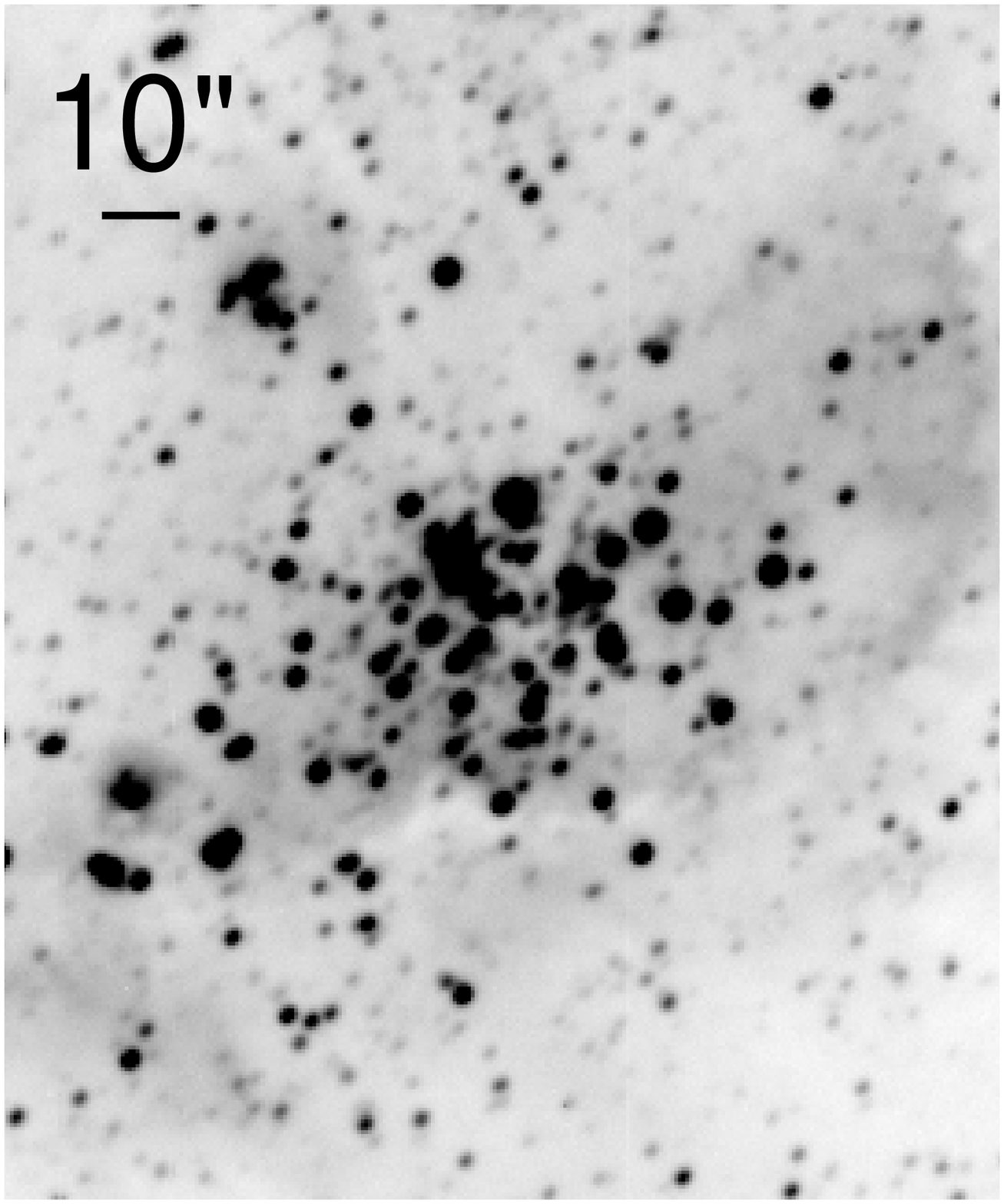}}
   \fbox{\includegraphics[scale=0.155,viewport=0 0 480 660,clip]{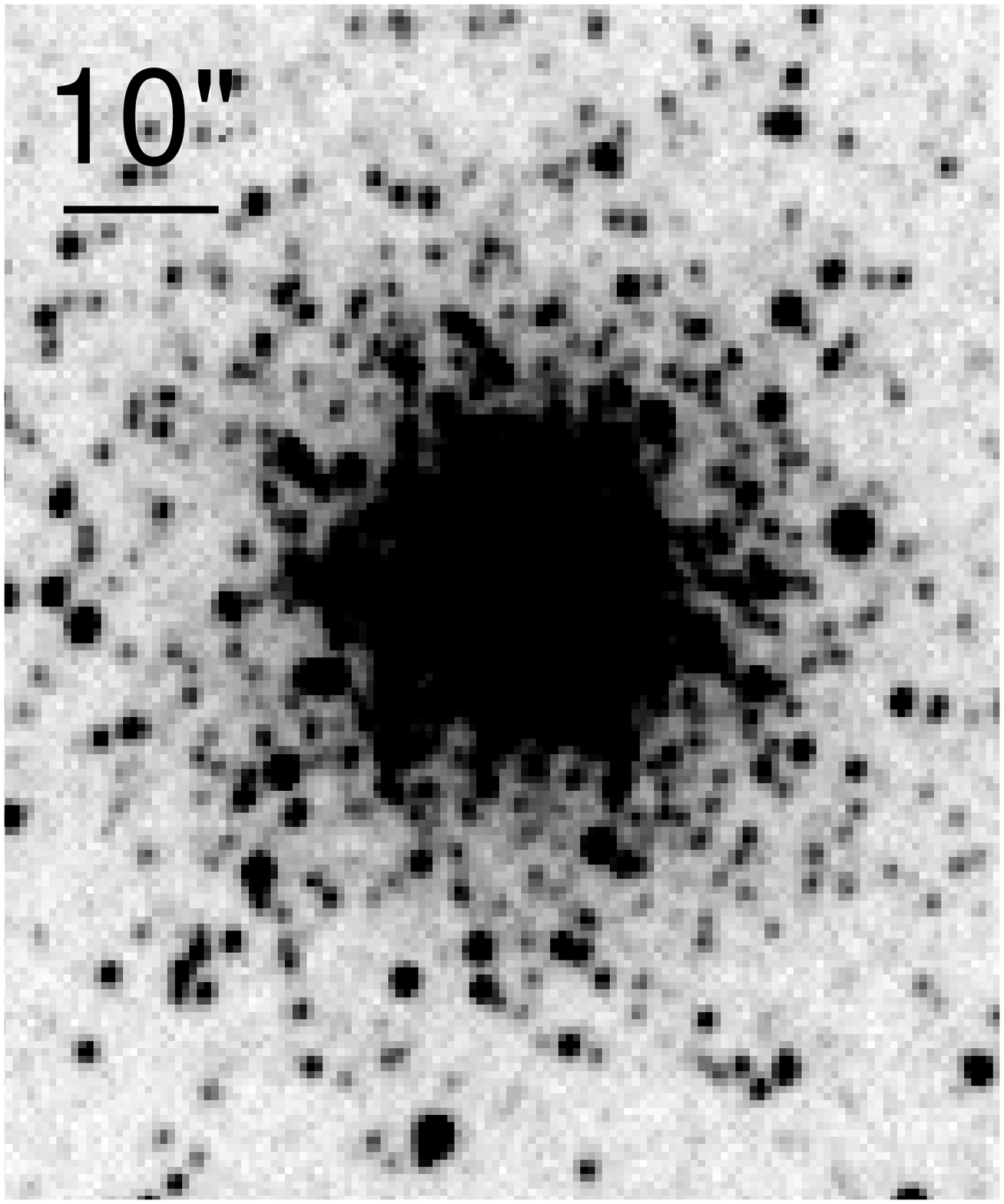}}
   \fbox{\includegraphics[scale=0.155,viewport=0 0 480 660,clip]{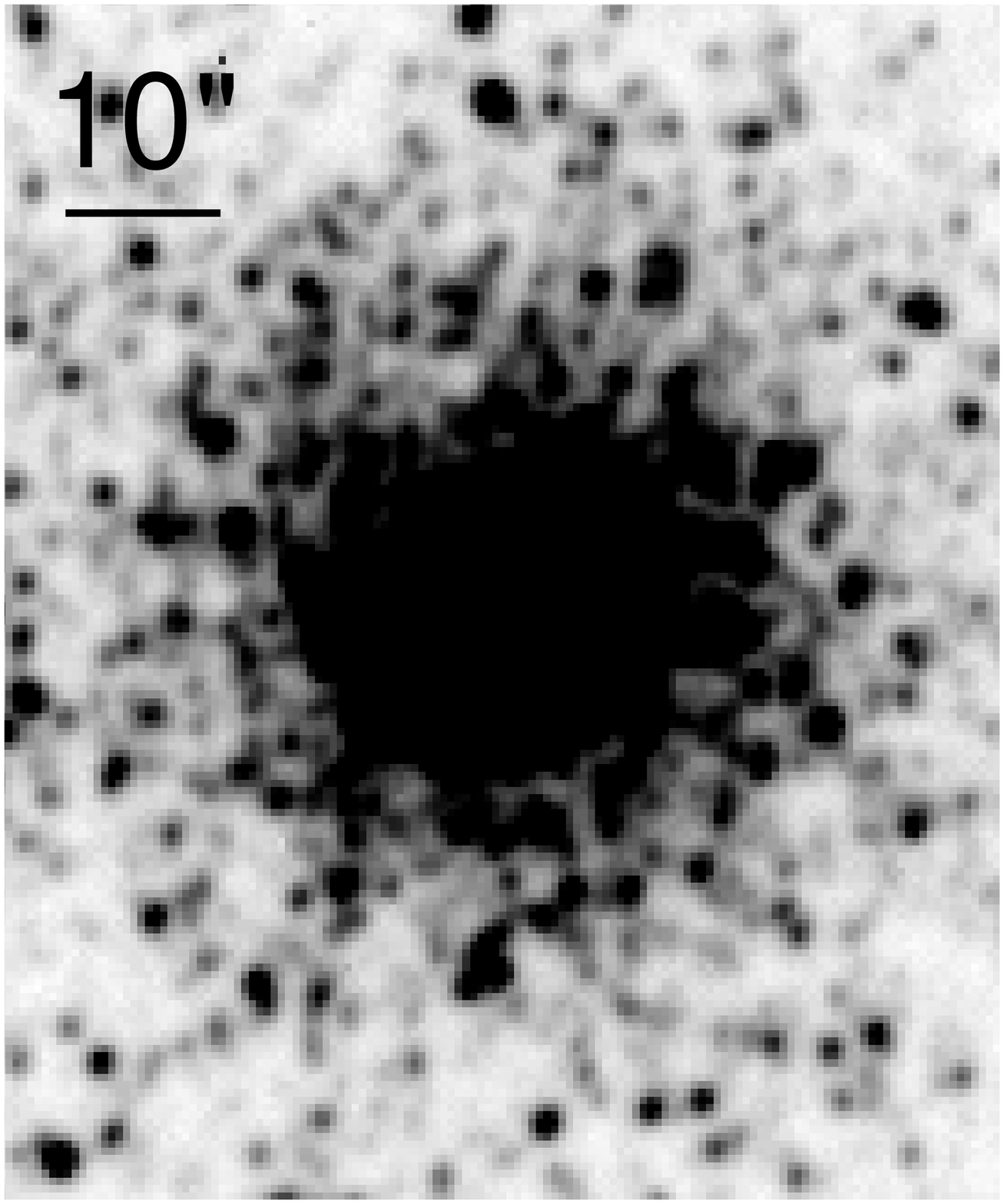}}
   \caption[]{Examples of CCD V band images of the sample clusters. Top panels (left to right):
   double clusters NGC\,241+242, IC\,1612 and NGC\,2011. Middle panels: merger
   candidates NGC\,376, K\,50 and NGC\,1810. Bottom panels: the young cluster
   NGC\,346 embedded in an HII region, and the populous clusters NGC\,416 and
   NGC\,1856. East to the left and North to the top.}
   \label{fig:img}
\end{figure}

Figure \ref{fig:img} shows examples of clusters of the present sample. The first
three (NGC\,241+242, IC\,1612, NGC\,2011) are double clusters (Sect.
\ref{sec:int}). The next three (NGC\,376, K\,50, NGC\,1810) are
candidates to mergers (Sect. \ref{sec:dis}). NGC\,346 is a young star cluster
embedded in an HII region. NGC\,416 is a populous intermediate age SMC cluster,
and NGC\,1856 is a populous blue LMC cluster.

\subsection{Data reduction}
\label{sec:red}

All frames were reduced using standard {\sc iraf} routines for bias and dark
current subtration and flat-field division. Bad and hot pixels (cosmic rays)
were replaced by linear interpolation along lines or columns using the nearest
good pixels.

Photometric calibration was performed with a least-squares fit routine that uses
errors as weights, with standard stars of \citet{Graham} and \citet{Land92},
resulting in a mean error smaller than 0.01\,mag at a 66\% confidence level. The
transformations used in the photometric calibration were
\[V = (0.998 \pm 0.006) v + (21.906 \pm 0.074){\rm ,}\]
for the first run, and
\[B = (1.008 \pm 0.013) b + (22.128 \pm 0.118){\rm ,}\]
\[V = (0.985 \pm 0.003) v + (22.304 \pm 0.034){\rm ,}\]
for the second run, where $b$ and $v$ are the instrumental magnitudes. The mean
errors are 0.008, 0.014, and 0.006\,mag, respectively.

\subsubsection{Cluster center determination}
\label{sec:cen}

Accurate cluster centers are fundamental, because errors will introduce
additional uncertainties in the SBPs. In Fig. \ref{fig:cen} we
present brightness profiles for an artificial image where we displaced the
center of the surface photometry by 5$''$, 10$''$, 15$''$ and 20$''$ of the
correct center, successively. It is possible to see that apparently acceptable
SBPs can be obtained, although the cluster center was not
determined correctly. However, resulting structural parameters especially $\mu_0$ present offsets.

\begin{figure}[!h]
   \begin{center}
    \caption[]{\small Example of the effect of the centering error of an object.
    The solid line represents the brightness distribution of an image
    artificially constructed. The points represented by crosses were obtained
    for the center of the surface photometry displaced by 5$''$ relatively to
    the center of the artificial image; squares by 10$''$; triangles by 15$''$
    and circles by 20$''$.}
    \label{fig:cen}
   \end{center}
\end{figure}

The coordinates of the symmetry center for each cluster have been obtained,
with a mean error smaller than 0.5$''$, using an implementation of the
mirror-autocorrelation algorithm (\citealt{Djorg88}). We obtained for each
cluster a set of autocorrelation amplitudes that have been fitted with an
elliptical paraboloid. The optimal center is the vertex of this paraboloid.

For some objects we determined the center by a heuristic method because of their
non-symmetrical stellar distribution as a whole, or presence of bright stars.
This was the case of NGC\,241+242, NGC\,339, K\,47, K\,50, NGC\,1810, and
NGC\,2004.

The derived cluster center positions in equatorial coordinates (J2000) are given
in Table \ref{tab:cen}.
\begin{table}[!h]
\begin{center}
\caption[]{Cluster center positions (J2000) determined in this study.}
\label{tab:cen}
\renewcommand{\tabcolsep}{2.5mm}
\renewcommand{\arraystretch}{1.0}
\begin{tabular}{lrrrr}
\hline
\hline
\multicolumn{1}{l}{Cluster} &\multicolumn{1}{c}{$\alpha$} &\multicolumn{1}{c}{$\sigma_{\alpha}$} &\multicolumn{1}{c}{$\delta$} &\multicolumn{1}{c}{$\sigma_{\delta}$}\\
\multicolumn{1}{l}{(1)} &\multicolumn{1}{c}{(2)} &\multicolumn{1}{c}{(3)} &\multicolumn{1}{c}{(4)} &\multicolumn{1}{c}{(5)}\\
\hline
\multicolumn{5}{c}{SMC} \\
\hline
NGC\,121  & $00^{\rm h}26^{\rm m}47^{\rm s}_{\upf{.}}89$ & $0^{\rm s}_{\upf{.}}01$ & $-71^{\circ}32'04''_{\upf{.}}8$ & $0''_{\upf{.}}4$ \\
NGC\,176  & $00^{\rm h}35^{\rm m}58^{\rm s}_{\upf{.}}18$ & $0^{\rm s}_{\upf{.}}03$ & $-73^{\circ}09'57''_{\upf{.}}7$ & $0''_{\upf{.}}5$ \\
K\,17     & $00^{\rm h}41^{\rm m}00^{\rm s}_{\upf{.}}82$ & $0^{\rm s}_{\upf{.}}02$ & $-72^{\circ}34'20''_{\upf{.}}8$ & $0''_{\upf{.}}3$ \\
NGC\,241  & $00^{\rm h}43^{\rm m}31^{\rm s}_{\upf{.}}61$ & $0^{\rm s}_{\upf{.}}03$ & $-73^{\circ}26'26''_{\upf{.}}8$ & $0''_{\upf{.}}5$ \\
NGC\,290  & $00^{\rm h}51^{\rm m}14^{\rm s}_{\upf{.}}91$ & $0^{\rm s}_{\upf{.}}02$ & $-73^{\circ}09'40''_{\upf{.}}5$ & $0''_{\upf{.}}4$ \\
L\,48     & $00^{\rm h}53^{\rm m}27^{\rm s}_{\upf{.}}60$ & $0^{\rm s}_{\upf{.}}02$ & $-71^{\circ}23'55''_{\upf{.}}1$ & $0''_{\upf{.}}3$ \\
K\,34     & $00^{\rm h}55^{\rm m}33^{\rm s}_{\upf{.}}15$ & $0^{\rm s}_{\upf{.}}01$ & $-72^{\circ}49'57''_{\upf{.}}4$ & $0''_{\upf{.}}1$ \\
NGC\,330  & $00^{\rm h}56^{\rm m}18^{\rm s}_{\upf{.}}30$ & $0^{\rm s}_{\upf{.}}01$ & $-72^{\circ}27'47''_{\upf{.}}9$ & $0''_{\upf{.}}3$ \\
L\,56     & $00^{\rm h}57^{\rm m}29^{\rm s}_{\upf{.}}82$ & $0^{\rm s}_{\upf{.}}03$ & $-72^{\circ}15'53''_{\upf{.}}1$ & $0''_{\upf{.}}6$ \\
NGC\,339  & $00^{\rm h}57^{\rm m}46^{\rm s}_{\upf{.}}04$ & $0^{\rm s}_{\upf{.}}03$ & $-74^{\circ}28'13''_{\upf{.}}2$ & $0''_{\upf{.}}5$ \\
NGC\,346  & $00^{\rm h}59^{\rm m}05^{\rm s}_{\upf{.}}18$ & $0^{\rm s}_{\upf{.}}04$ & $-72^{\circ}10'38''_{\upf{.}}0$ & $0''_{\upf{.}}4$ \\
IC\,1611  & $00^{\rm h}59^{\rm m}48^{\rm s}_{\upf{.}}08$ & $0^{\rm s}_{\upf{.}}02$ & $-72^{\circ}20'03''_{\upf{.}}4$ & $0''_{\upf{.}}4$ \\
IC\,1612  & $00^{\rm h}59^{\rm m}55^{\rm s}_{\upf{.}}31$ & $0^{\rm s}_{\upf{.}}01$ & $-72^{\circ}22'18''_{\upf{.}}6$ & $0''_{\upf{.}}1$ \\
L\,66     & $01^{\rm h}01^{\rm m}44^{\rm s}_{\upf{.}}44$ & $0^{\rm s}_{\upf{.}}05$ & $-72^{\circ}33'51''_{\upf{.}}9$ & $0''_{\upf{.}}5$ \\
NGC\,361  & $01^{\rm h}02^{\rm m}10^{\rm s}_{\upf{.}}48$ & $0^{\rm s}_{\upf{.}}03$ & $-71^{\circ}36'23''_{\upf{.}}3$ & $0''_{\upf{.}}5$ \\
K\,47     & $01^{\rm h}03^{\rm m}11^{\rm s}_{\upf{.}}34$ & $0^{\rm s}_{\upf{.}}03$ & $-72^{\circ}16'18''_{\upf{.}}7$ & $0''_{\upf{.}}5$ \\
NGC\,376  & $01^{\rm h}03^{\rm m}53^{\rm s}_{\upf{.}}63$ & $0^{\rm s}_{\upf{.}}01$ & $-72^{\circ}49'32''_{\upf{.}}9$ & $0''_{\upf{.}}2$ \\
K\,50     & $01^{\rm h}04^{\rm m}37^{\rm s}_{\upf{.}}03$ & $0^{\rm s}_{\upf{.}}03$ & $-72^{\circ}09'39''_{\upf{.}}7$ & $1''_{\upf{.}}5$ \\
IC\,1624  & $01^{\rm h}05^{\rm m}20^{\rm s}_{\upf{.}}92$ & $0^{\rm s}_{\upf{.}}03$ & $-72^{\circ}02'37''_{\upf{.}}3$ & $0''_{\upf{.}}3$ \\
K\,54     & $01^{\rm h}06^{\rm m}47^{\rm s}_{\upf{.}}99$ & $0^{\rm s}_{\upf{.}}04$ & $-72^{\circ}16'23''_{\upf{.}}9$ & $0''_{\upf{.}}8$ \\
NGC\,411  & $01^{\rm h}07^{\rm m}54^{\rm s}_{\upf{.}}34$ & $0^{\rm s}_{\upf{.}}03$ & $-71^{\circ}46'05''_{\upf{.}}6$ & $0''_{\upf{.}}5$ \\
NGC\,416  & $01^{\rm h}07^{\rm m}59^{\rm s}_{\upf{.}}18$ & $0^{\rm s}_{\upf{.}}02$ & $-72^{\circ}21'19''_{\upf{.}}8$ & $0''_{\upf{.}}2$ \\
NGC\,419  & $01^{\rm h}08^{\rm m}17^{\rm s}_{\upf{.}}39$ & $0^{\rm s}_{\upf{.}}01$ & $-72^{\circ}53'00''_{\upf{.}}7$ & $0''_{\upf{.}}1$ \\
NGC\,458  & $01^{\rm h}14^{\rm m}52^{\rm s}_{\upf{.}}29$ & $0^{\rm s}_{\upf{.}}01$ & $-71^{\circ}33'00''_{\upf{.}}2$ & $0''_{\upf{.}}3$ \\
L\,114    & $01^{\rm h}50^{\rm m}19^{\rm s}_{\upf{.}}63$ & $0^{\rm s}_{\upf{.}}02$ & $-74^{\circ}21'23''_{\upf{.}}9$ & $0''_{\upf{.}}1$ \\
\hline
\multicolumn{5}{c}{LMC} \\
\hline
NGC\,1783 & $04^{\rm h}59^{\rm m}08^{\rm s}_{\upf{.}}78$ & $0^{\rm s}_{\upf{.}}03$ & $-65^{\circ}59'17''_{\upf{.}}1$ & $0''_{\upf{.}}3$ \\
NGC\,1810 & $05^{\rm h}03^{\rm m}23^{\rm s}_{\upf{.}}31$ & $0^{\rm s}_{\upf{.}}03$ & $-66^{\circ}22'57''_{\upf{.}}3$ & $0''_{\upf{.}}5$ \\
NGC\,1818 & $05^{\rm h}04^{\rm m}13^{\rm s}_{\upf{.}}92$ & $0^{\rm s}_{\upf{.}}01$ & $-66^{\circ}26'03''_{\upf{.}}4$ & $0''_{\upf{.}}2$ \\
NGC\,1831 & $05^{\rm h}06^{\rm m}16^{\rm s}_{\upf{.}}17$ & $0^{\rm s}_{\upf{.}}01$ & $-64^{\circ}55'10''_{\upf{.}}1$ & $0''_{\upf{.}}2$ \\
NGC\,1847 & $05^{\rm h}07^{\rm m}08^{\rm s}_{\upf{.}}16$ & $0^{\rm s}_{\upf{.}}01$ & $-68^{\circ}58'22''_{\upf{.}}6$ & $0''_{\upf{.}}1$ \\
NGC\,1856 & $05^{\rm h}09^{\rm m}30^{\rm s}_{\upf{.}}32$ & $0^{\rm s}_{\upf{.}}01$ & $-69^{\circ}07'44''_{\upf{.}}2$ & $0''_{\upf{.}}2$ \\
NGC\,1866 & $05^{\rm h}13^{\rm m}38^{\rm s}_{\upf{.}}82$ & $0^{\rm s}_{\upf{.}}03$ & $-65^{\circ}27'55''_{\upf{.}}7$ & $0''_{\upf{.}}2$ \\
NGC\,1868 & $05^{\rm h}14^{\rm m}36^{\rm s}_{\upf{.}}09$ & $0^{\rm s}_{\upf{.}}03$ & $-63^{\circ}57'14''_{\upf{.}}8$ & $0''_{\upf{.}}1$ \\
NGC\,1870 & $05^{\rm h}13^{\rm m}10^{\rm s}_{\upf{.}}71$ & $0^{\rm s}_{\upf{.}}01$ & $-69^{\circ}07'04''_{\upf{.}}0$ & $0''_{\upf{.}}1$ \\
NGC\,1978 & $05^{\rm h}28^{\rm m}44^{\rm s}_{\upf{.}}99$ & $0^{\rm s}_{\upf{.}}02$ & $-66^{\circ}14'10''_{\upf{.}}5$ & $0''_{\upf{.}}4$ \\
NGC\,2004 & $05^{\rm h}30^{\rm m}40^{\rm s}_{\upf{.}}34$ & $0^{\rm s}_{\upf{.}}03$ & $-67^{\circ}17'12''_{\upf{.}}8$ & $0''_{\upf{.}}5$ \\
NGC\,2011 & $05^{\rm h}32^{\rm m}19^{\rm s}_{\upf{.}}50$ & $0^{\rm s}_{\upf{.}}01$ & $-67^{\circ}31'19''_{\upf{.}}8$ & $0''_{\upf{.}}2$ \\
NGC\,2100 & $05^{\rm h}42^{\rm m}07^{\rm s}_{\upf{.}}82$ & $0^{\rm s}_{\upf{.}}02$ & $-69^{\circ}12'40''_{\upf{.}}9$ & $0''_{\upf{.}}3$ \\
NGC\,2121 & $05^{\rm h}48^{\rm m}12^{\rm s}_{\upf{.}}71$ & $0^{\rm s}_{\upf{.}}03$ & $-71^{\circ}28'51''_{\upf{.}}7$ & $0''_{\upf{.}}7$ \\
NGC\,2157 & $05^{\rm h}57^{\rm m}35^{\rm s}_{\upf{.}}32$ & $0^{\rm s}_{\upf{.}}01$ & $-69^{\circ}11'47''_{\upf{.}}0$ & $0''_{\upf{.}}1$ \\
NGC\,2159 & $05^{\rm h}58^{\rm m}03^{\rm s}_{\upf{.}}03$ & $0^{\rm s}_{\upf{.}}03$ & $-68^{\circ}37'30''_{\upf{.}}1$ & $0''_{\upf{.}}5$ \\
NGC\,2164 & $05^{\rm h}58^{\rm m}55^{\rm s}_{\upf{.}}88$ & $0^{\rm s}_{\upf{.}}01$ & $-68^{\circ}30'58''_{\upf{.}}3$ & $0''_{\upf{.}}4$ \\
NGC\,2210 & $06^{\rm h}11^{\rm m}31^{\rm s}_{\upf{.}}09$ & $0^{\rm s}_{\upf{.}}01$ & $-69^{\circ}07'20''_{\upf{.}}5$ & $0''_{\upf{.}}2$ \\
NGC\,2213 & $06^{\rm h}10^{\rm m}42^{\rm s}_{\upf{.}}33$ & $0^{\rm s}_{\upf{.}}01$ & $-71^{\circ}31'45''_{\upf{.}}7$ & $0''_{\upf{.}}1$ \\
NGC\,2214 & $06^{\rm h}12^{\rm m}57^{\rm s}_{\upf{.}}24$ & $0^{\rm s}_{\upf{.}}03$ & $-68^{\circ}15'38''_{\upf{.}}1$ & $0''_{\upf{.}}5$ \\
H\,11     & $06^{\rm h}14^{\rm m}22^{\rm s}_{\upf{.}}72$ & $0^{\rm s}_{\upf{.}}01$ & $-69^{\circ}50'50''_{\upf{.}}5$ & $0''_{\upf{.}}2$ \\
HS\,314   & $05^{\rm h}28^{\rm m}26^{\rm s}_{\upf{.}}68$ & $0^{\rm s}_{\upf{.}}04$ & $-68^{\circ}58'56''_{\upf{.}}3$ & $0''_{\upf{.}}5$ \\
\hline
\end{tabular}
\begin{list}{Table Notes.}
\item Col. 2: Right ascension. Col. 3: Standard deviation. Col. 4: Declination.
Col. 5: Standard deviation.
\end{list}
\end{center}
\end{table}

\subsection{Surface photometry}
\label{sec:pho}
Surface photometry was performed with concentric annular apertures centered on
the cluster coordinates (Table \ref{tab:cen}) and subdivided in at least 4,
and at most, 32 sectors. An effective radius for each annulus was determined
iteractively by following the radial distribution of brightness for each annulus. The
counts in each anullus were determined and adopted as surface-brightness,
together with its standard deviation. The surface-brightness for each annulus is
the mean or median of the values measured for all sectors together with standard
deviation. In col. 6 of Table \ref{tab:par} we give the adopted statistics.
The approach was based on the median. However, for convergence arguments, in
some cases we adopted the mean. In general, the median was more suitable for red
clusters. A limiting radius of the photometry for each cluster was estimated, so
that the rings did not exceed the physical limits of the image.

For each profile, the points are the result of four sets of anulli with
radial steps of 1.5$''$, 2.0$''$, 3.0$''$ and 4.0$''$. Four anullus sets are
plotted in the same axes with the best-fitting EFF profile. This is particularly
useful to build SBPs with an adequate spatial resolution in the inner regions of
the cluster. Conversely, large steps are more suitable for the outer regions. We
find that the simultaneous use of four steps yields comparable average
profiles.

For each cluster we tested if the mean or the median of the flux in the sectors
provided profiles with minimal errors in the surface-brightness and resulting
fits. In the luminosity and mass estimates this choice was also important.

The sky level was evaluated considering four peripheral regions in each image
that were not affected by the background and field stars. These regions have
been merged into a single one, of larger dimension, to compute a self-
consistent histogram of sky counts. They were truncated at 3$\sigma$ of the
average until the histogram did not vary anymore. Finally, we chose the mode of
this distribution as the value of the brightness of the sky and subtracted it
from the surface-brightness of the clusters.

Similarly to \citet{EFF87}, the young LMC clusters dealt with in this work do
not appear to be tidally truncated. Indeed, their profiles are well
represented by the following model

\begin{equation}
   \mu(r) = \mu_0 + 1.25 \gamma \log\bigg( 1 + \frac{r^2}{a^2}\bigg),
   \label{eq:eff}
\end{equation}
where $\mu_0$ is the central surface-brightness in magnitude scale, $\gamma$ is
a dimensionless power-law and $a$ is a parameter that is related to the core
radius ($r_c$) in arcsecs by

\begin{equation}
   r_c = a \sqrt{2^{2/\gamma} - 1}.
   \label{eq:rc}
\end{equation}

Fits of Eq. \ref{eq:eff} were performed with a nonlinear least-squares routine
that uses SBP errors as weights. For most of the sample this procedure
converged, and the resulting fit parameters are given in Table \ref{tab:par}.
Exceptions are NGC\,376, K\,50, K\,54 and NGC\,1810, for which there was no
convergence.

\subsubsection{Surface brightness profiles}
\label{sec:sbp}

Surface brightness profiles in magnitude scale for the 47 clusters in the sample
are shown in Fig. \ref{fig:1}. The maximum fit radius ($r_m$) is indicated
for each object and these values are listed in Table \ref{tab:lum}. The clusters
NGC\,376, K\,50, K\,54, NGC\,1810, could not be fitted by Eq. \ref{eq:eff}. All
profiles were obtained in the V band, except NGC 2159, for which a B image was used.

The variations seen in the profiles do not appear to depend on cluster bright
stars, which arise in different age ranges, nor on bright field stars. In order
to test this, we cleaned the images of e.g. NGC\,330 and NGC\,2004 of
supergiants, and NGC\,2121 of AGB stars. The original and the clean profiles are
essentially the same providing the same parameters, within uncertainties. No
bright field star is superimposed on the present clusters.

In addition, since we are dealing with populous clusters, statistical
uncertainties are small.

\begin{figure*}
   \centering
   \includegraphics[scale=0.30,viewport=0 0 550 725,clip]{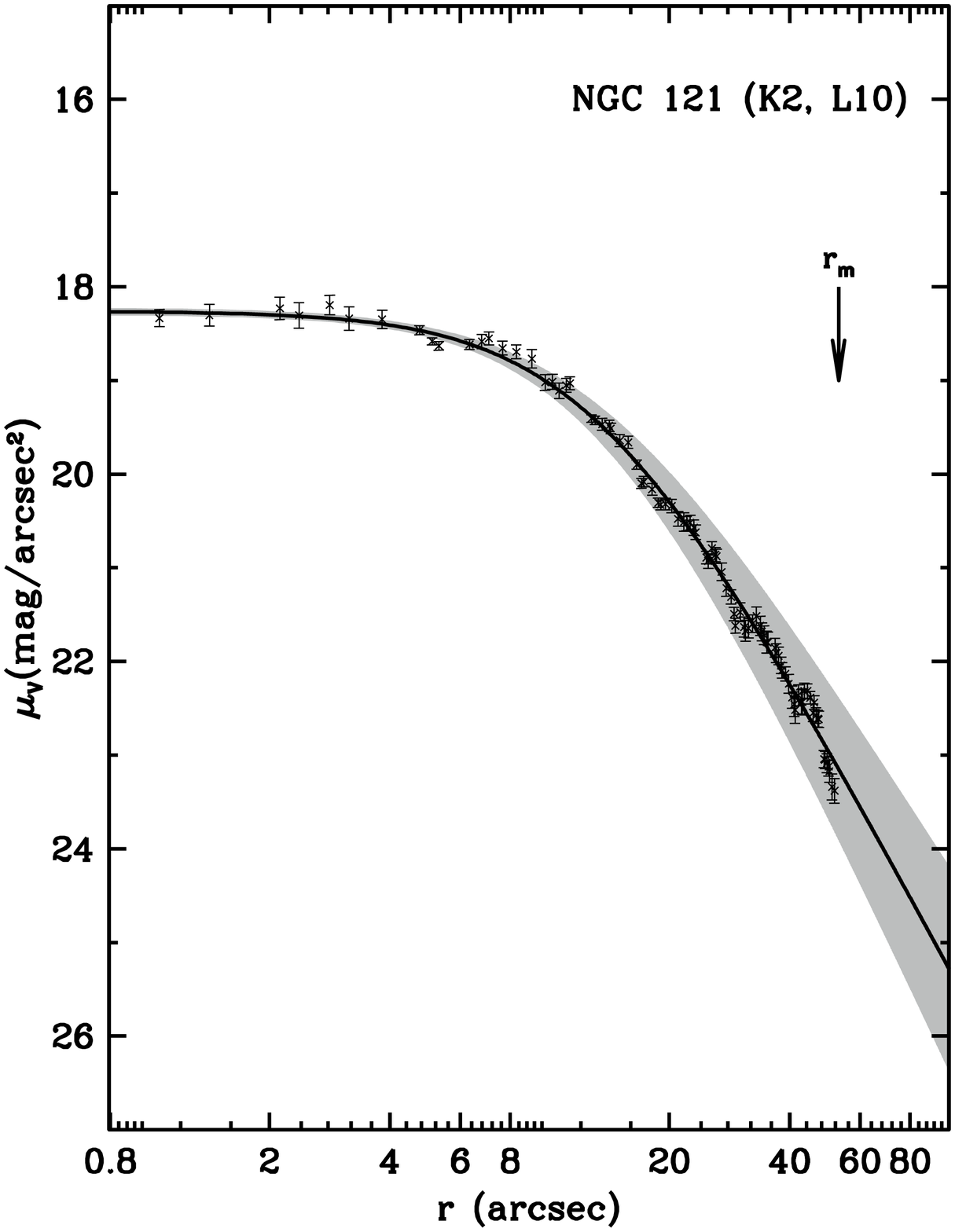}
   \includegraphics[scale=0.30,viewport=0 0 550 725,clip]{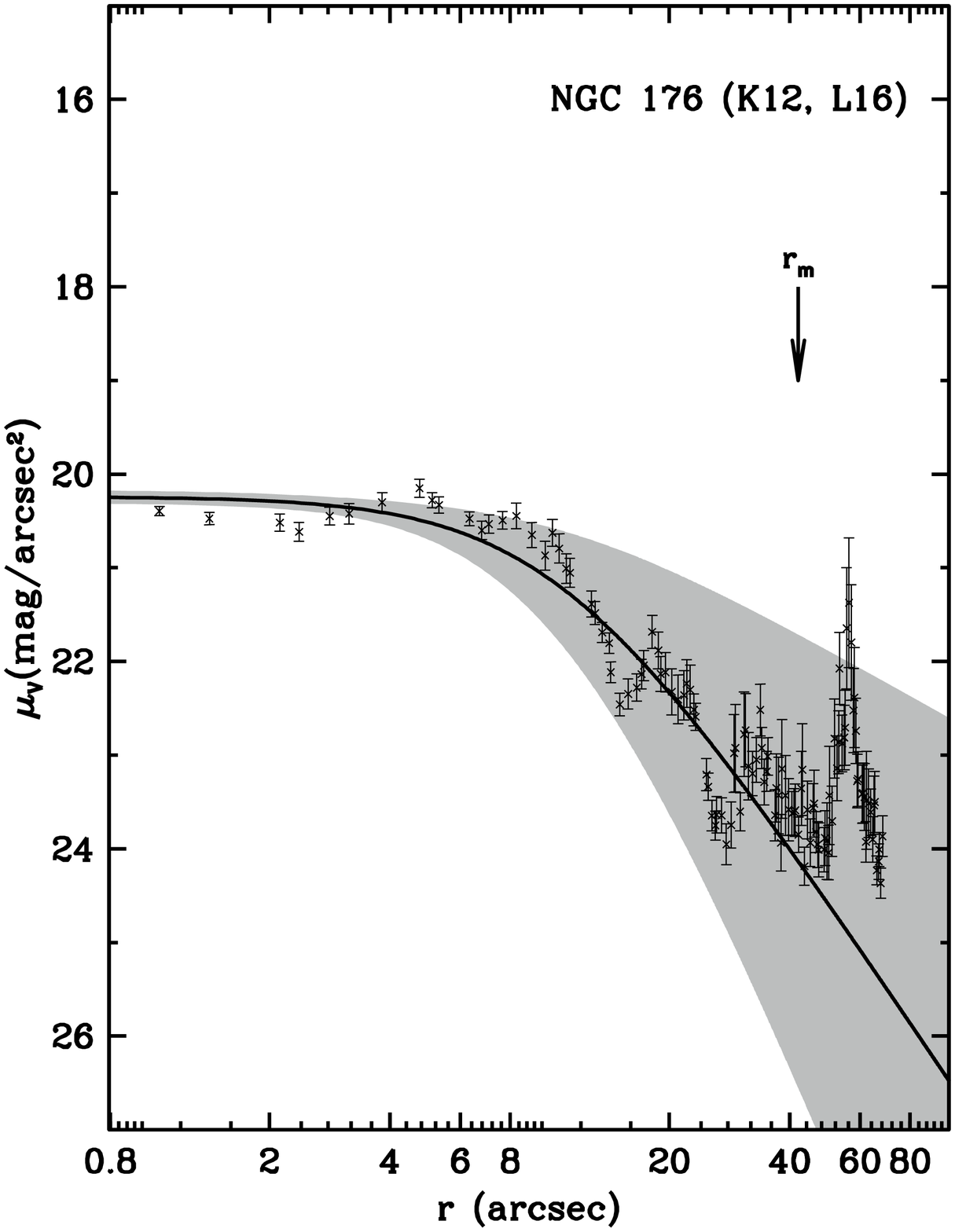}
   \includegraphics[scale=0.30,viewport=0 0 550 725,clip]{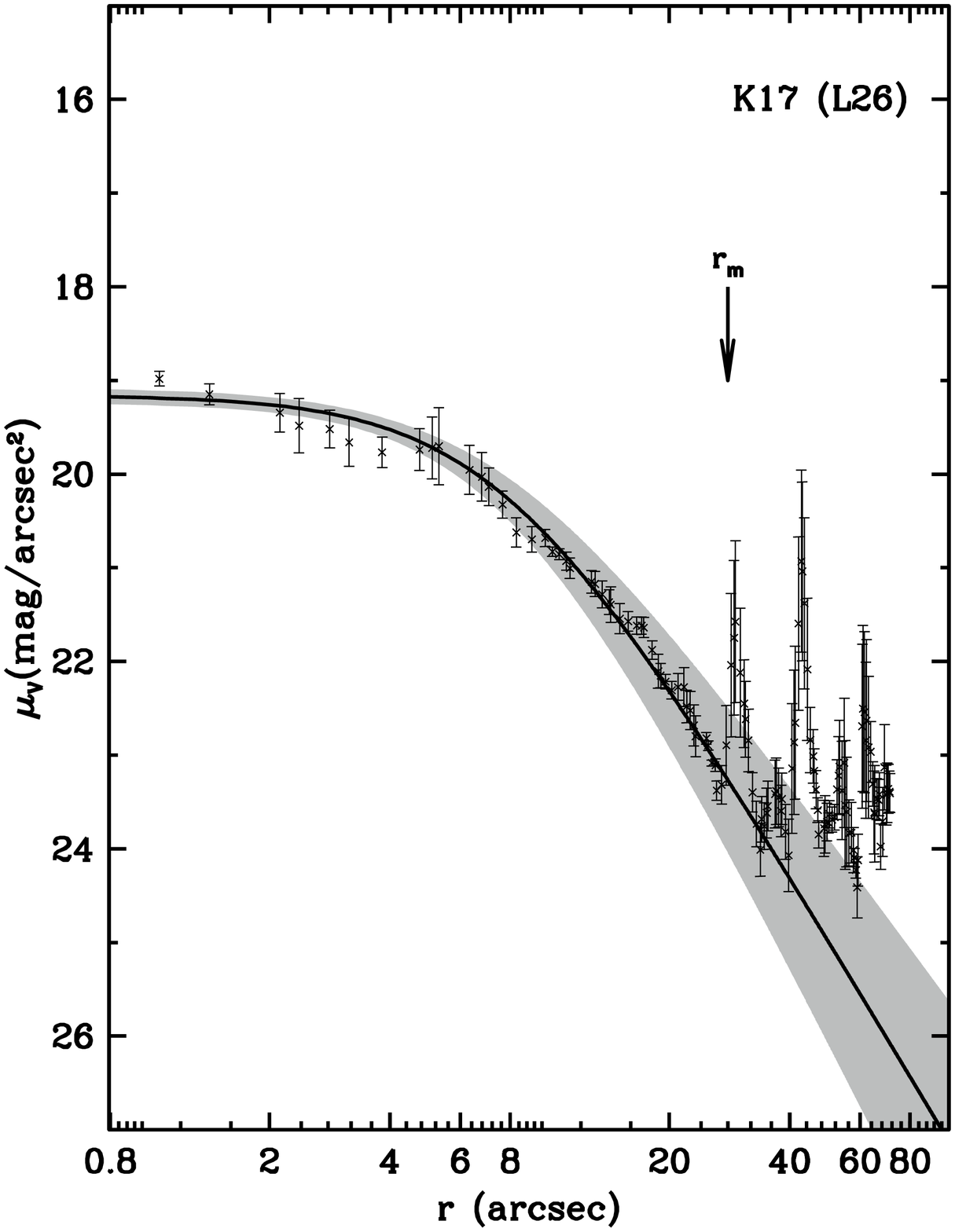}
   \includegraphics[scale=0.30,viewport=0 0 550 725,clip]{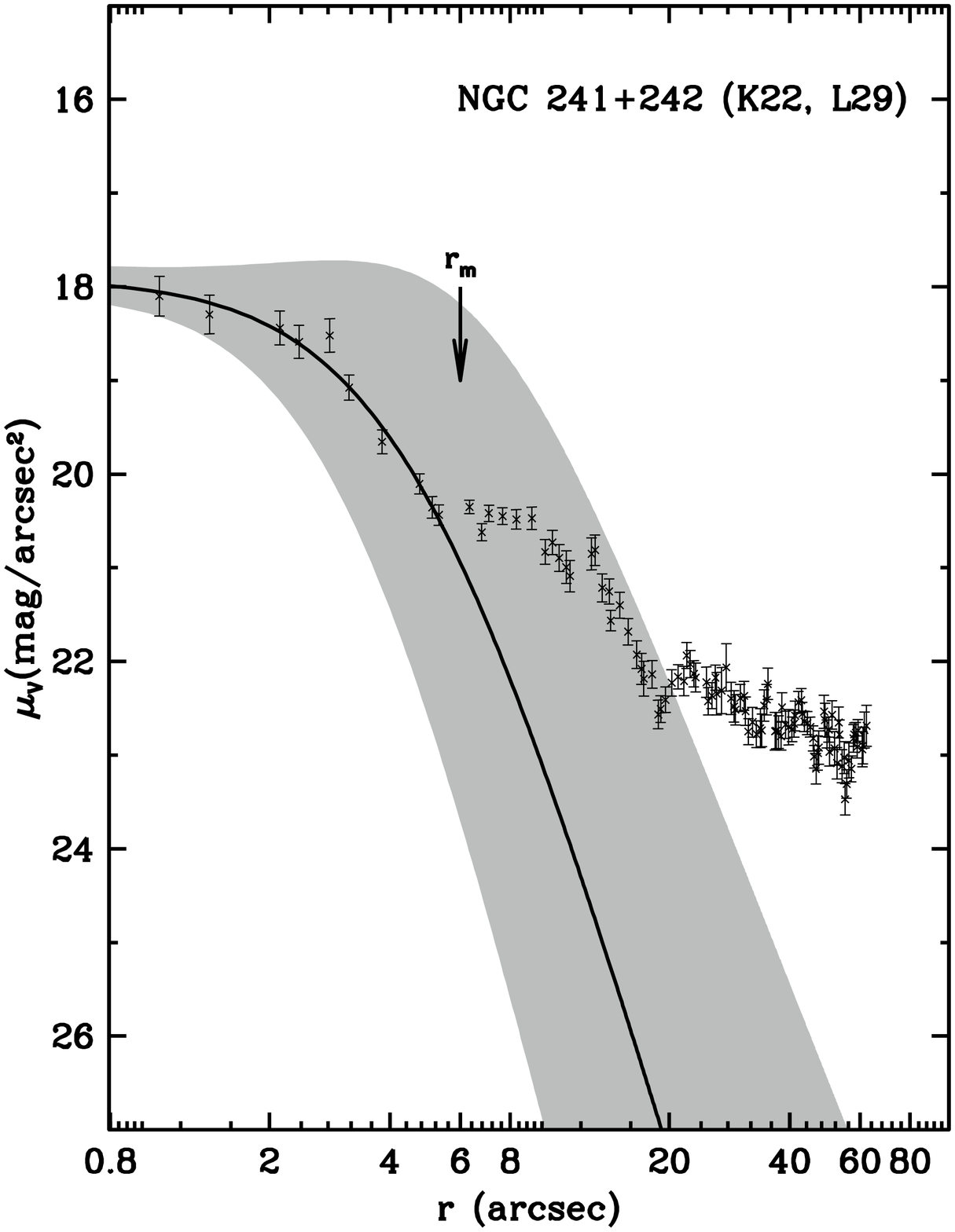}
   \includegraphics[scale=0.30,viewport=0 0 550 725,clip]{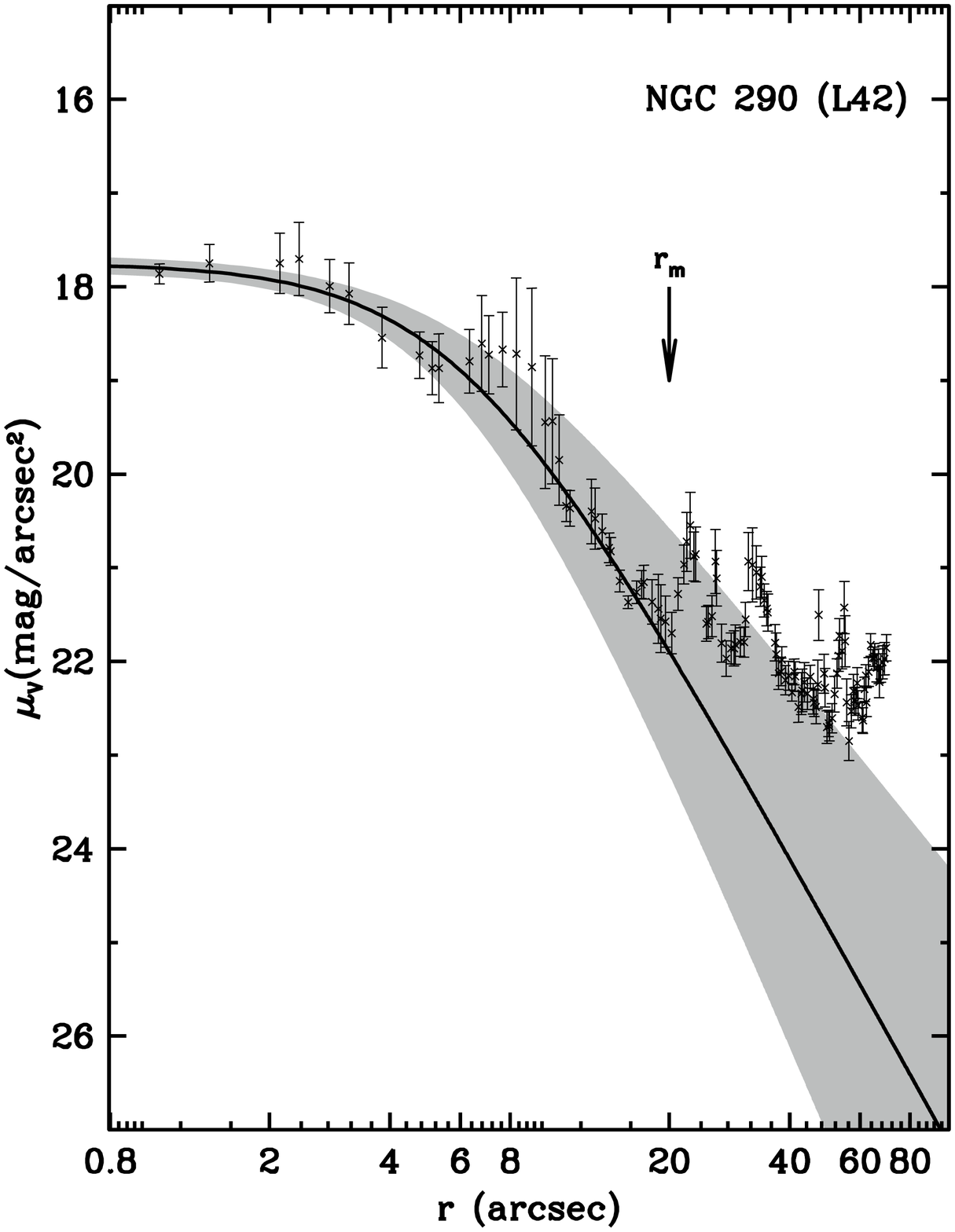}
   \includegraphics[scale=0.30,viewport=0 0 550 725,clip]{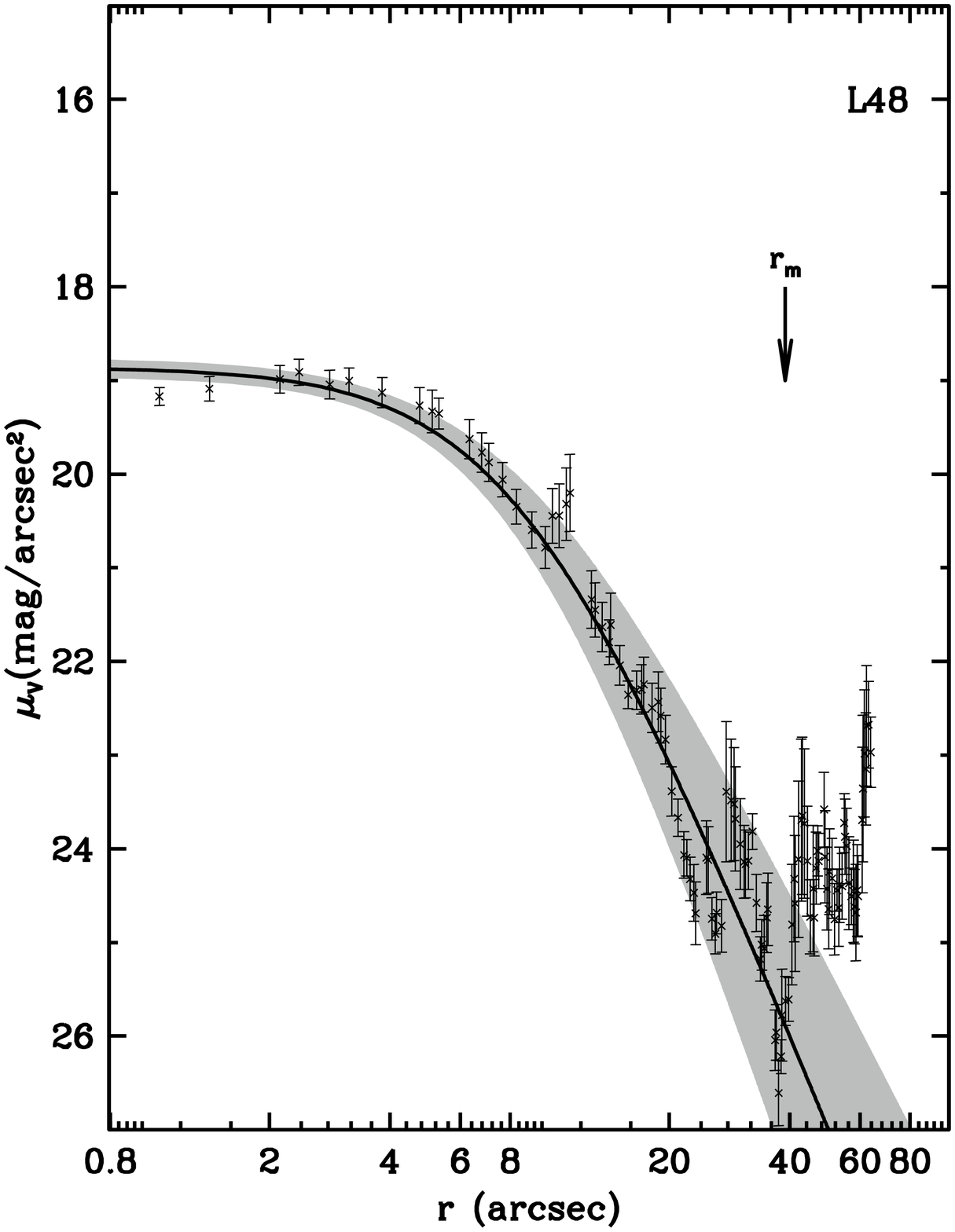}
   \includegraphics[scale=0.30,viewport=0 0 550 725,clip]{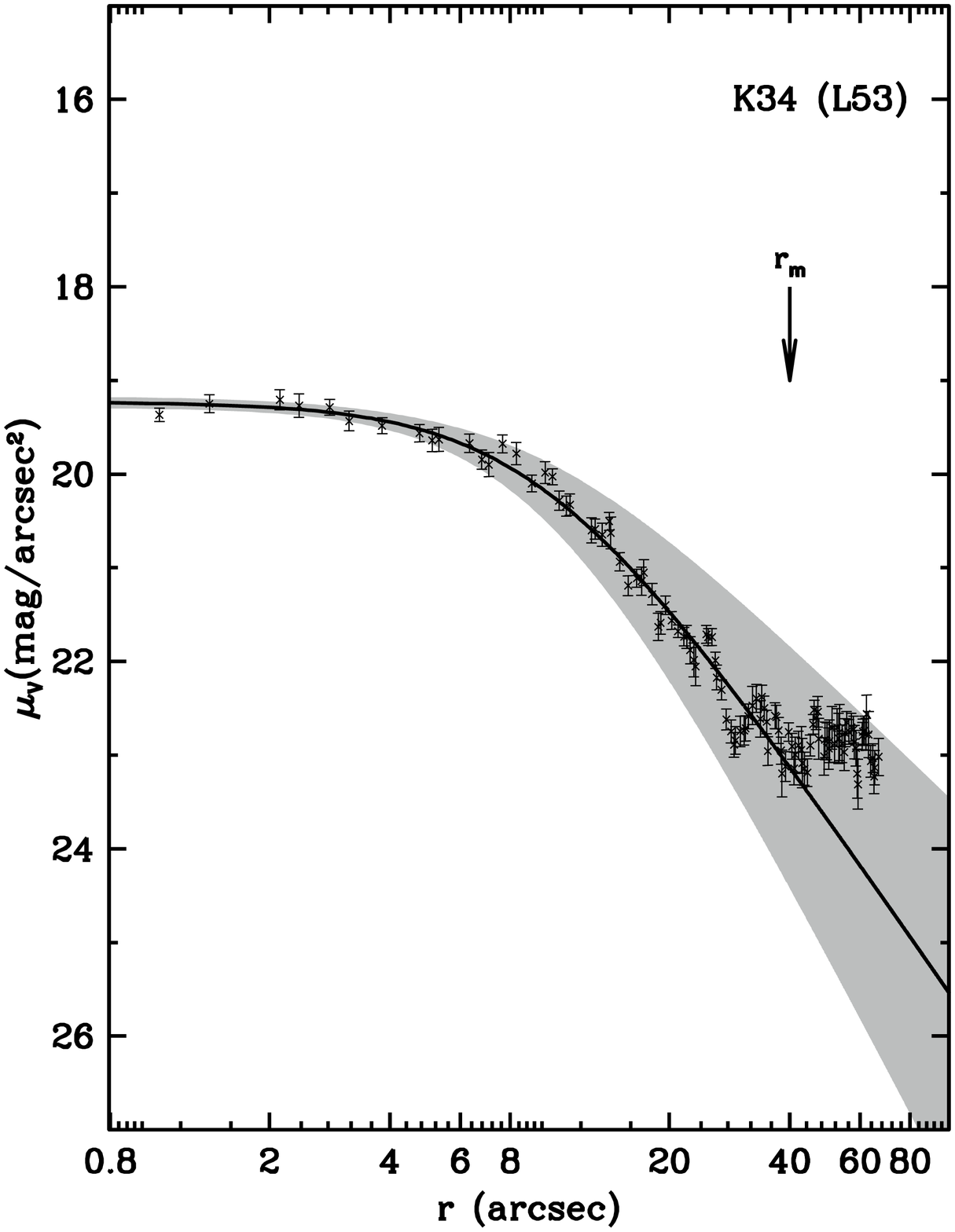}
   \includegraphics[scale=0.30,viewport=0 0 550 725,clip]{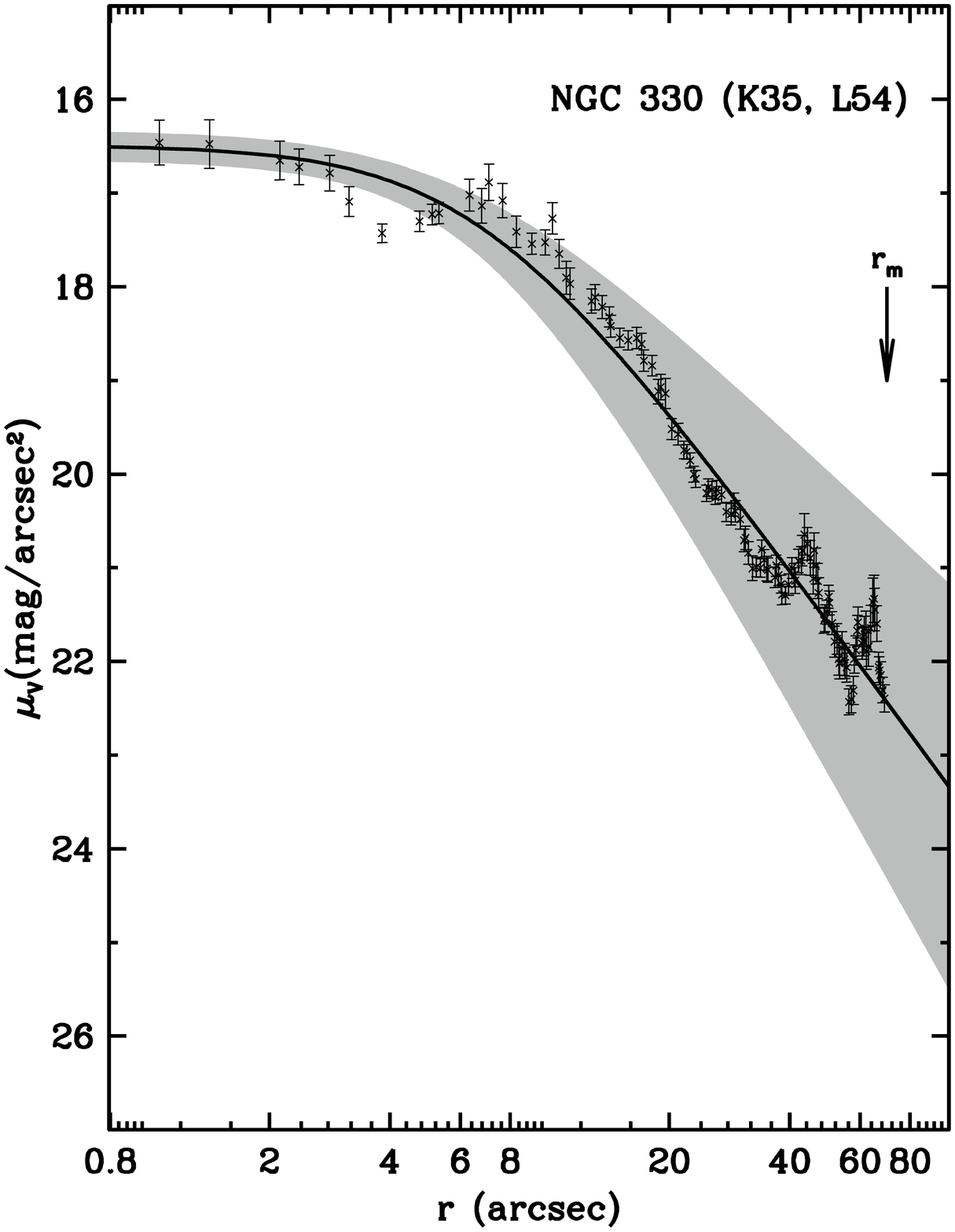}
   \includegraphics[scale=0.30,viewport=0 0 550 725,clip]{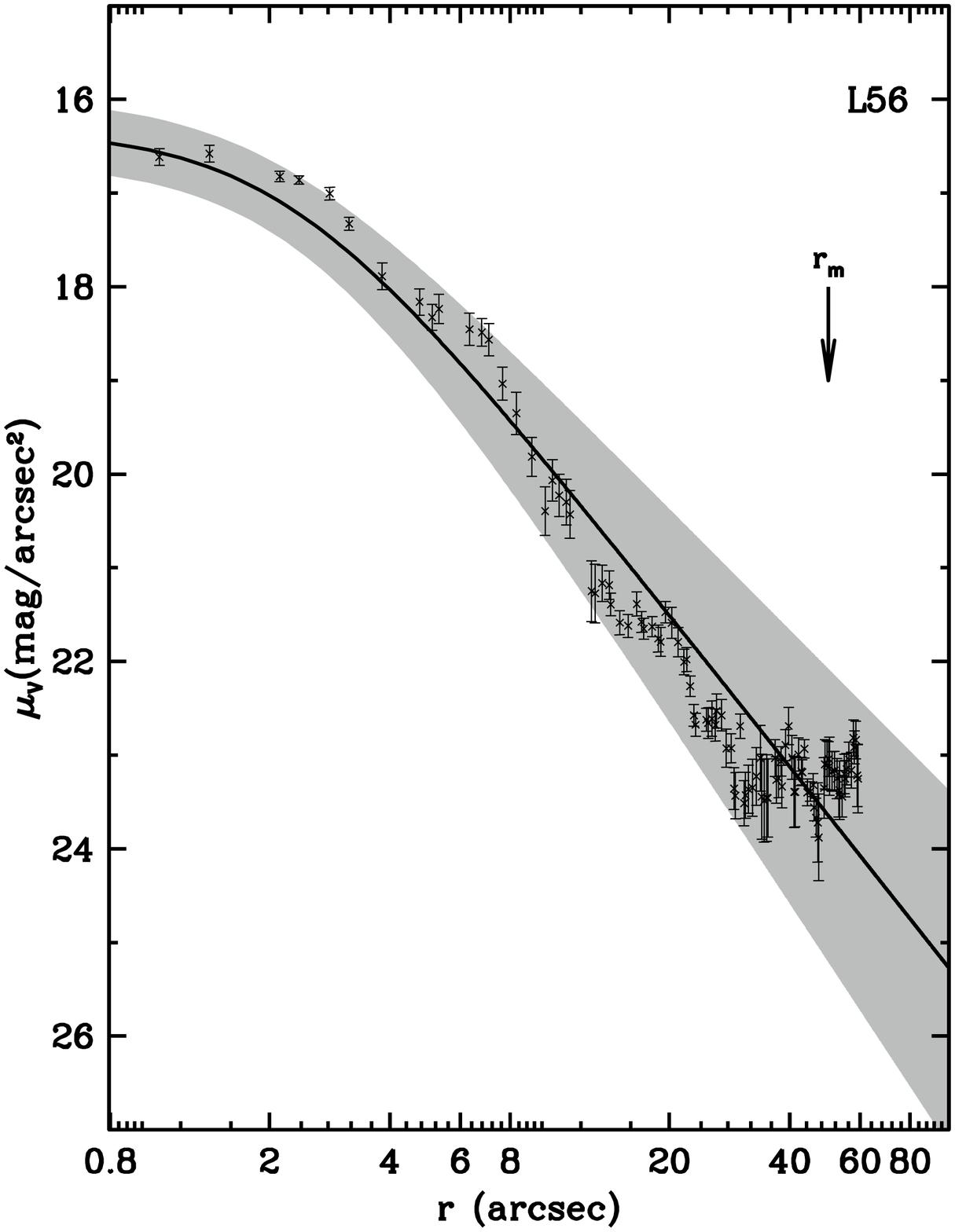}
   \caption[]{Surface brightness profiles for the star clusters in the sample.
   The solid line represents the best fitting EFF model. The shaded line
   represents the uncertainty of the fit. For each cluster the maximum fitting
   radius ($r_m$) is indicated.}
   \label{fig:1}
\end{figure*}

\begin{figure*}
\addtocounter{figure}{-1}
   \centering
   \includegraphics[scale=0.30,viewport=0 0 550 725,clip]{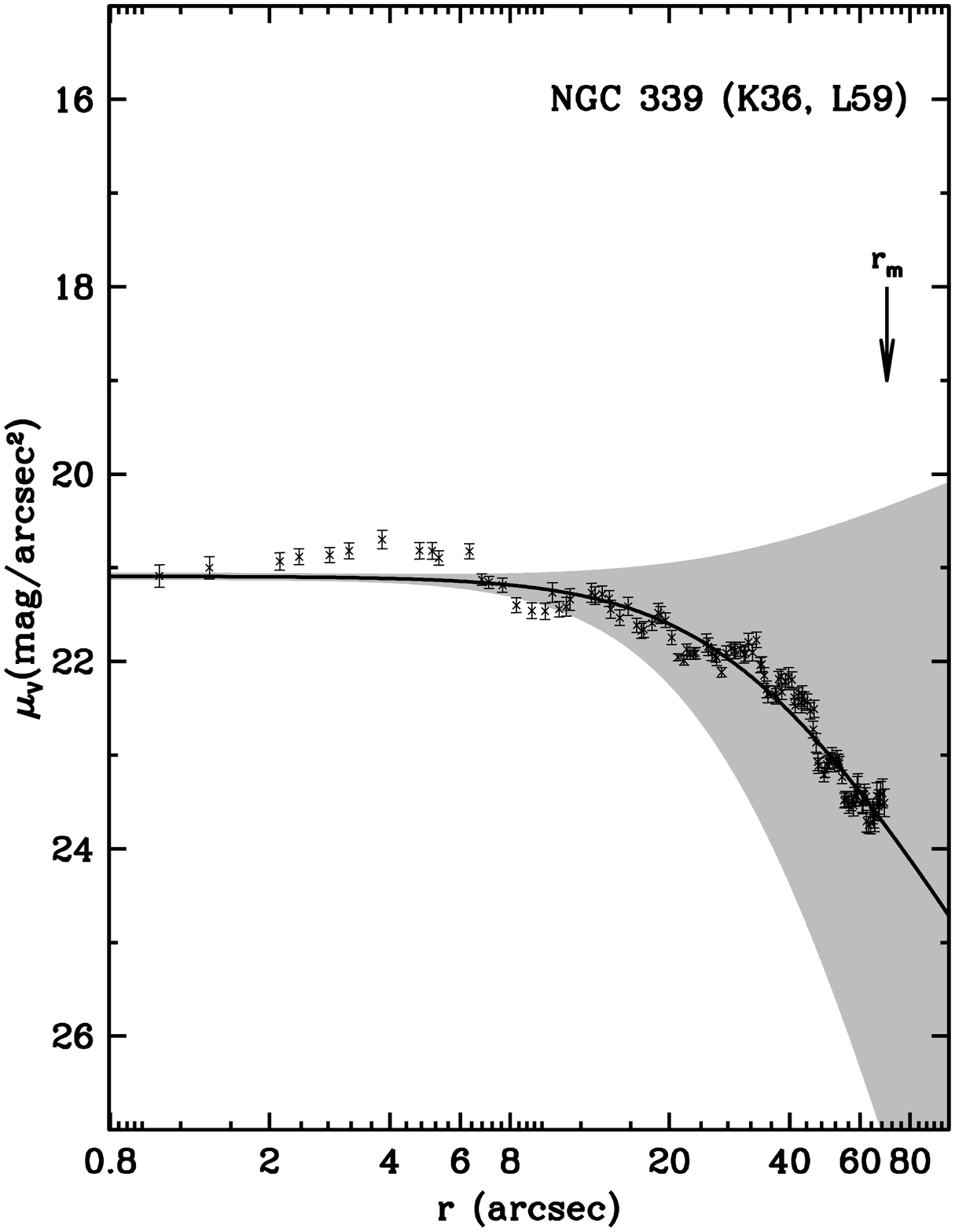}
   \includegraphics[scale=0.30,viewport=0 0 550 725,clip]{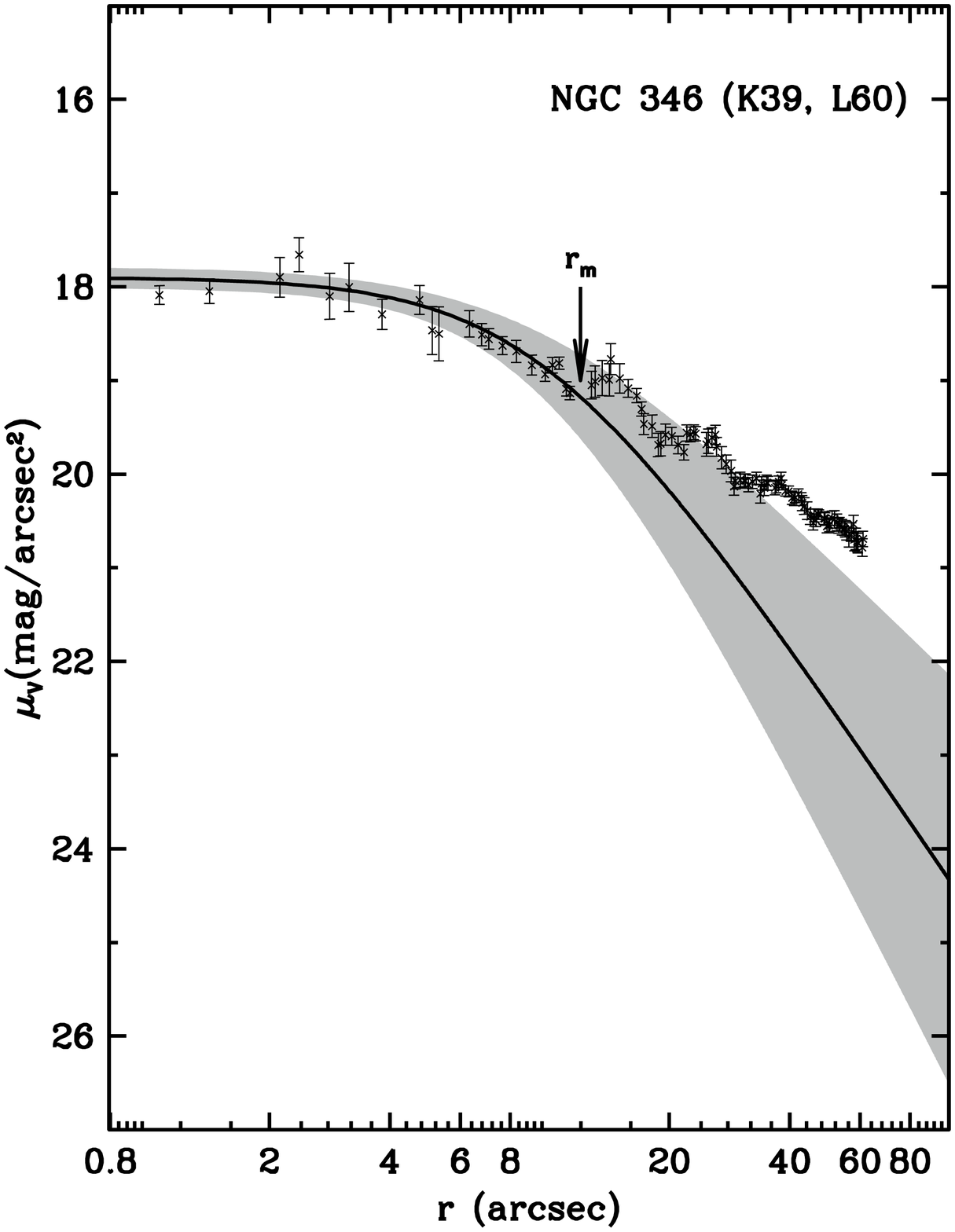}
   \includegraphics[scale=0.30,viewport=0 0 550 725,clip]{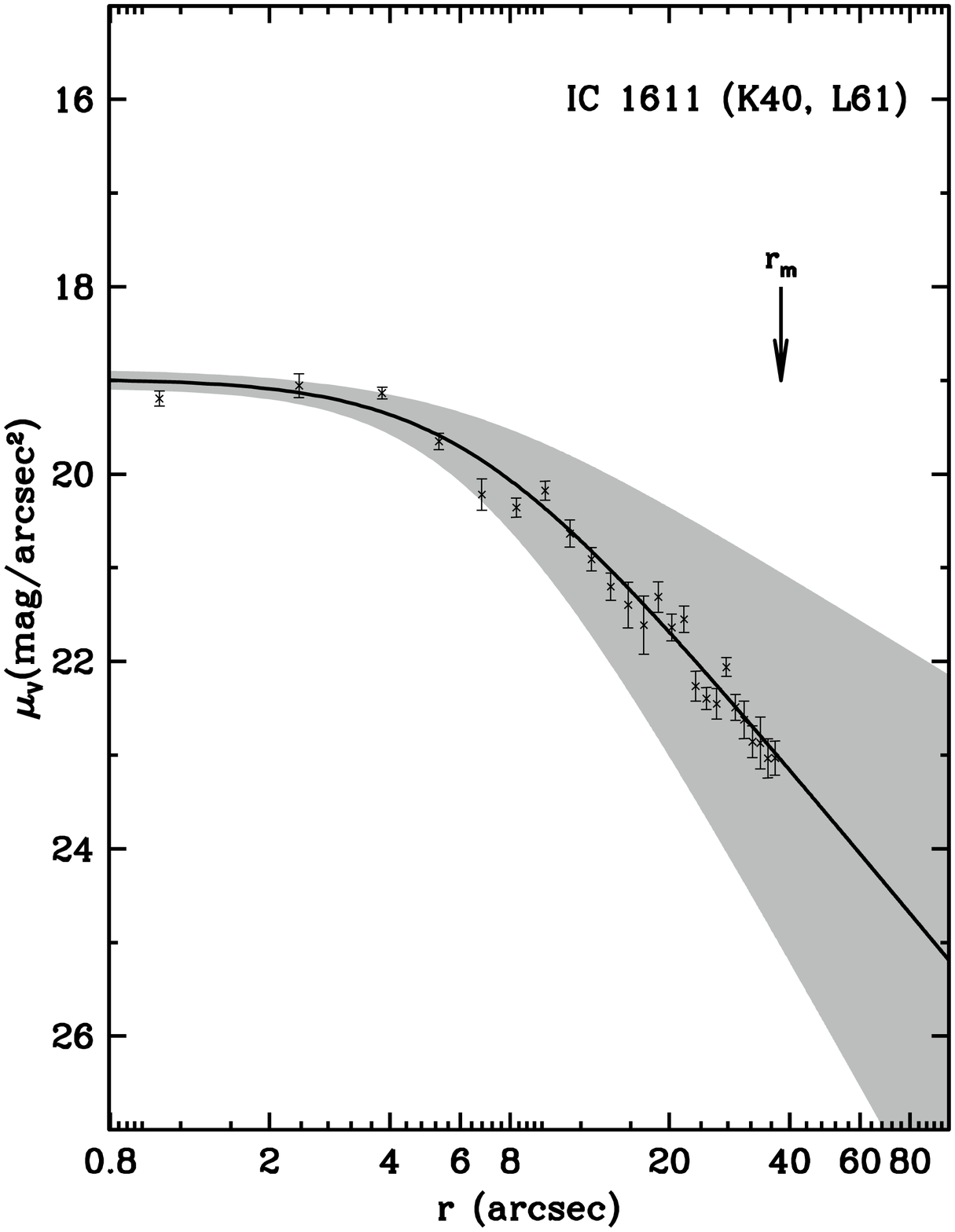}
   \includegraphics[scale=0.30,viewport=0 0 550 725,clip]{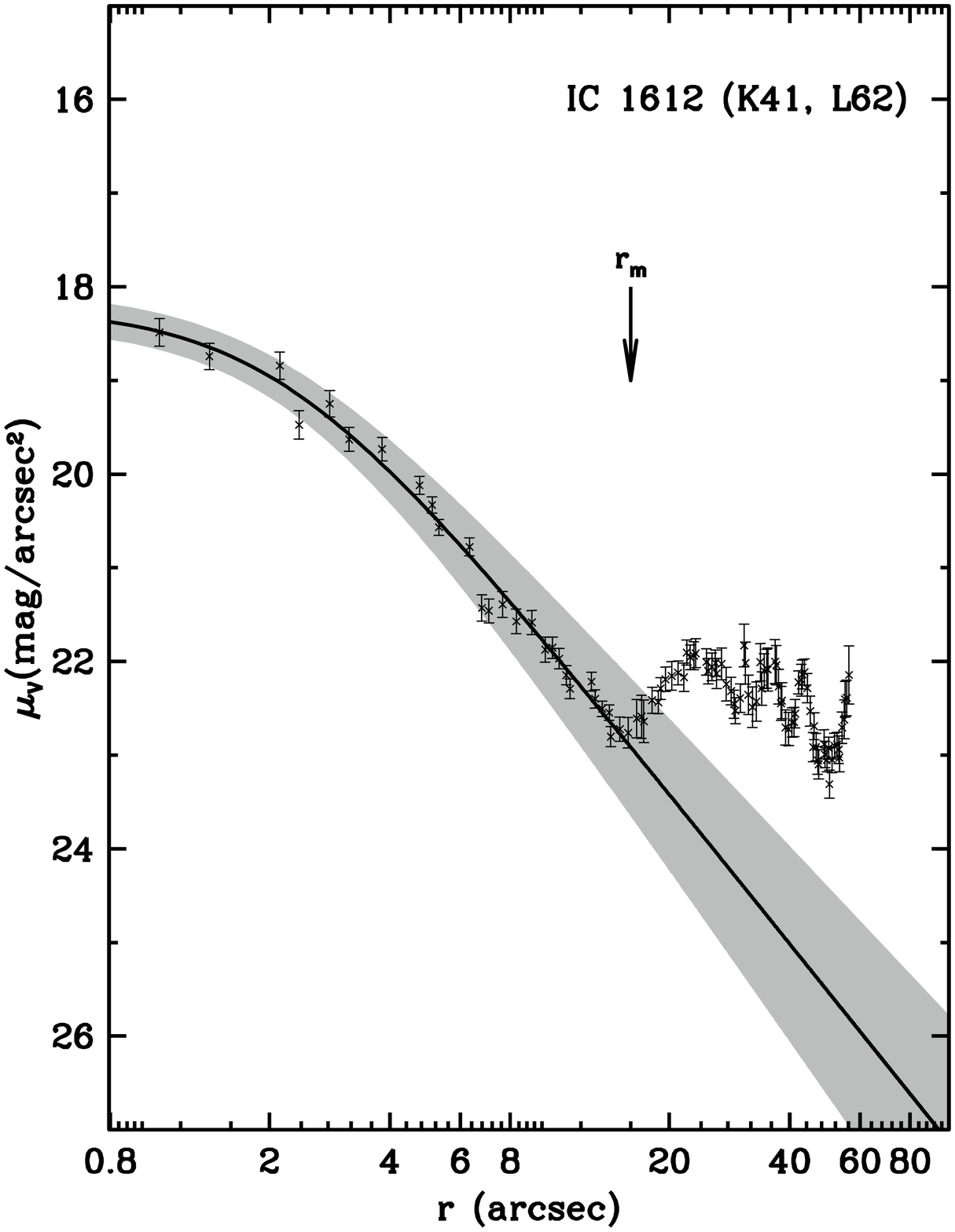}
   \includegraphics[scale=0.30,viewport=0 0 550 725,clip]{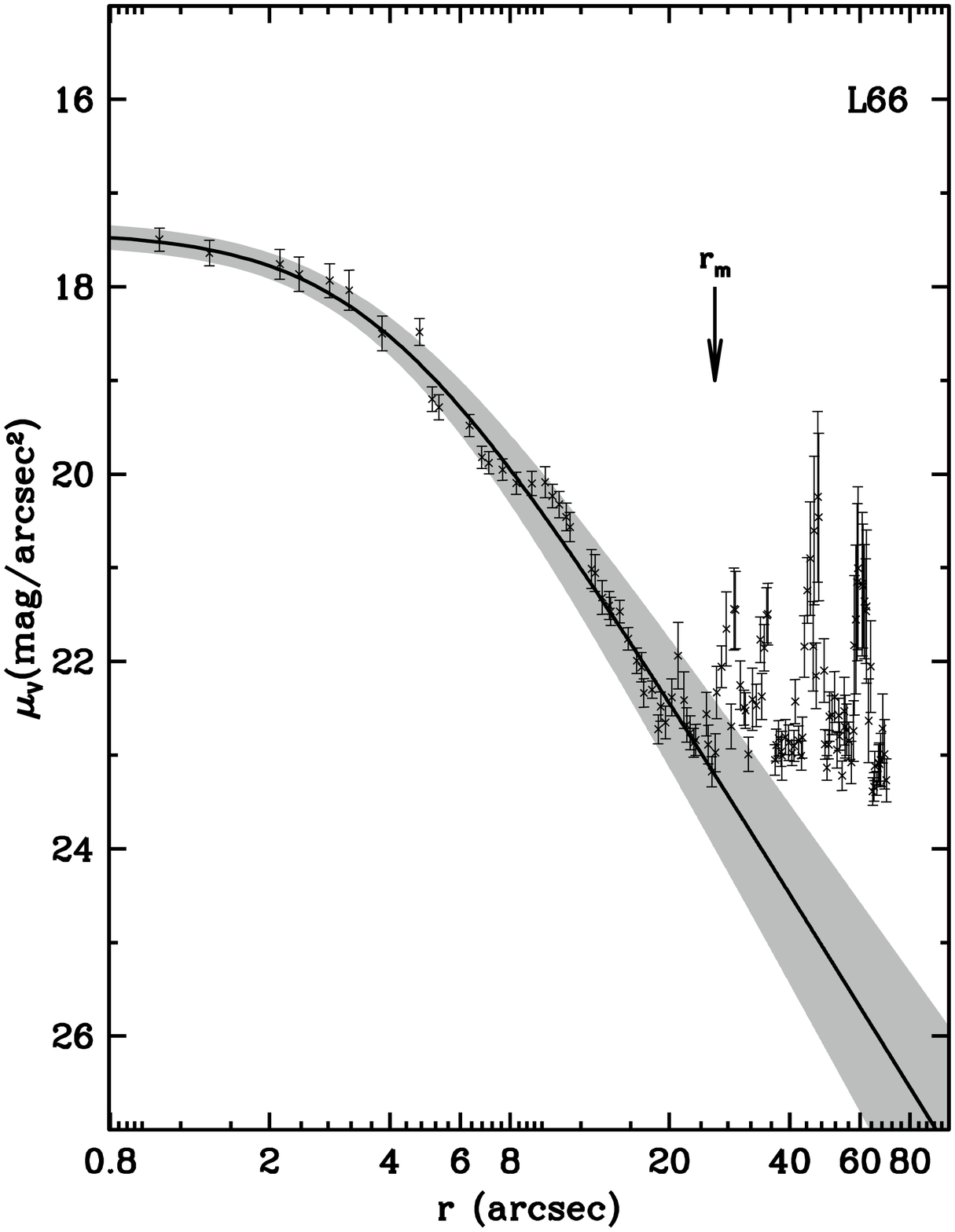}
   \includegraphics[scale=0.30,viewport=0 0 550 725,clip]{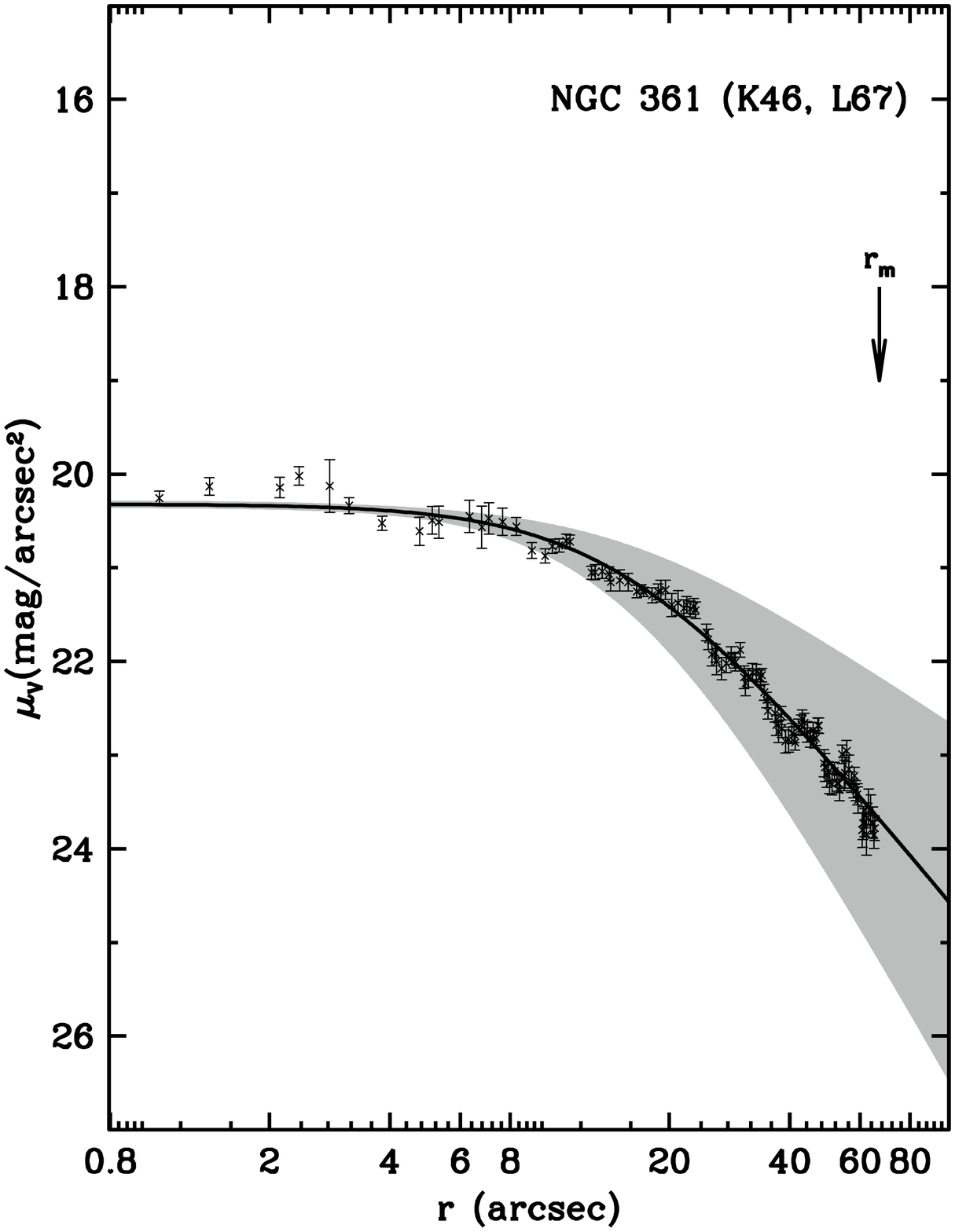}
   \includegraphics[scale=0.30,viewport=0 0 550 725,clip]{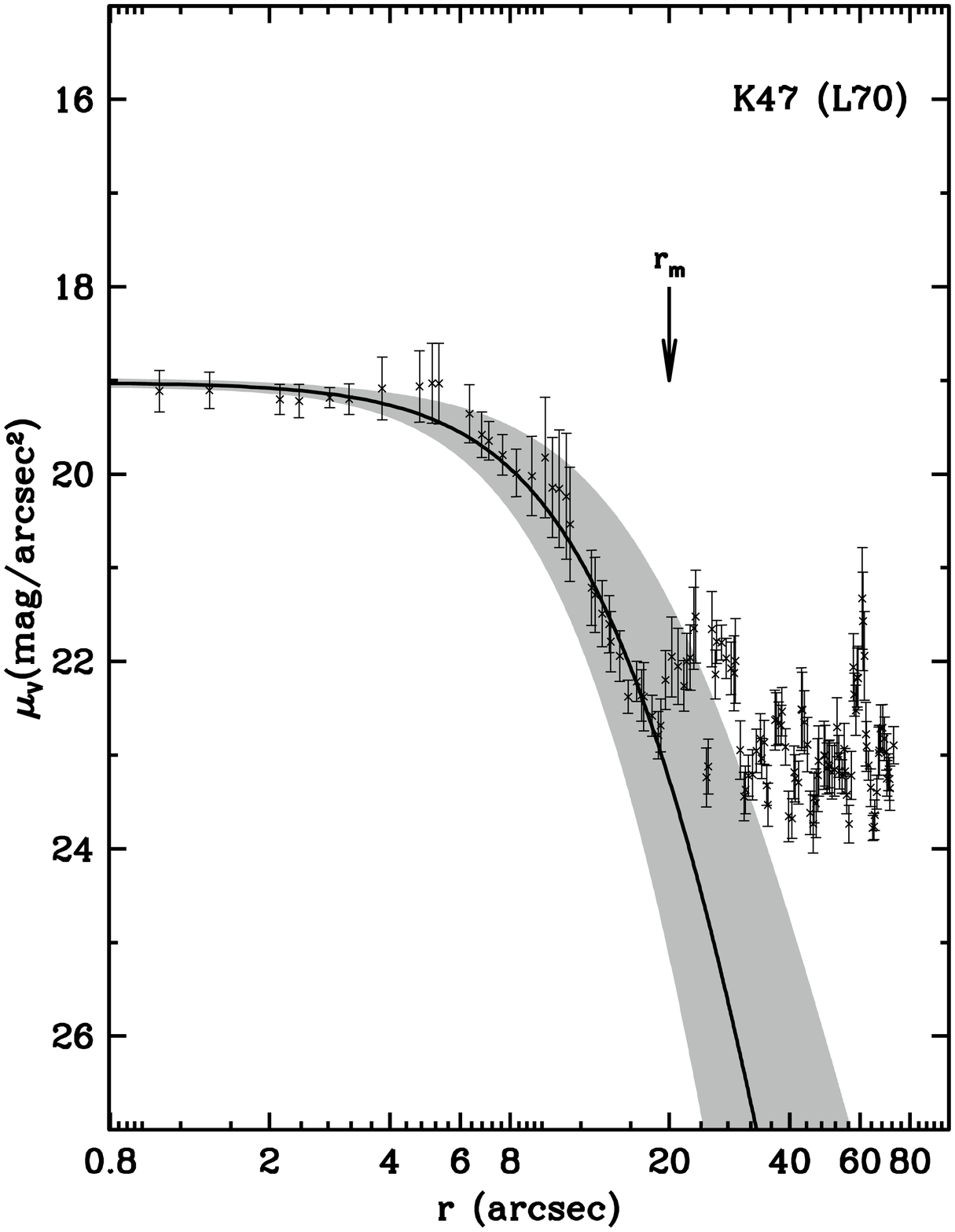}
   \includegraphics[scale=0.30,viewport=0 0 550 725,clip]{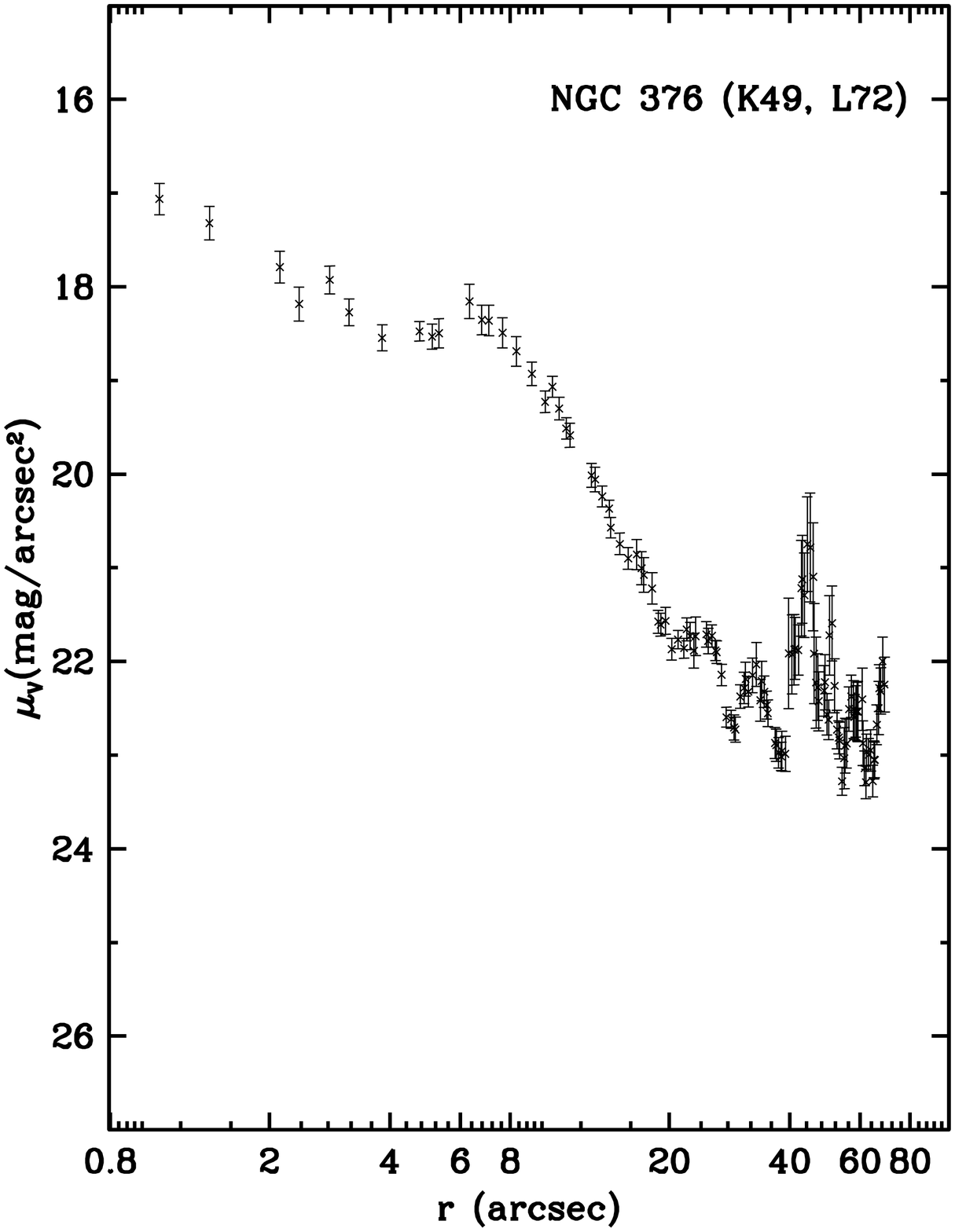}
   \includegraphics[scale=0.30,viewport=0 0 550 725,clip]{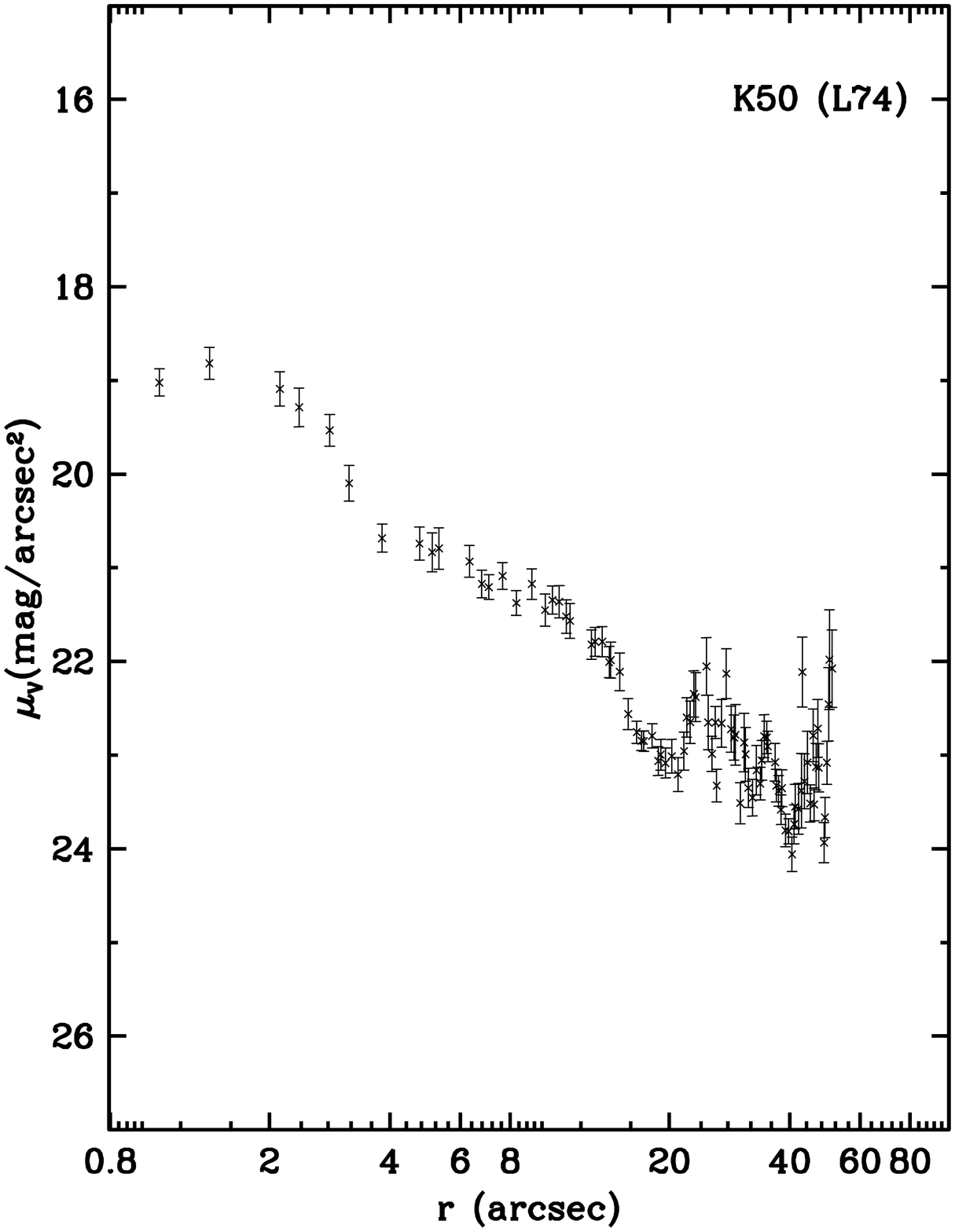}
   \caption[]{continued}
\end{figure*}
\begin{figure*}
\addtocounter{figure}{-1}
   \centering
   \includegraphics[scale=0.30,viewport=0 0 550 725,clip]{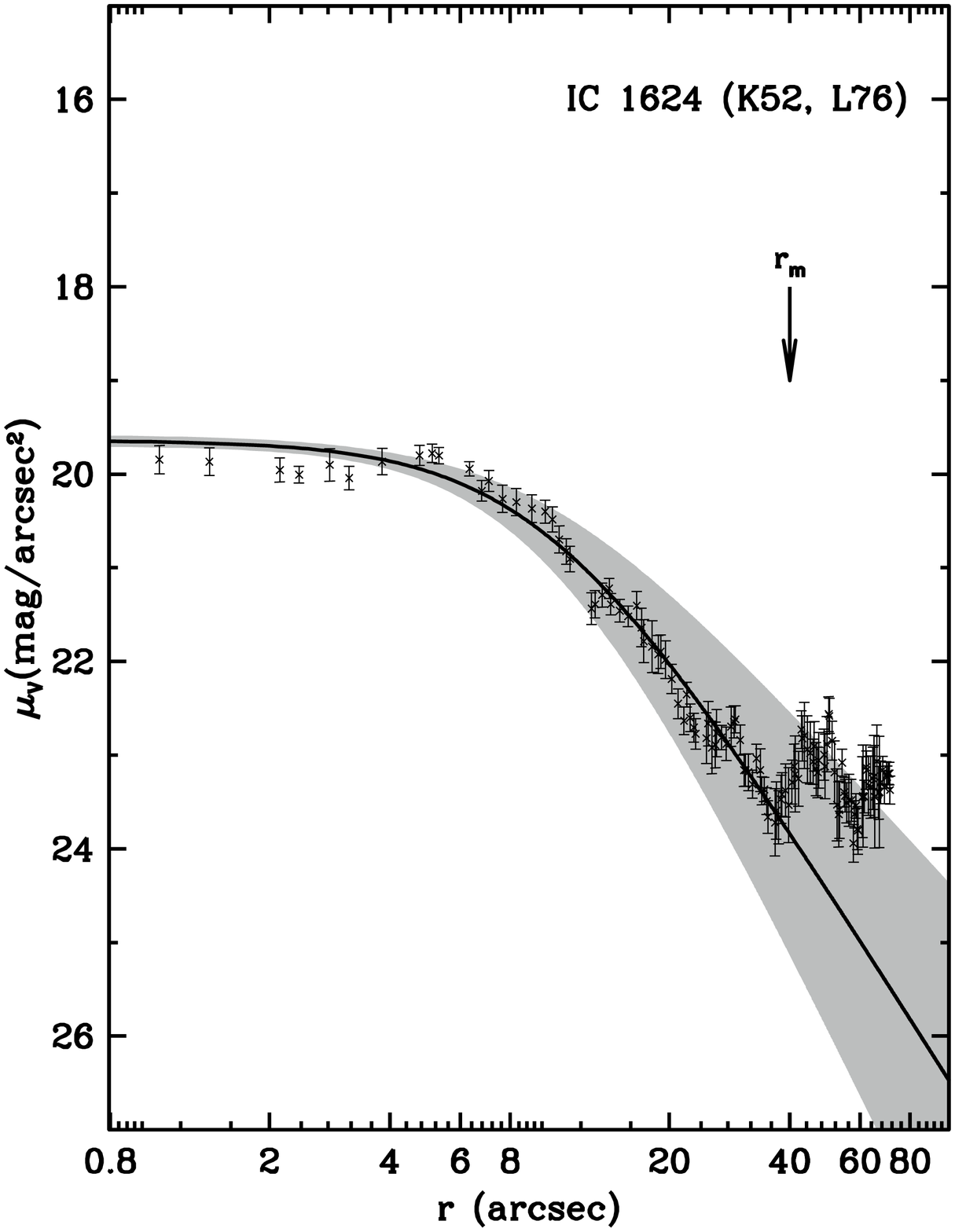}
   \includegraphics[scale=0.30,viewport=0 0 550 725,clip]{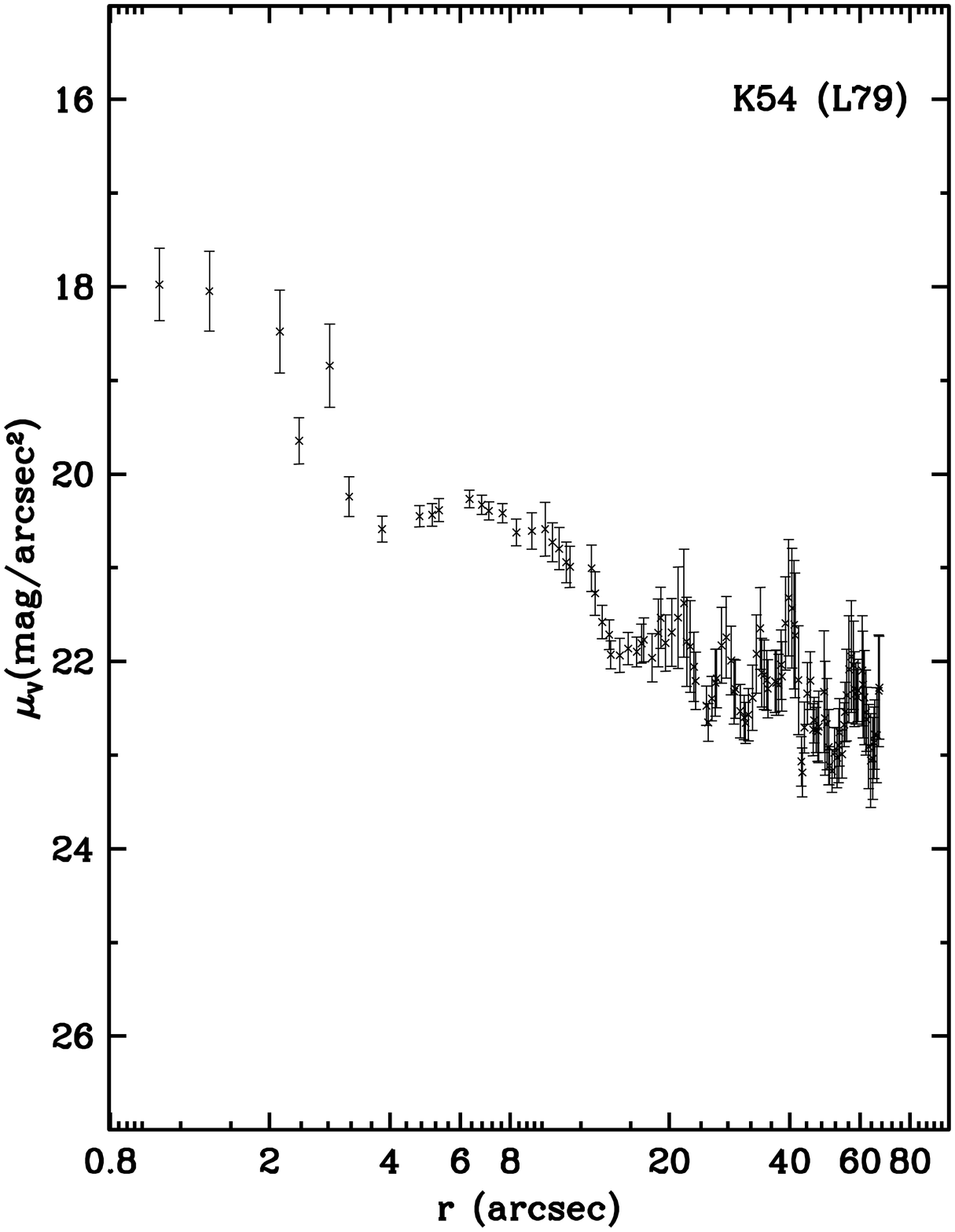}
   \includegraphics[scale=0.30,viewport=0 0 550 725,clip]{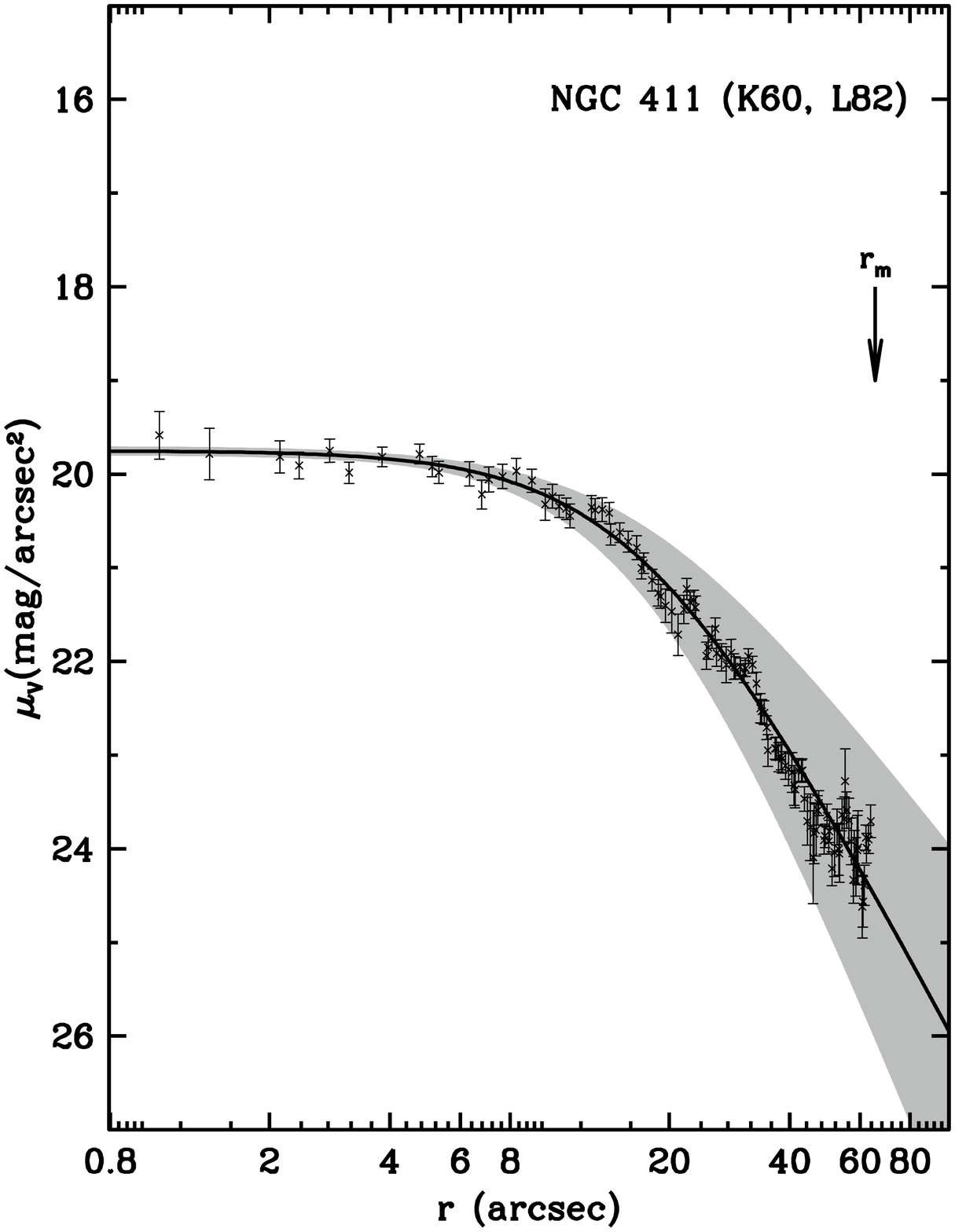}
   \includegraphics[scale=0.30,viewport=0 0 550 725,clip]{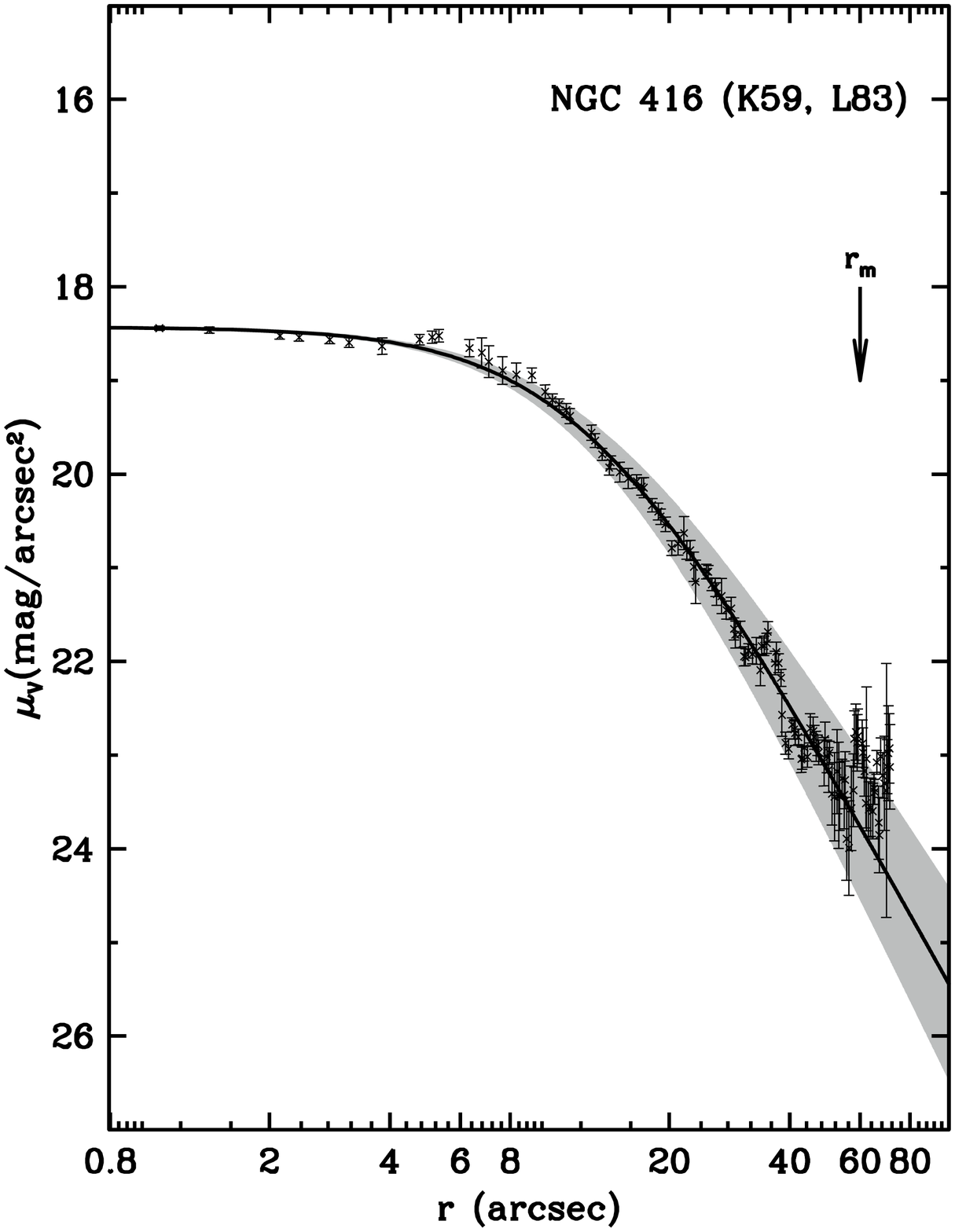}
   \includegraphics[scale=0.30,viewport=0 0 550 725,clip]{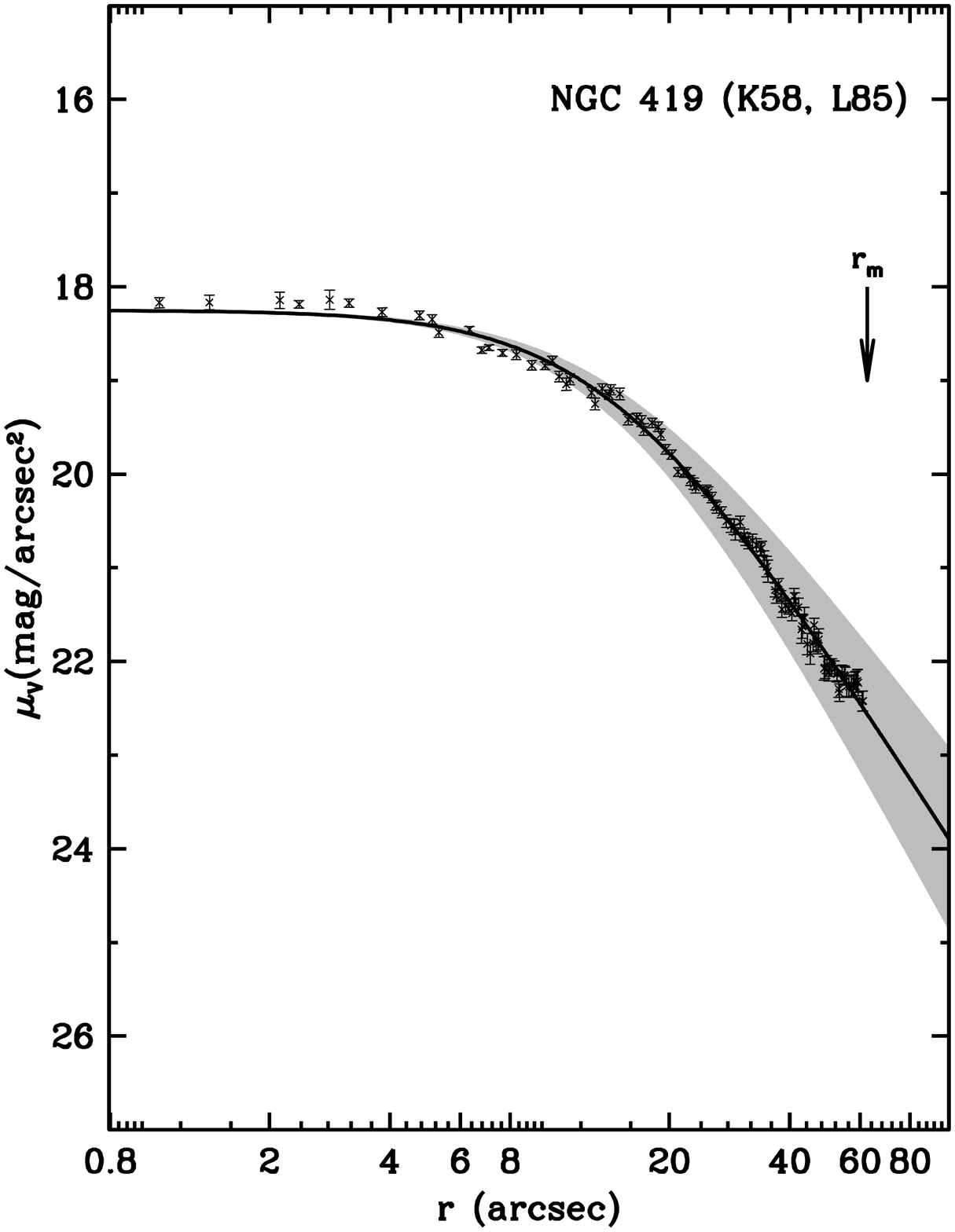}
   \includegraphics[scale=0.30,viewport=0 0 550 725,clip]{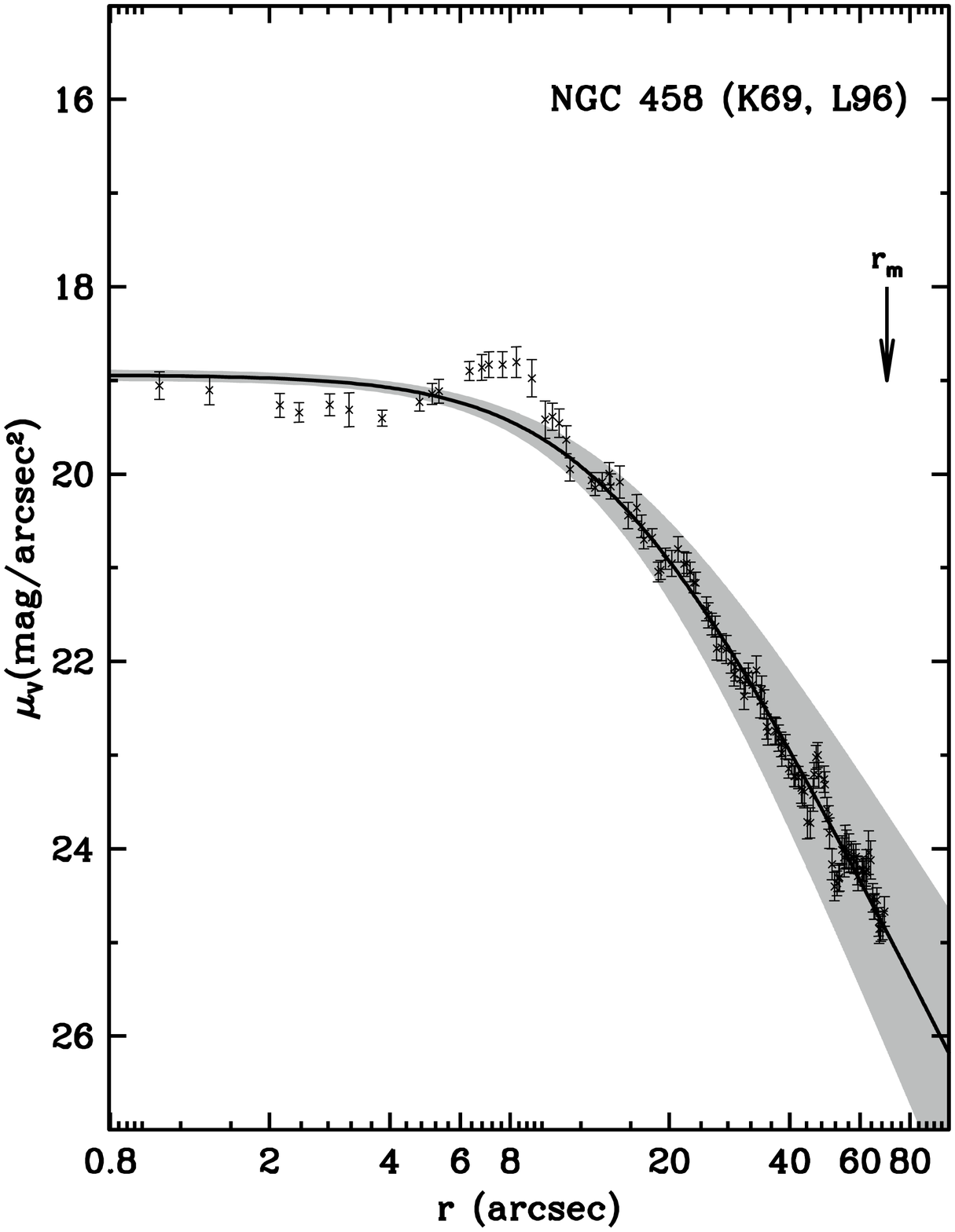}
   \includegraphics[scale=0.30,viewport=0 0 550 725,clip]{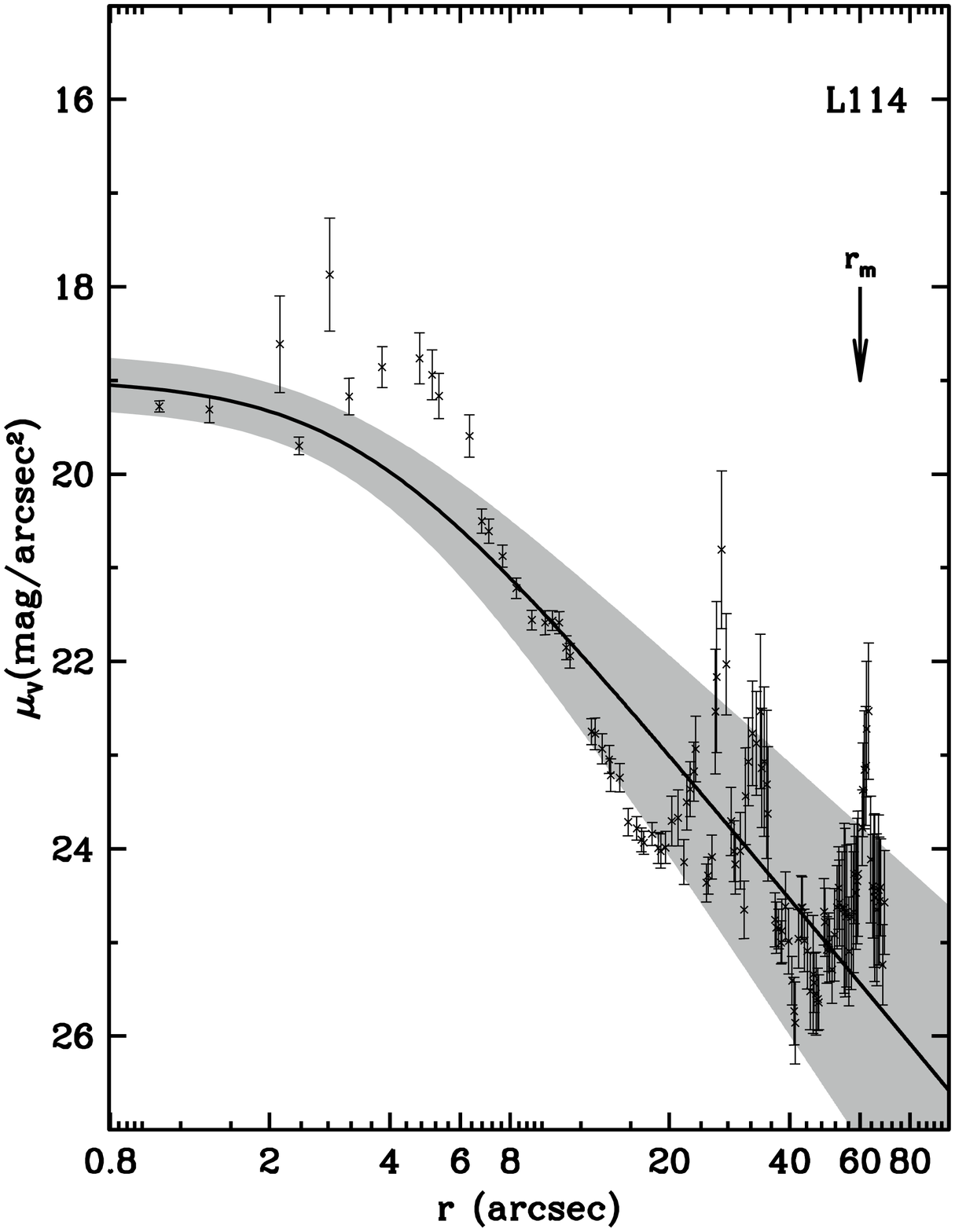}
   \includegraphics[scale=0.30,viewport=0 0 550 725,clip]{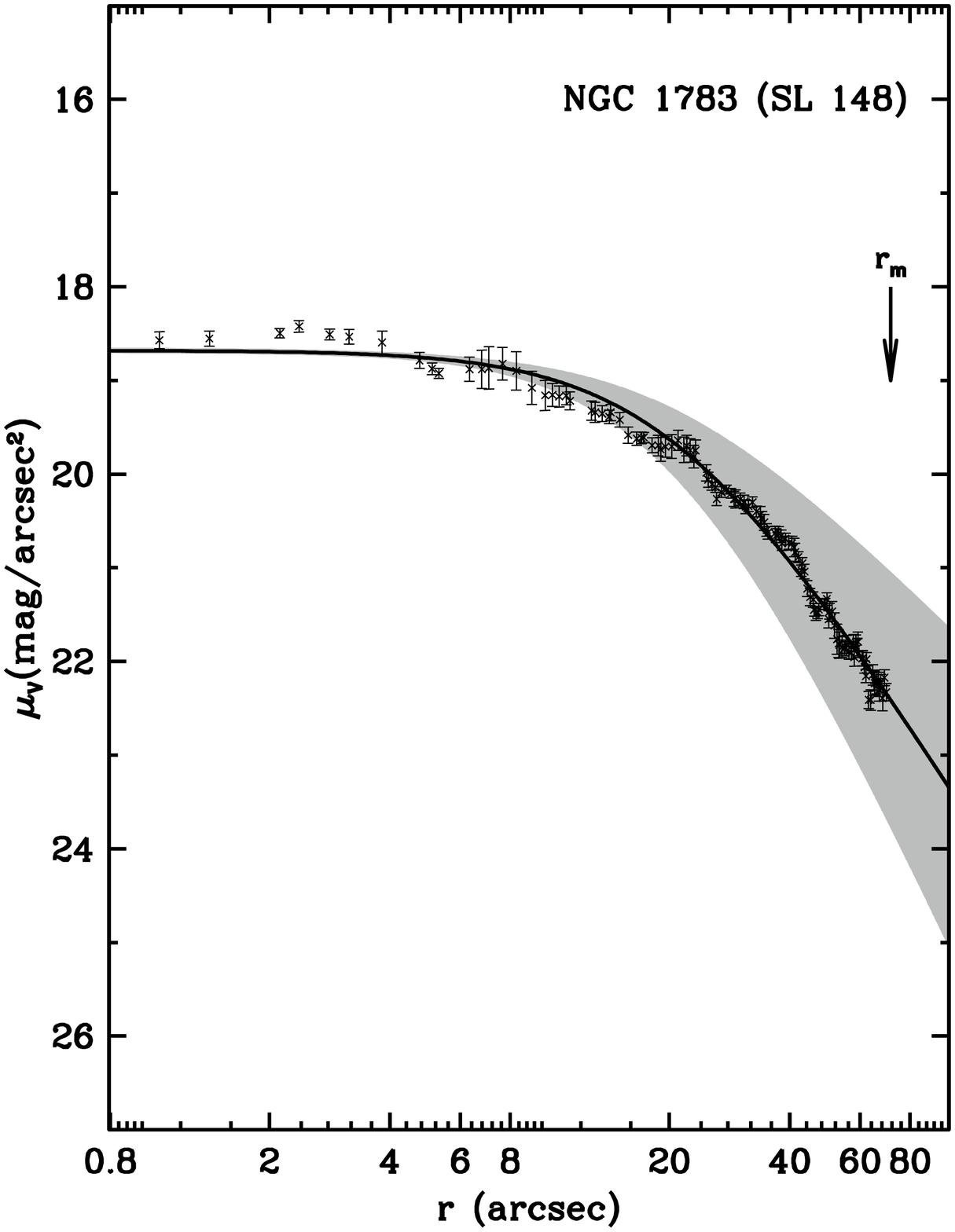}
   \includegraphics[scale=0.30,viewport=0 0 550 725,clip]{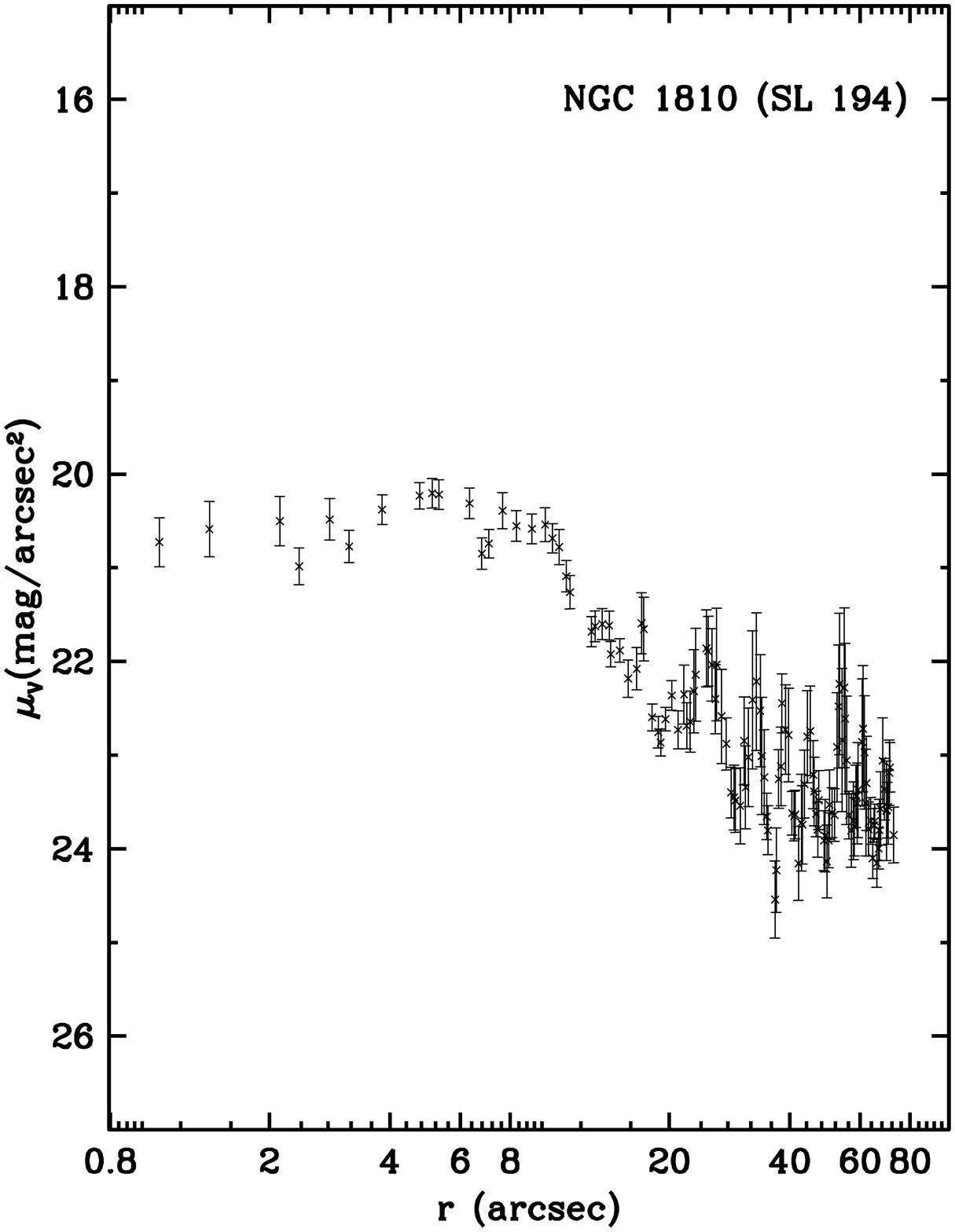}
   \caption[]{continued}
\end{figure*}
\begin{figure*}
\addtocounter{figure}{-1}
   \centering
   \includegraphics[scale=0.30,viewport=0 0 550 725,clip]{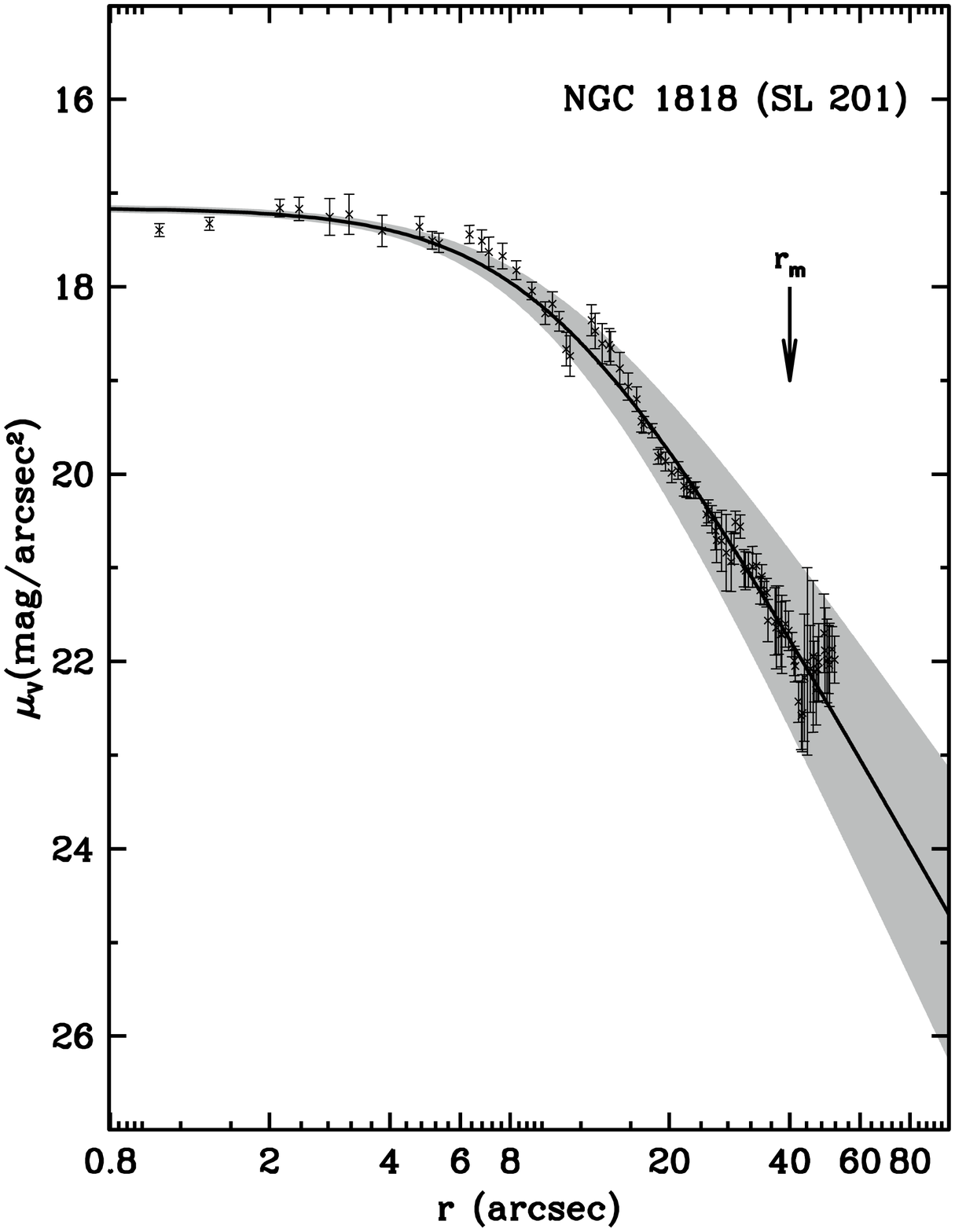}
   \includegraphics[scale=0.30,viewport=0 0 550 725,clip]{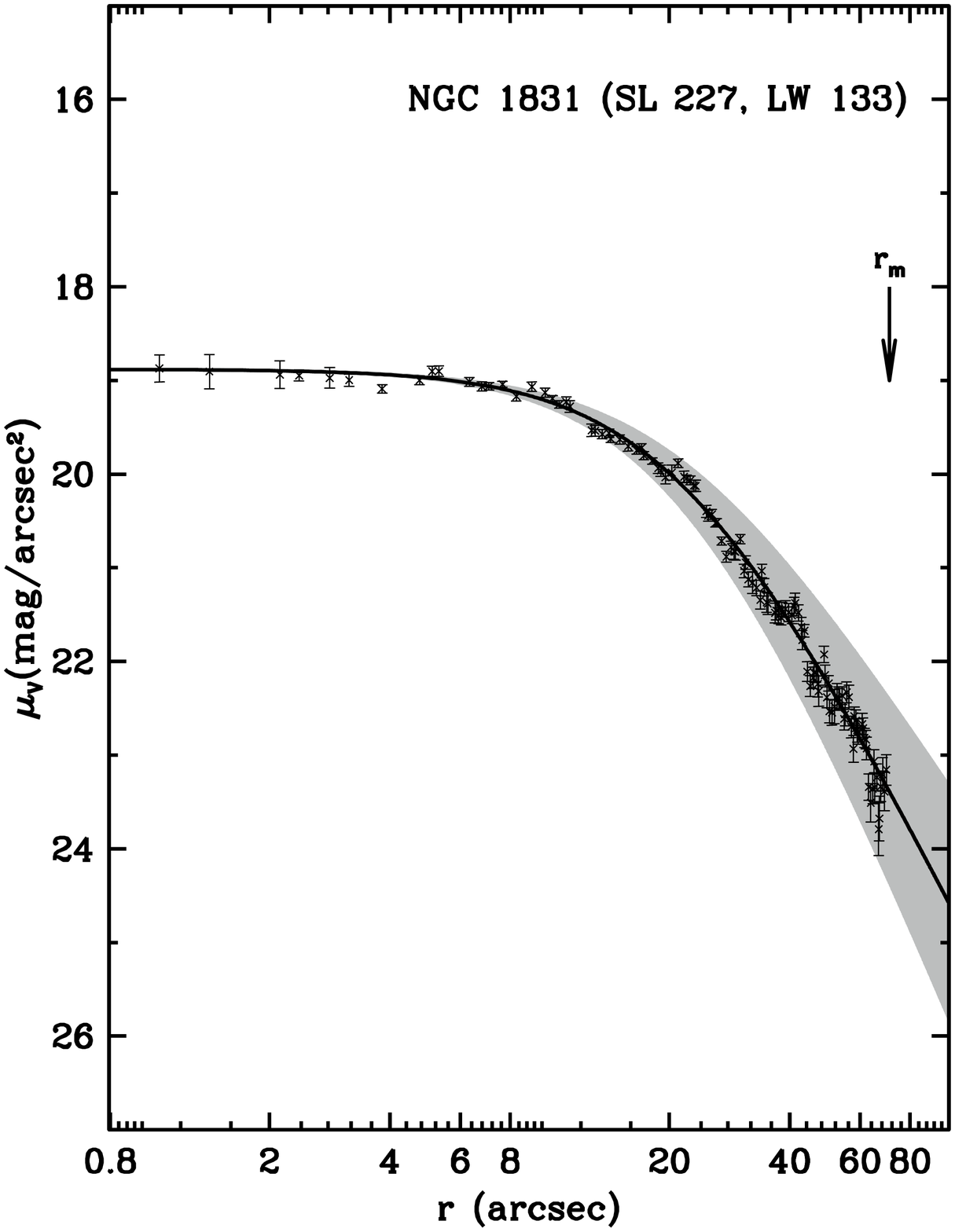}
   \includegraphics[scale=0.30,viewport=0 0 550 725,clip]{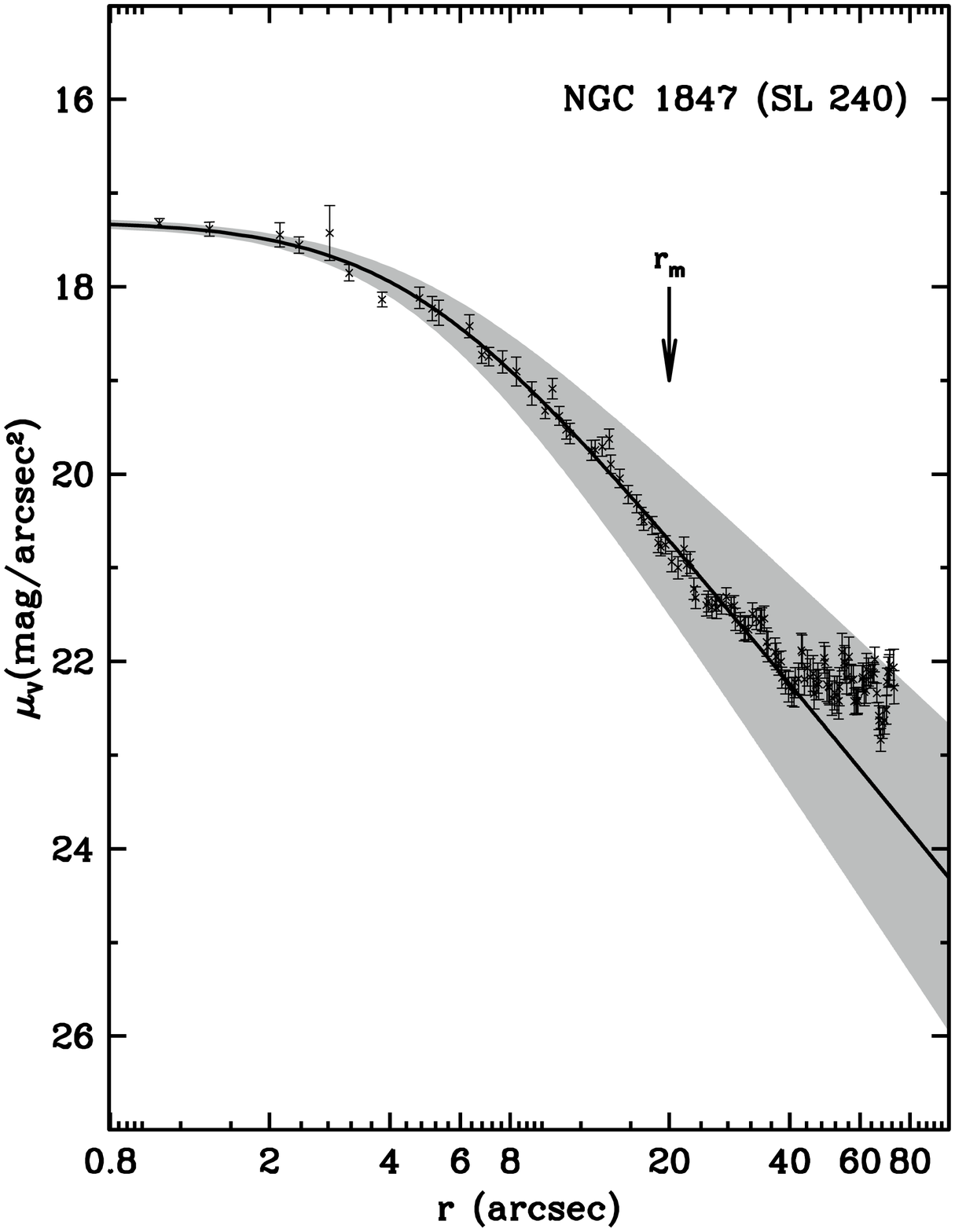}
   \includegraphics[scale=0.30,viewport=0 0 550 725,clip]{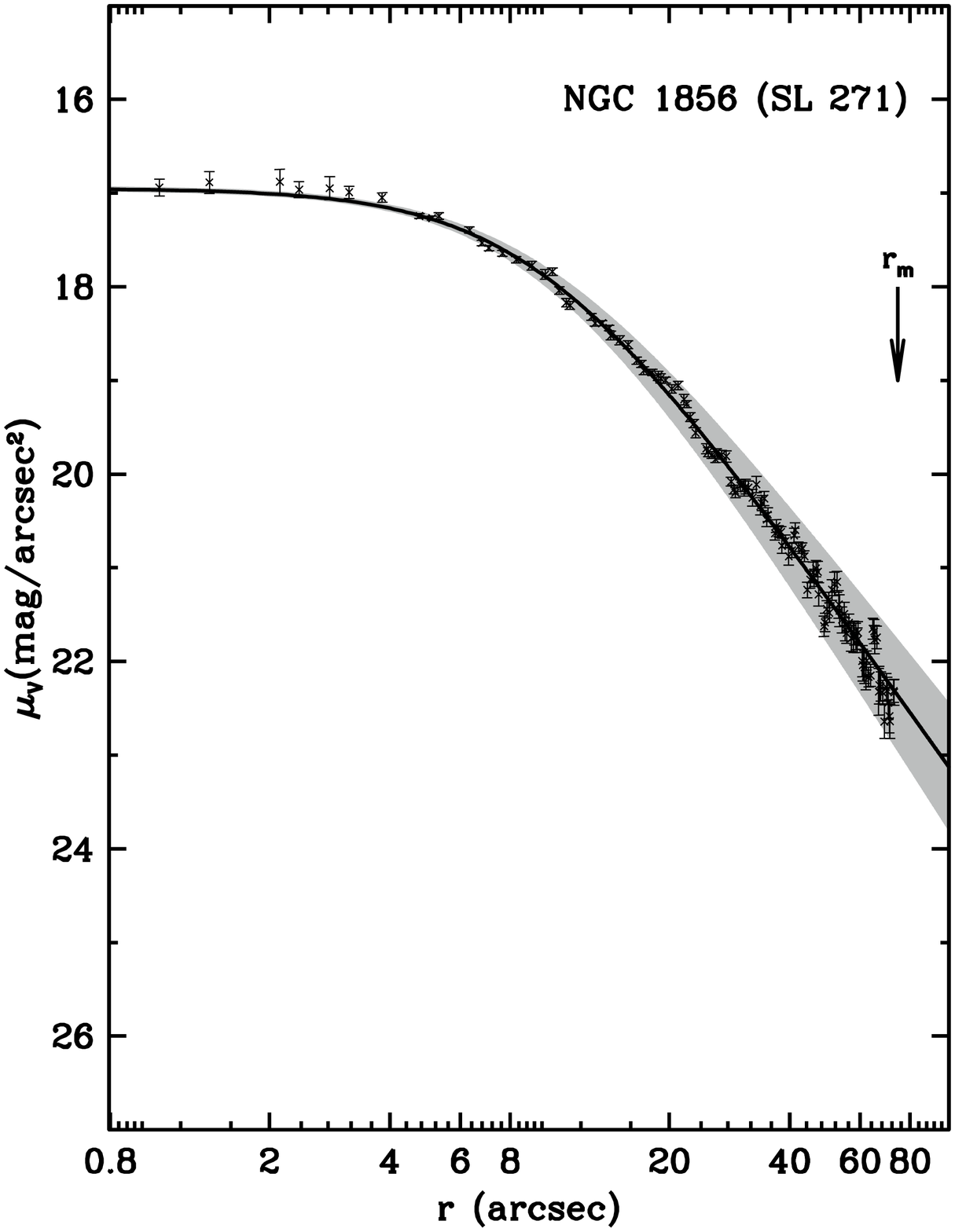}
   \includegraphics[scale=0.30,viewport=0 0 550 725,clip]{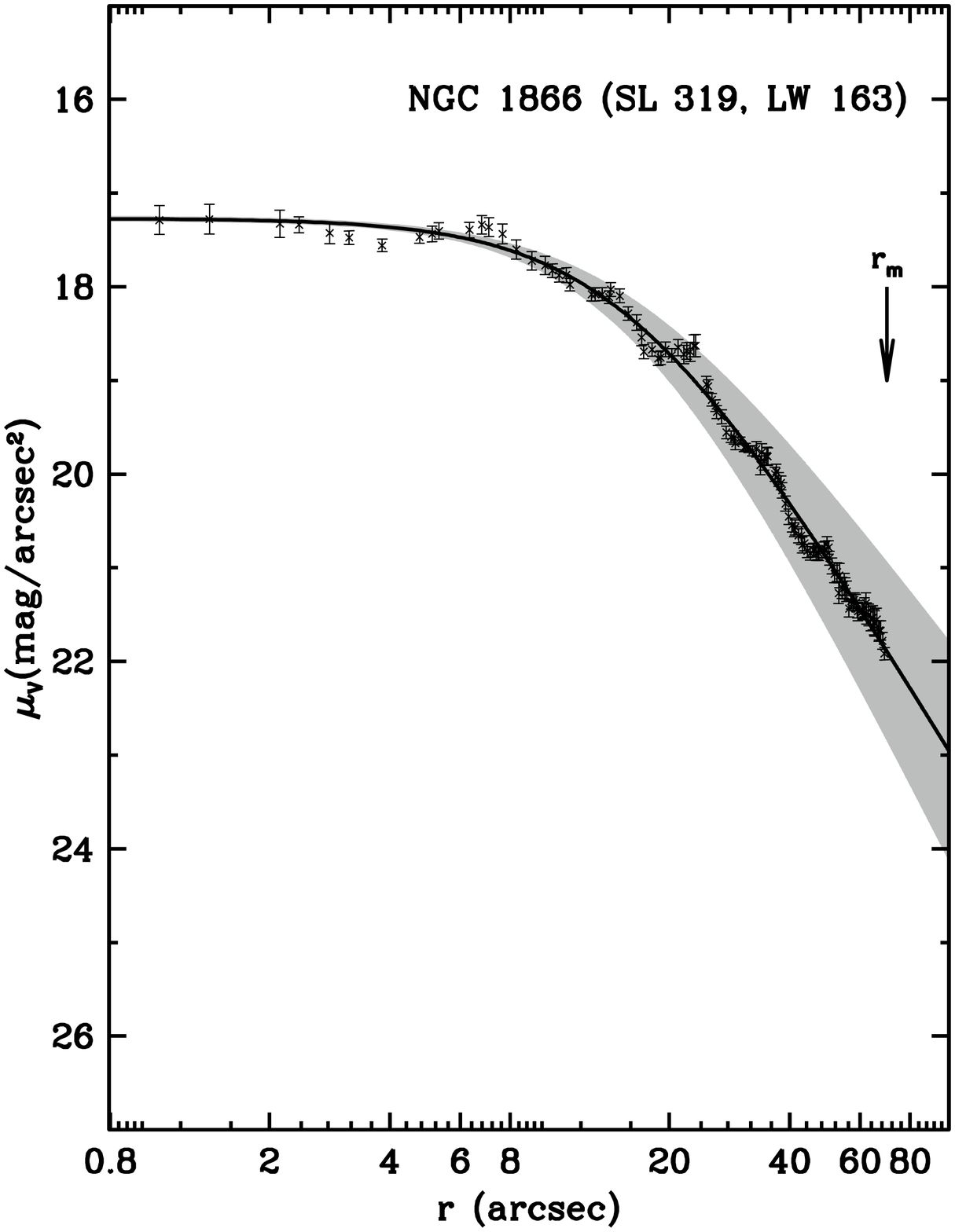}
   \includegraphics[scale=0.30,viewport=0 0 550 725,clip]{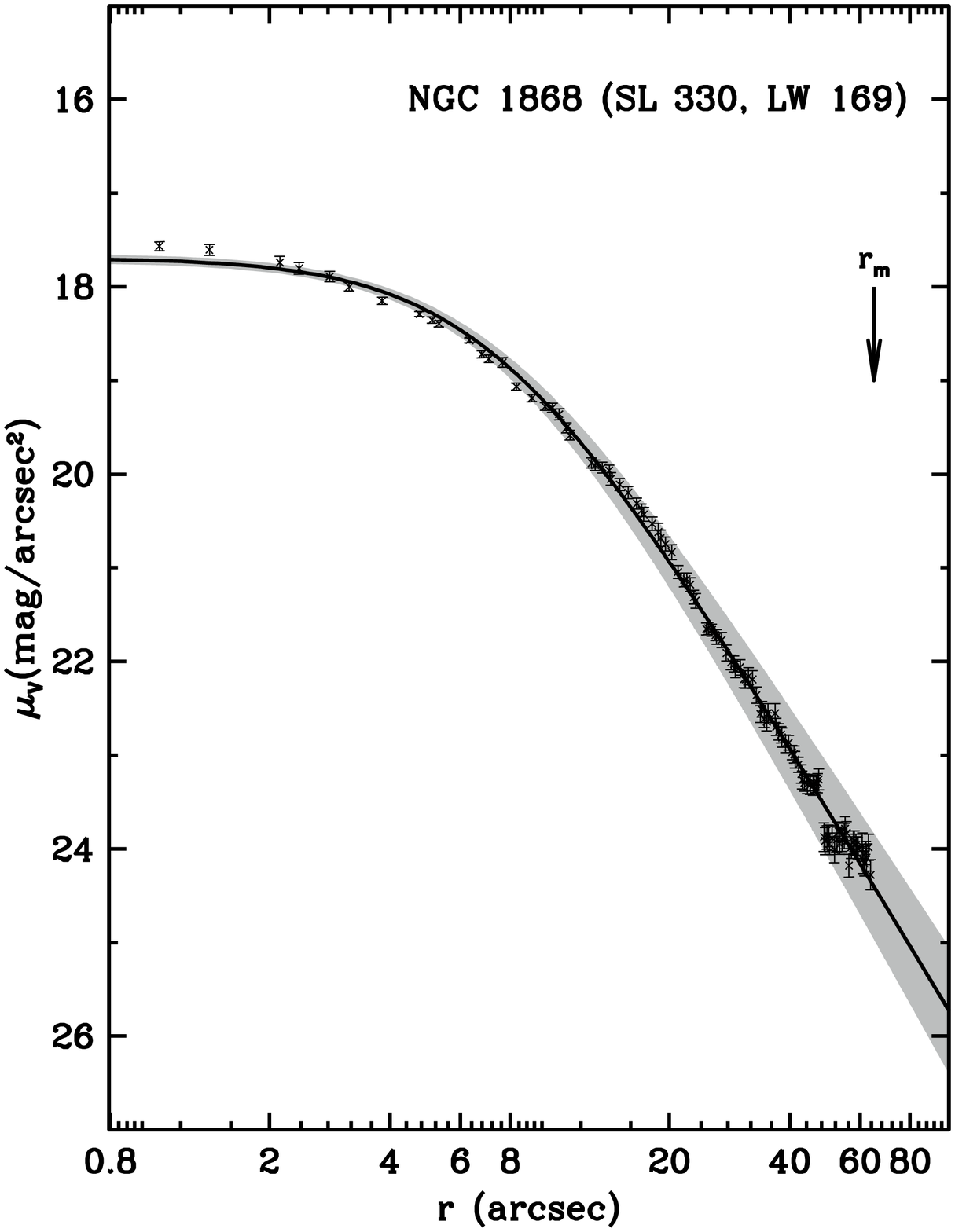}
   \includegraphics[scale=0.30,viewport=0 0 550 725,clip]{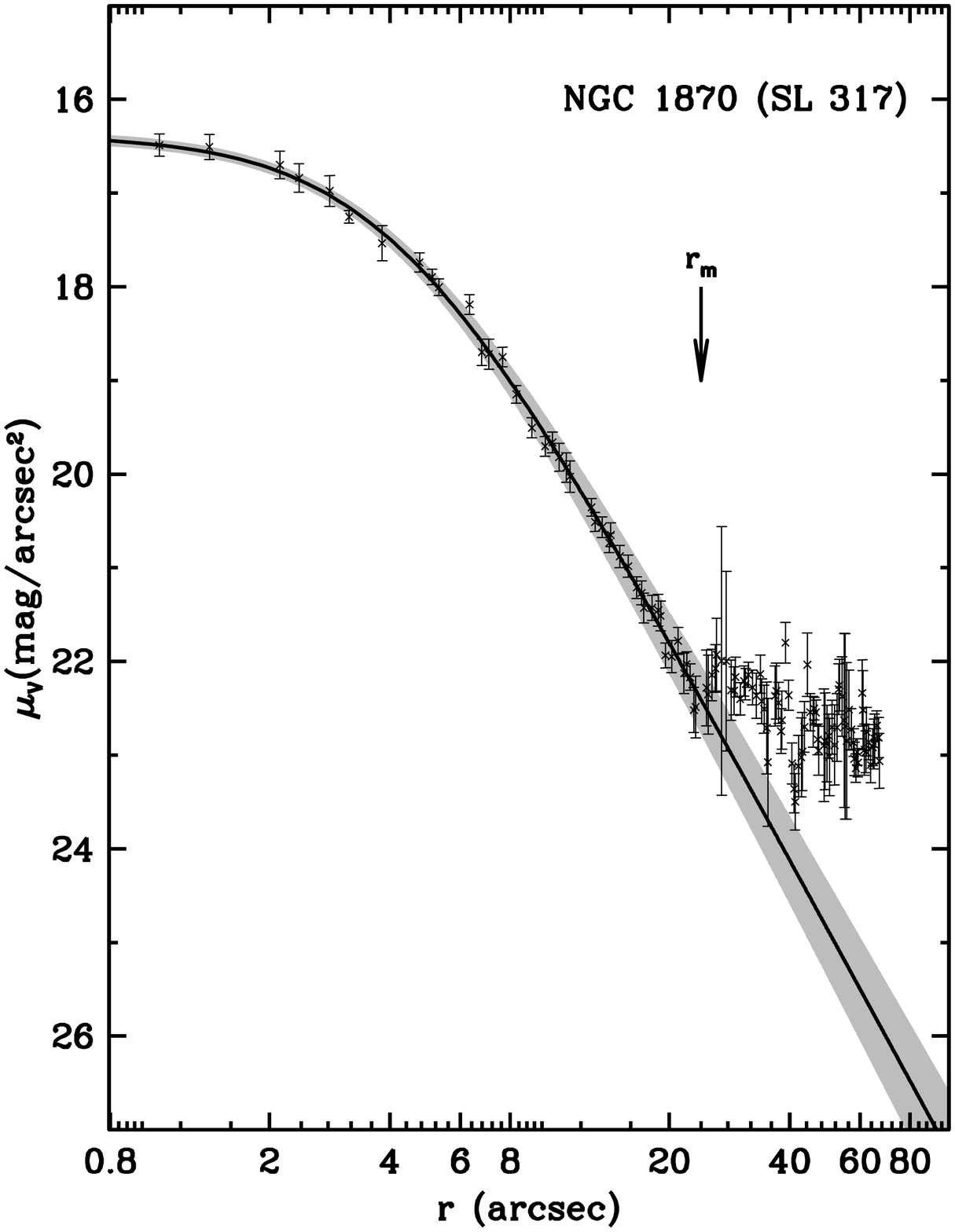}
   \includegraphics[scale=0.30,viewport=0 0 550 725,clip]{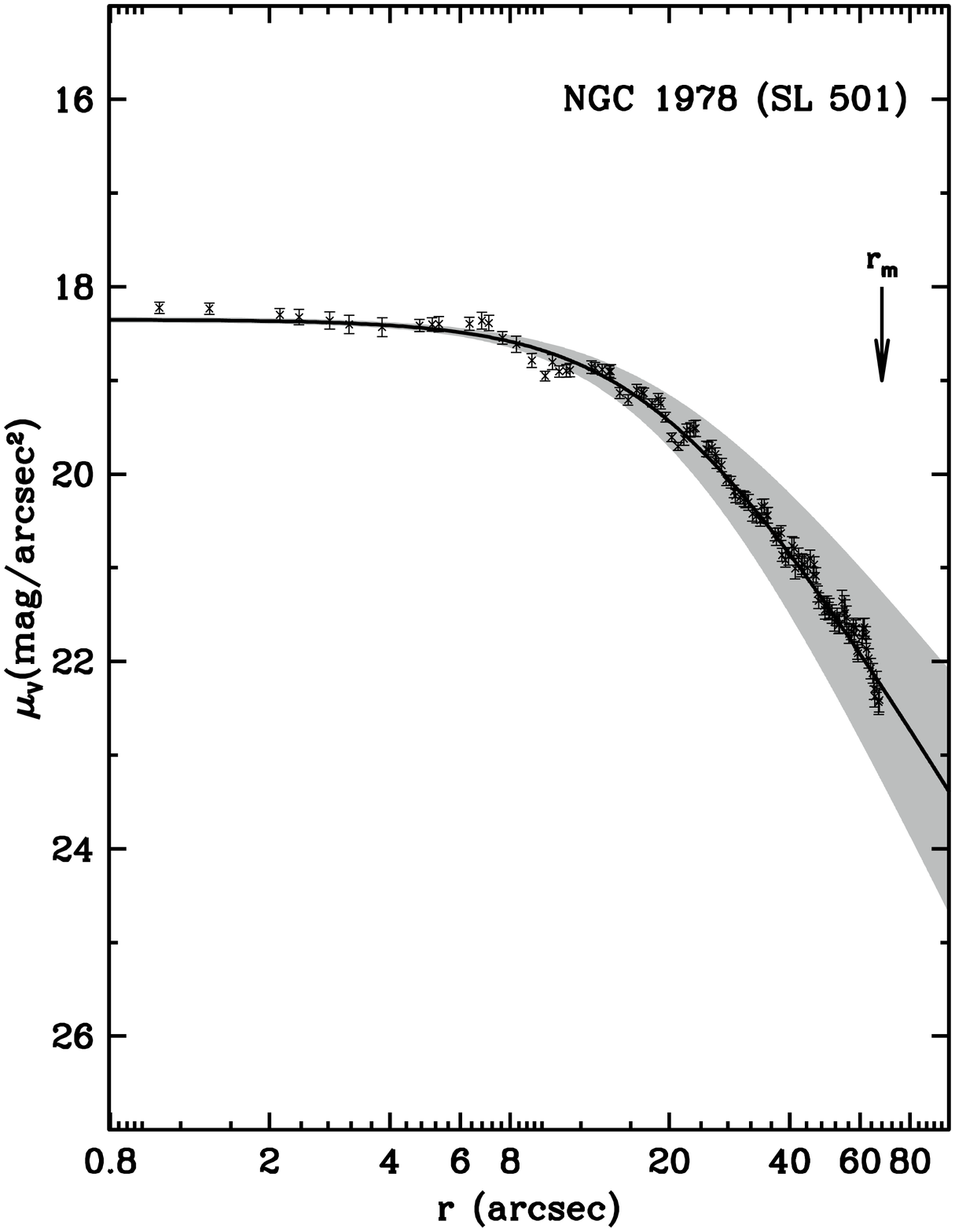}
   \includegraphics[scale=0.30,viewport=0 0 550 725,clip]{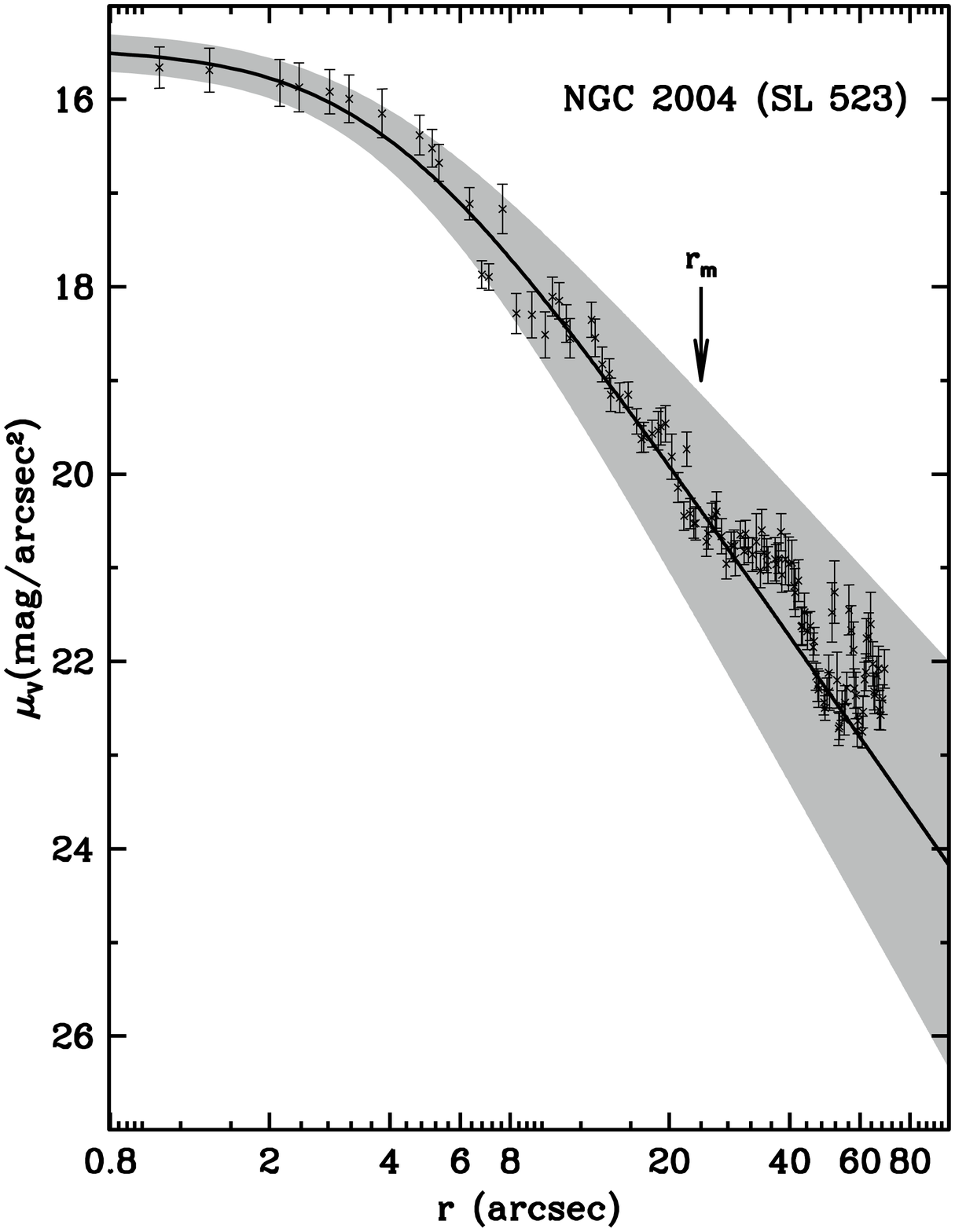}
   \caption[]{continued}
\end{figure*}
\begin{figure*}
\addtocounter{figure}{-1}
   \centering
   \includegraphics[scale=0.30,viewport=0 0 550 725,clip]{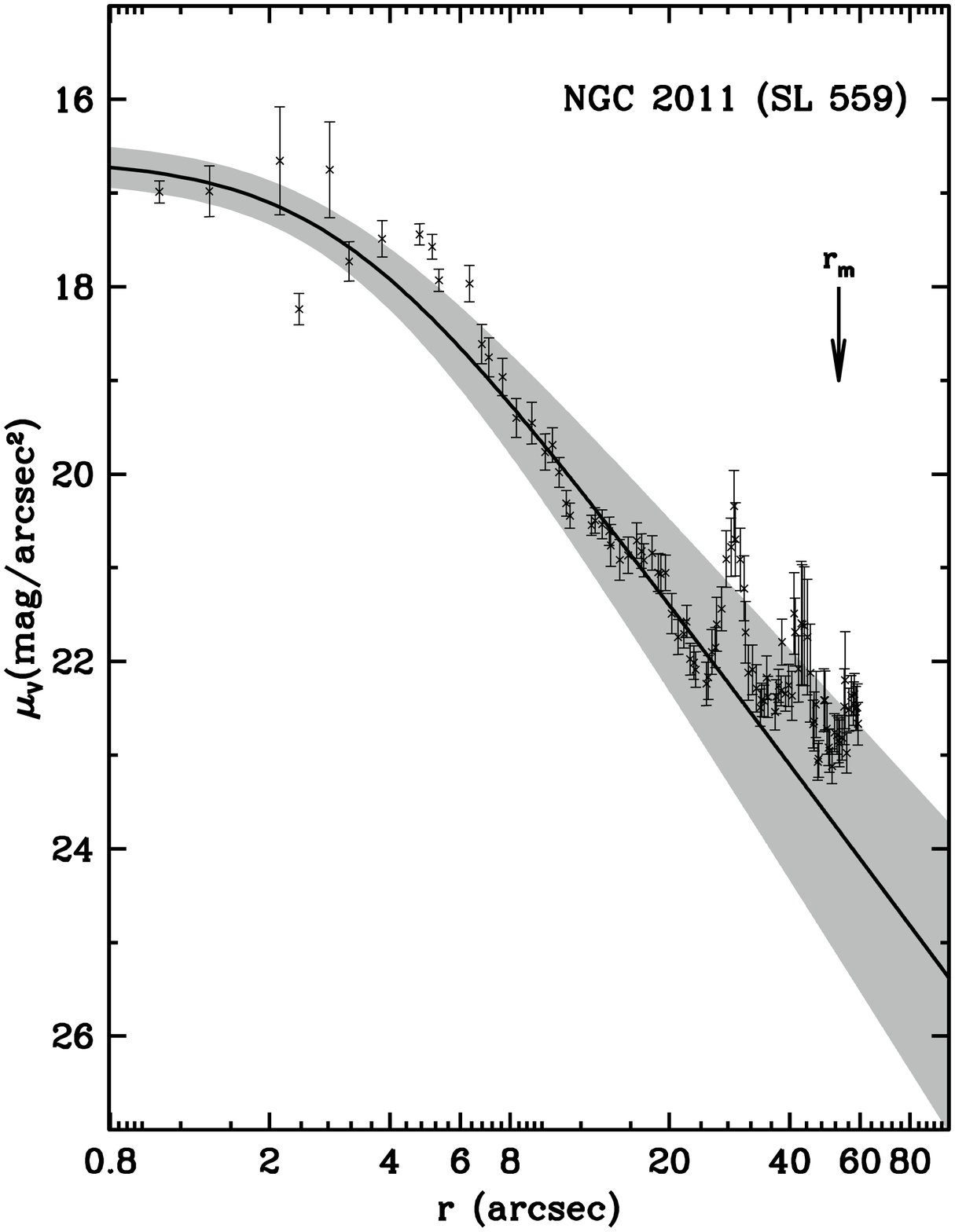}
   \includegraphics[scale=0.30,viewport=0 0 550 725,clip]{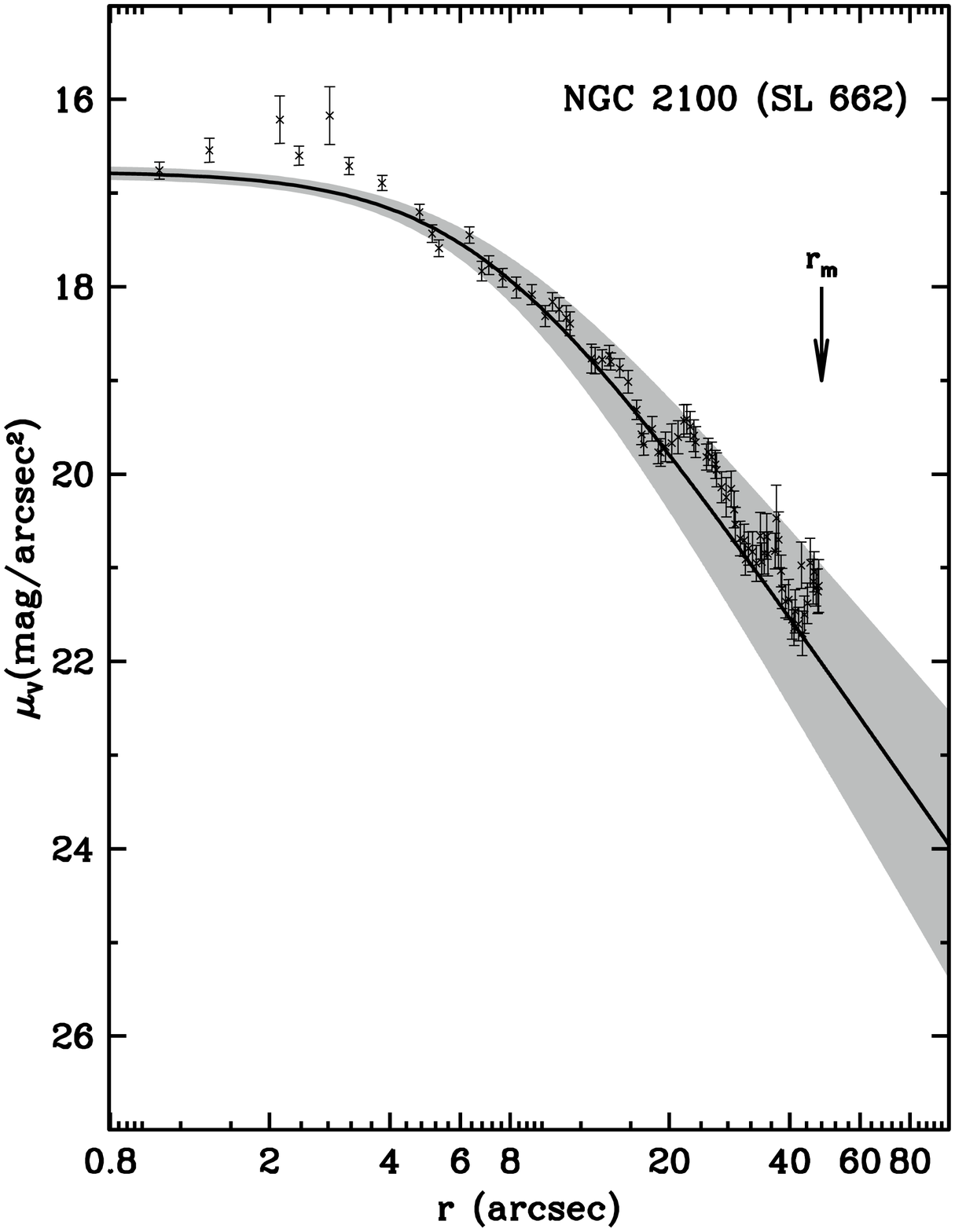}
   \includegraphics[scale=0.30,viewport=0 0 550 725,clip]{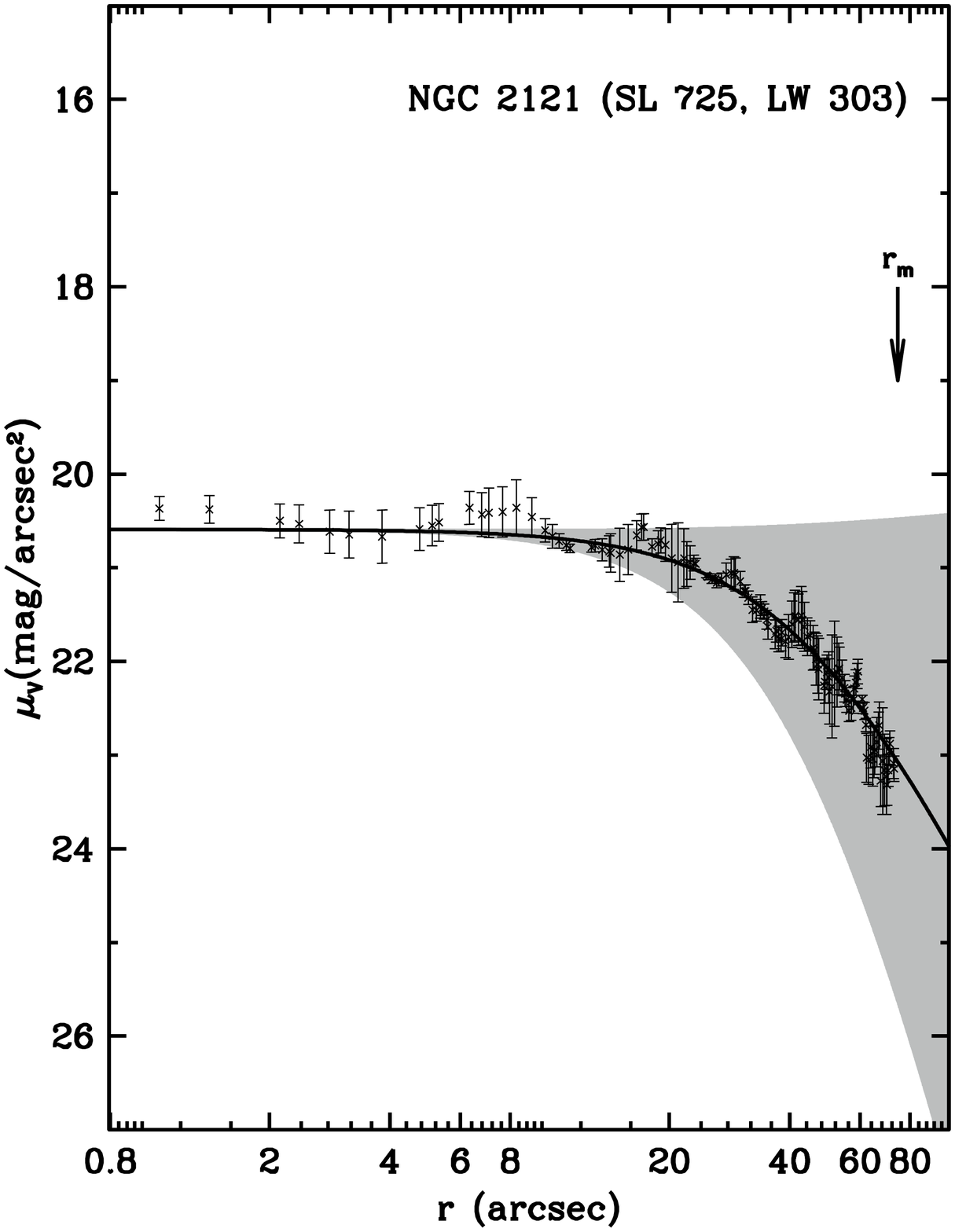}
   \includegraphics[scale=0.30,viewport=0 0 550 725,clip]{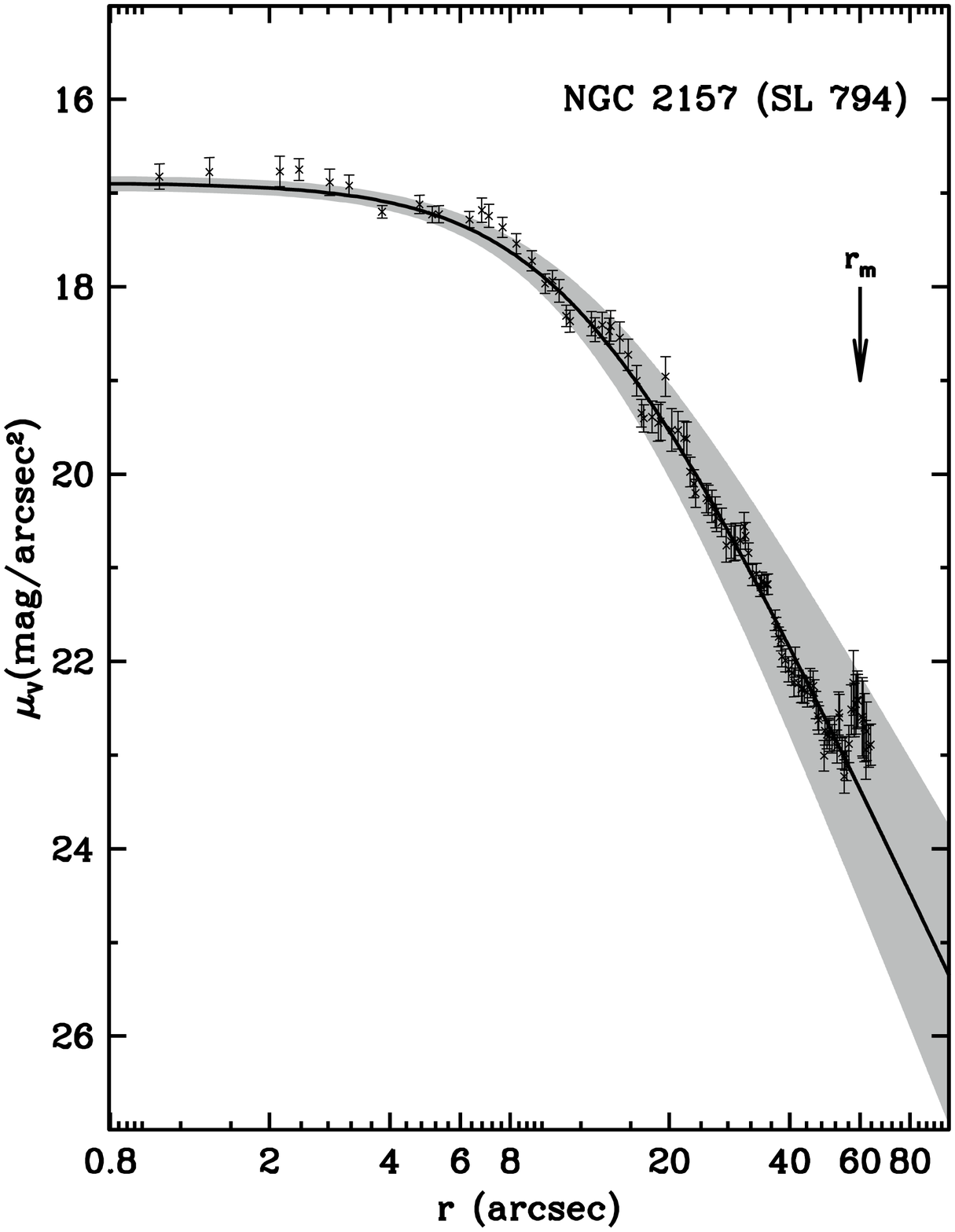}
   \includegraphics[scale=0.30,viewport=0 0 550 725,clip]{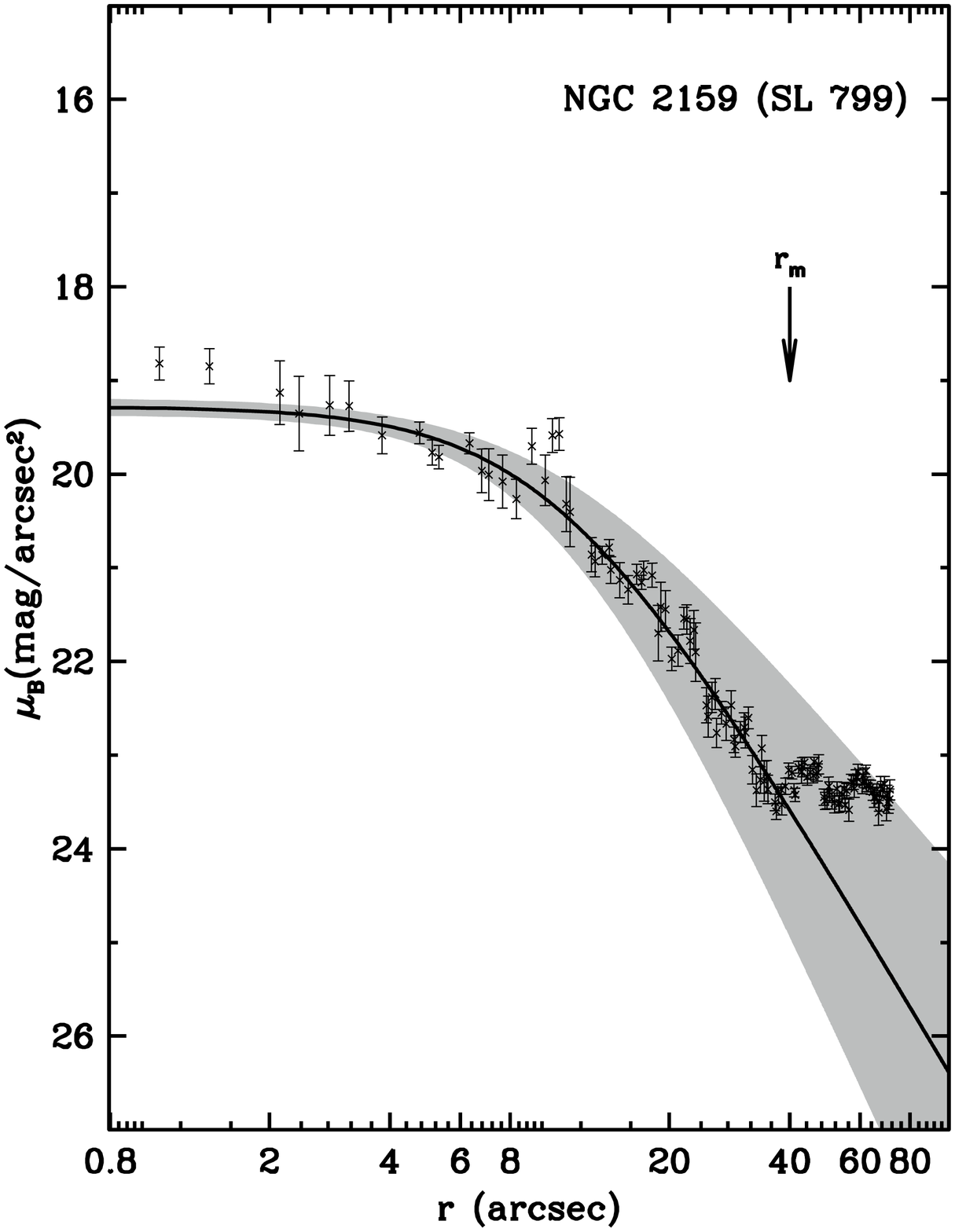}
   \includegraphics[scale=0.30,viewport=0 0 550 725,clip]{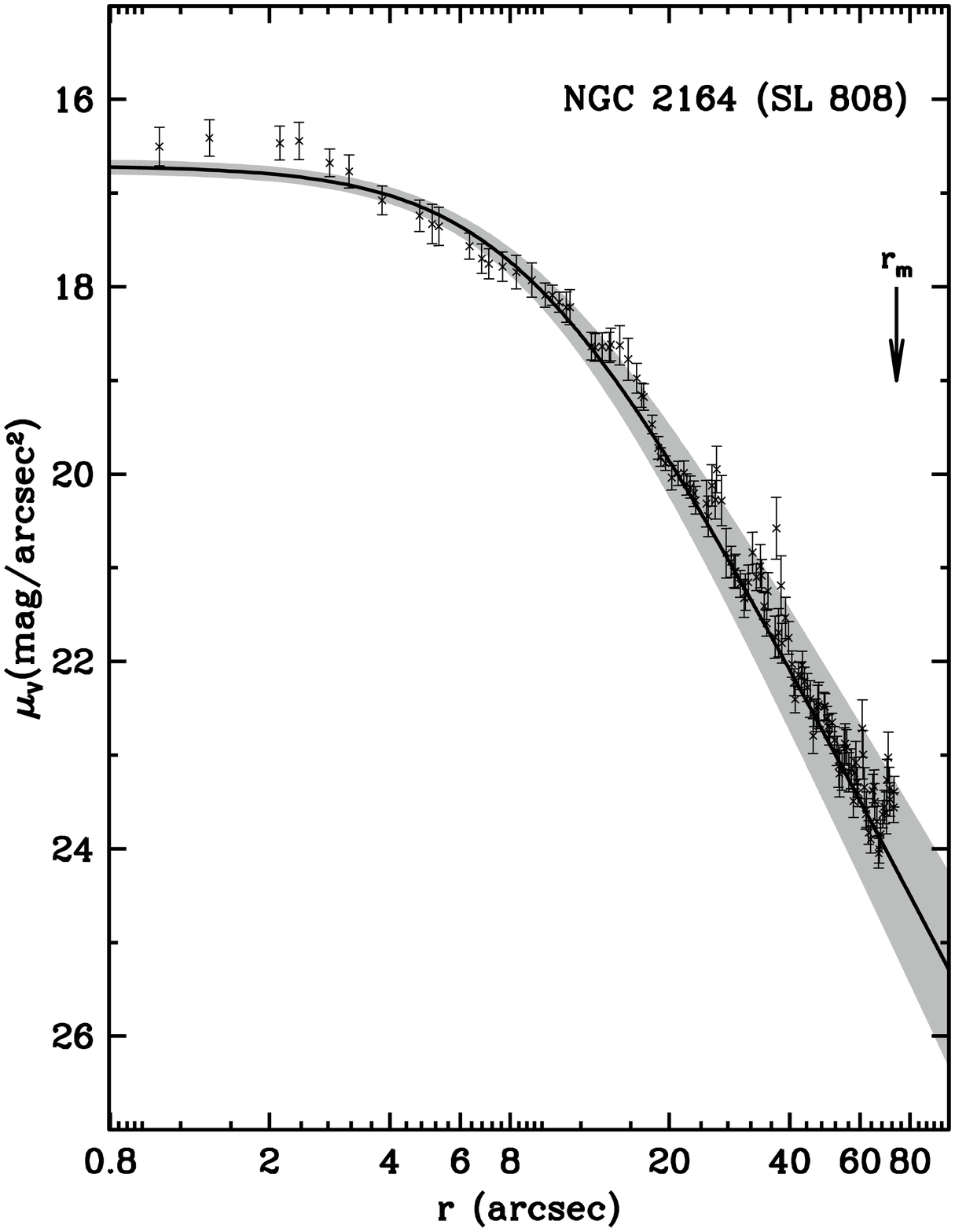}
   \includegraphics[scale=0.30,viewport=0 0 550 725,clip]{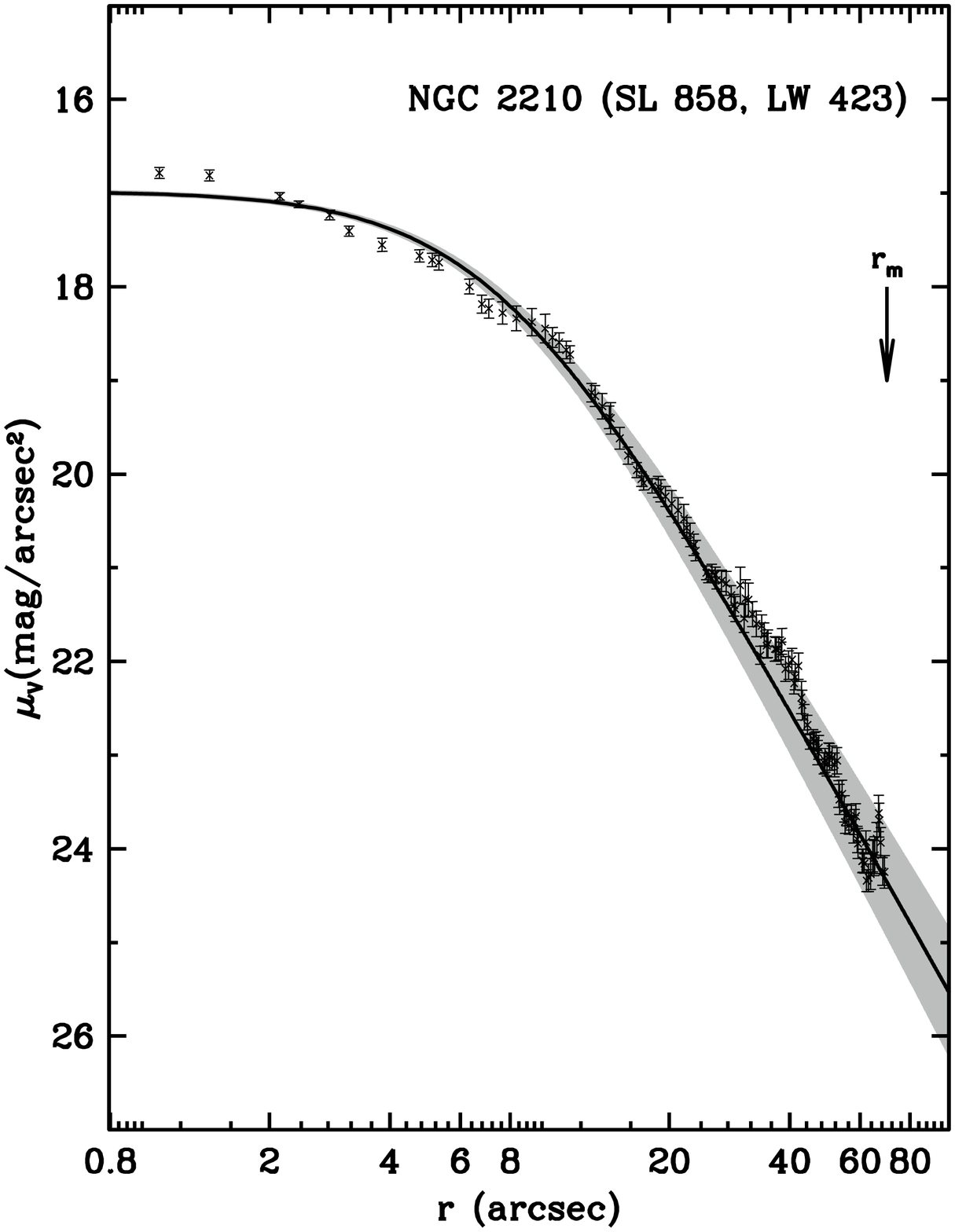}
   \includegraphics[scale=0.30,viewport=0 0 550 725,clip]{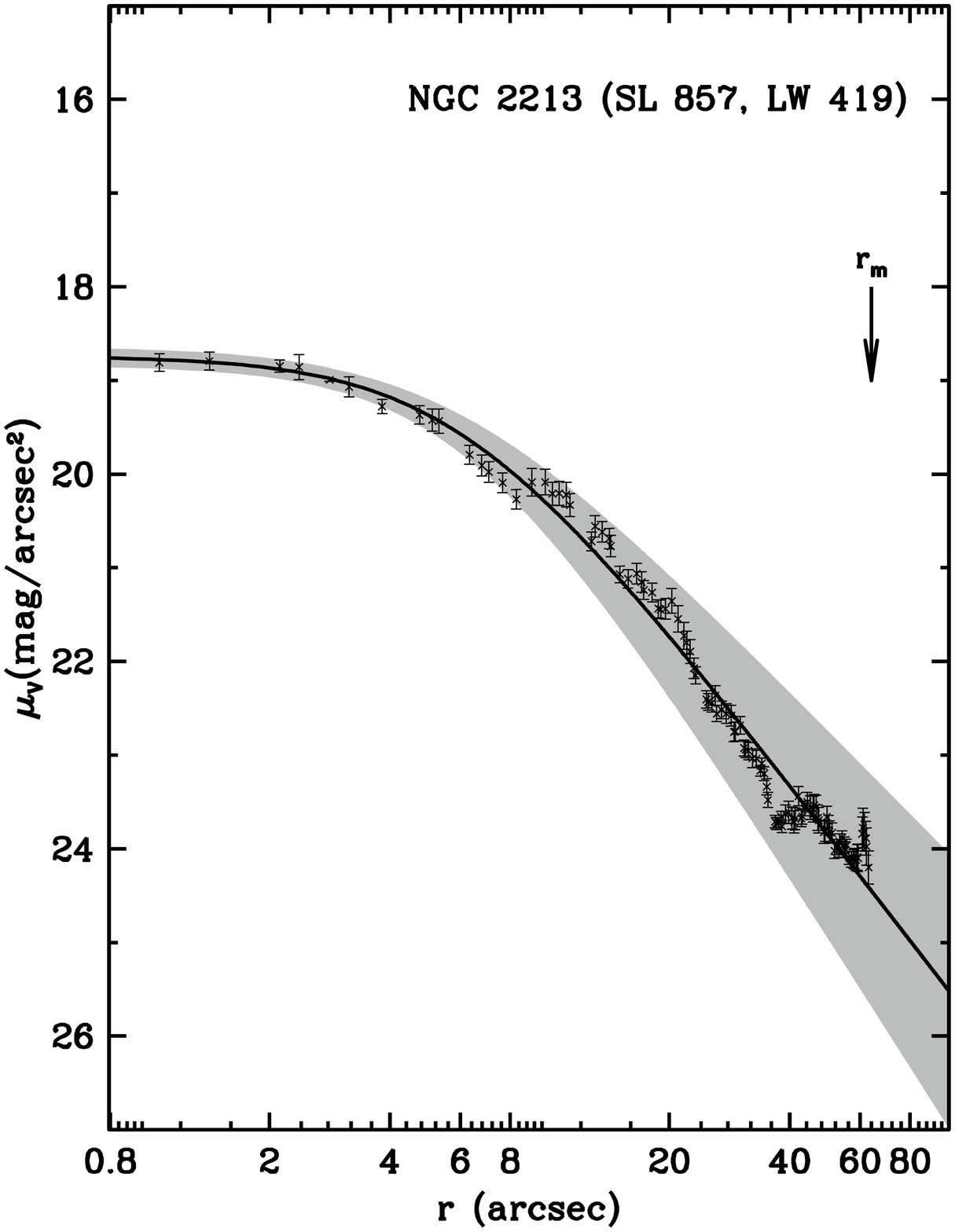}
   \includegraphics[scale=0.30,viewport=0 0 550 725,clip]{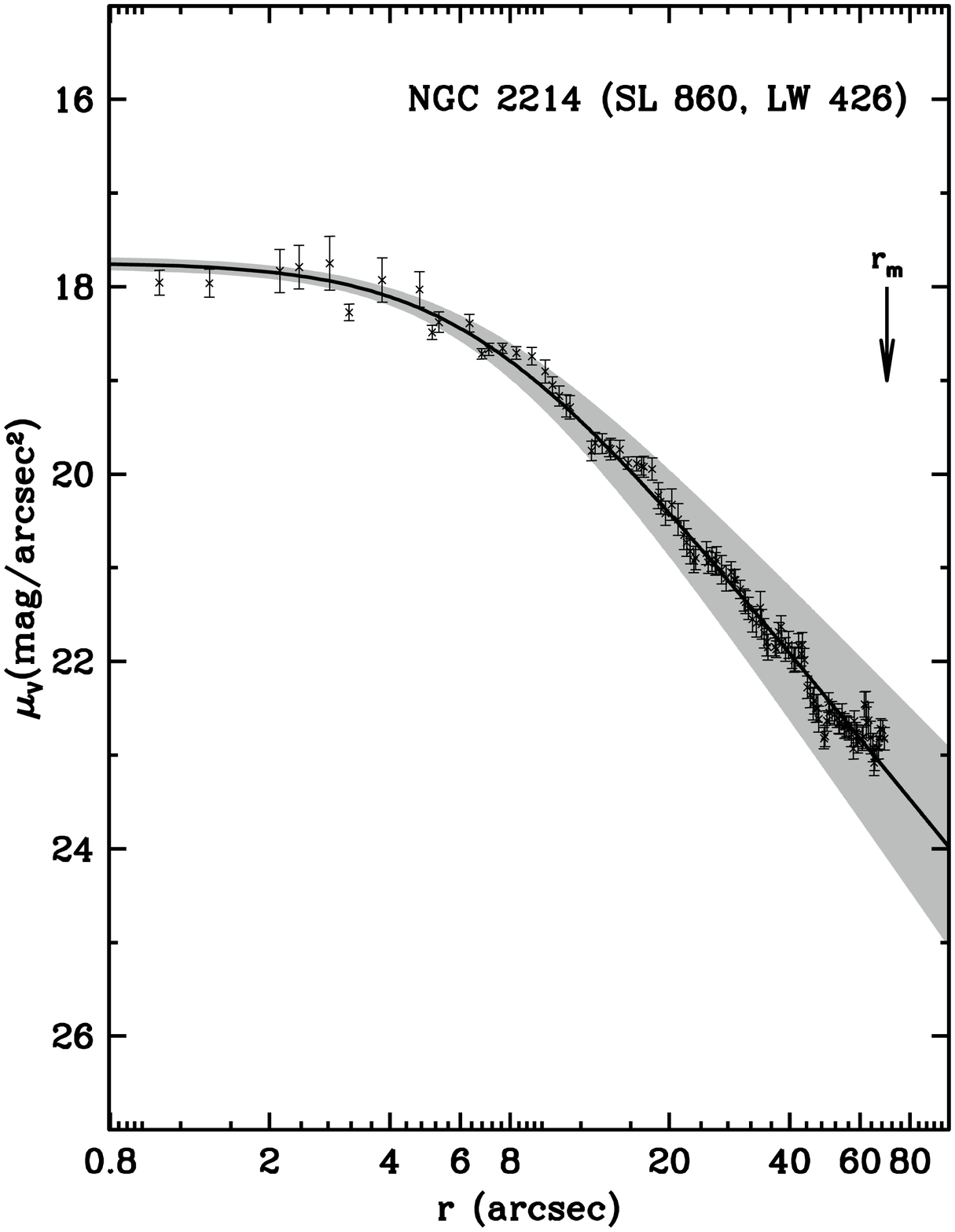}
   \caption[]{continued}
\end{figure*}
\begin{figure*}
\addtocounter{figure}{-1}
   \centering
   \includegraphics[scale=0.30,viewport=0 0 550 725,clip]{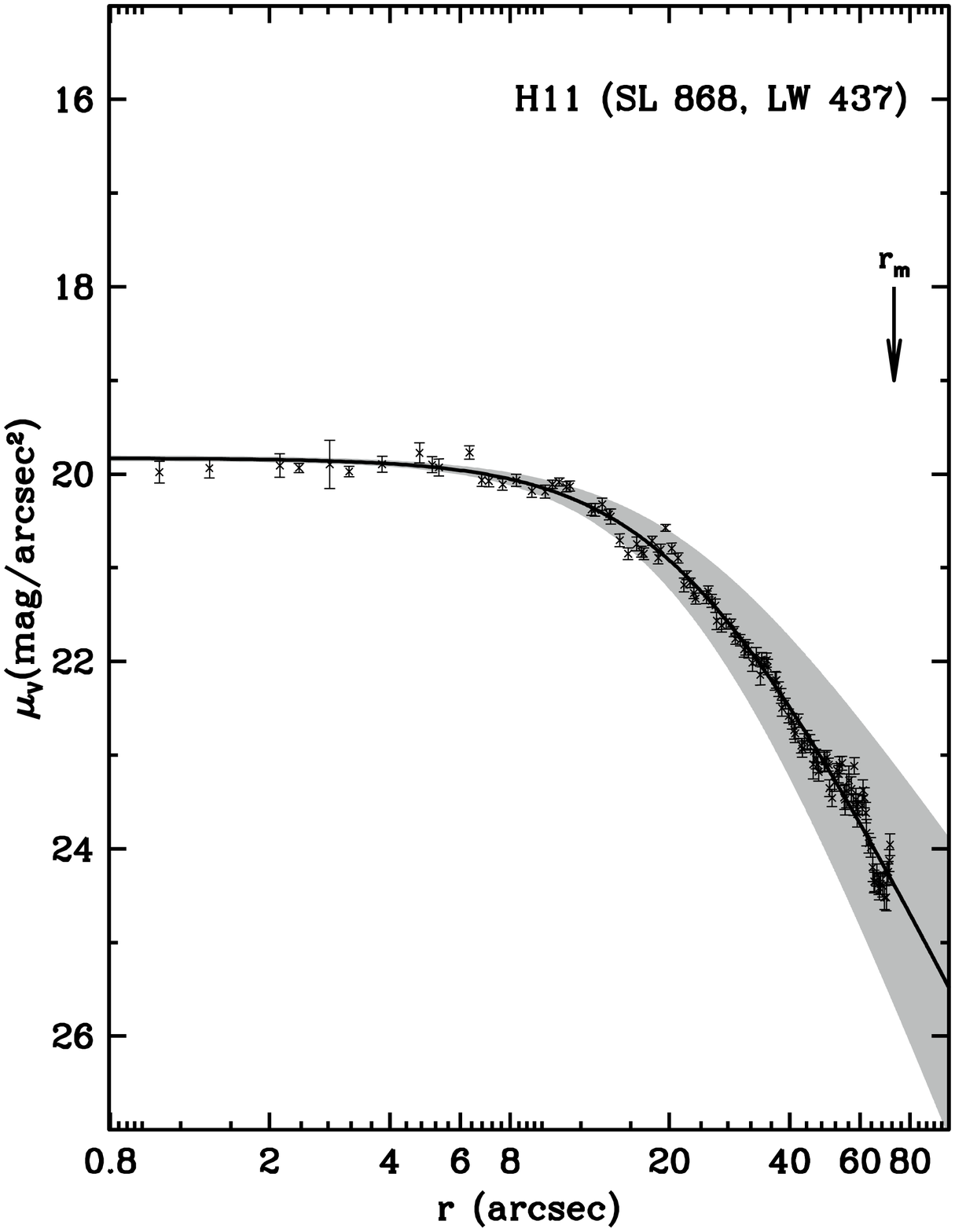}
   \includegraphics[scale=0.30,viewport=0 0 550 725,clip]{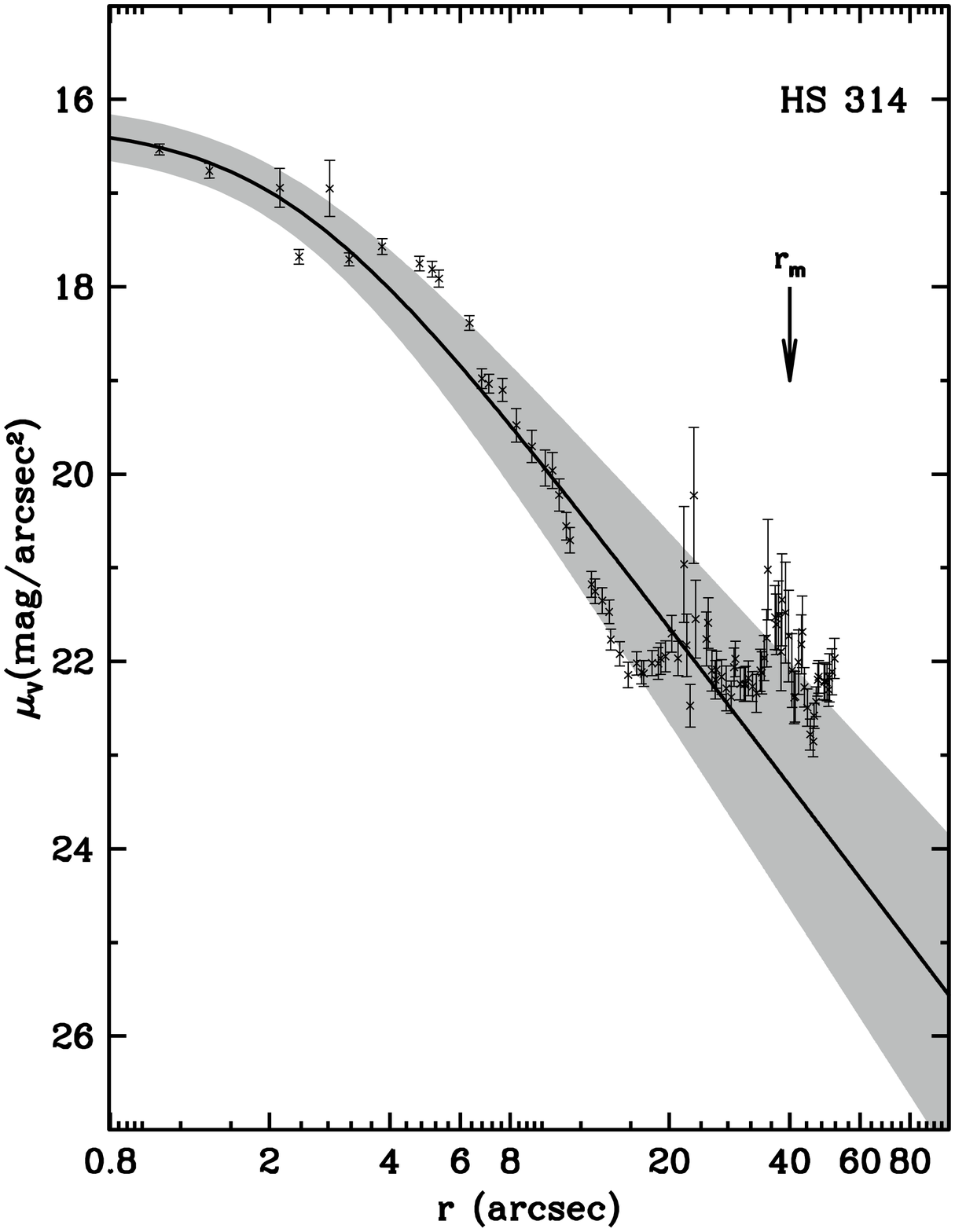}
   \caption[]{continued}
\end{figure*}

\subsection{Luminosity}
\label{sec:lum}

Integration of Eq. \ref{eq:eff} from the center to the maximum fit radius $r_m$
yields the luminosity $L(r_m)$:

\begin{equation}
   L(r_m) = \frac{2 \pi 10^{-0.4\mu_0} a^2}{\gamma - 2}\Bigg[1 - \Bigg(1 + \frac{r^2_m}{a^2}\Bigg)^{1 - \gamma/2}\Bigg].
   \label{eq:lum}
\end{equation}

For $r_m \rightarrow \infty$ and $\gamma >$ 2, we have the asymptotic cluster
luminosity $L_{\infty}$:

\begin{equation}
   L_{\infty} = \frac{2 \pi 10^{-0.4\mu_0} a^2}{\gamma - 2}.
   \label{eq:linf}
\end{equation}

\section{Results}
\label{sec:res}

\subsection{Structural parameters, luminosity and mass}
\label{sec:par}

Table \ref{tab:par} presents the structural parameters obtained by fitting
Eq. \ref{eq:eff}, together with the core radius obtained from the {\it a} and
$\gamma$ parameters in Eq. \ref{eq:rc}.

Luminosity estimates can be obtained by means of Eqs. \ref{eq:lum} and
\ref{eq:linf}, both as a function of the fitting radius ($r_m$), and asymptotic
luminosity. Mass estimates have been calculated by multiplying the luminosity
by the appropriate mass-to-light ratio. The latter values were obtained (Table
\ref{tab:lum}) from the calibration in \citet{Mack03a,Mack03b}, which are based
on evolutionary synthesis code of \citet{Fioc97} (PEGASE v2.0, 1999).

The calculated values for $L(r_m)$, $L_{\infty}$, in the V band, $M(r_m)$ and $M_{\infty}$ are
listed in Table \ref{tab:lum}, together with the fit radius $r_m$.

\begin{table*}[!h]
\caption[]{Structural parameters for the cluster sample derived from the best
fitting EFF profiles and core radii.}
\label{tab:par}
\renewcommand{\tabcolsep}{5.4mm}
\renewcommand{\arraystretch}{1.1}
\begin{tabular}{lccrcc}
\hline
\hline
\multicolumn{1}{l}{Cluster} &\multicolumn{1}{c}{$\mu_{0,V}$} &\multicolumn{1}{c}{$\gamma$} &\multicolumn{1}{c}{$a$} &\multicolumn{1}{c}{$r_c$} &\multicolumn{1}{r}{}\\
\multicolumn{1}{l}{} &\multicolumn{1}{c}{(mag/arcsec$^2$)} &\multicolumn{1}{c}{} &\multicolumn{1}{c}{(arcsec)} &\multicolumn{1}{c}{(pc)} &\multicolumn{1}{r}{Adopted statistics}\\
\multicolumn{1}{l}{(1)} &\multicolumn{1}{c}{(2)} &\multicolumn{1}{c}{(3)} &\multicolumn{1}{c}{(4)} &\multicolumn{1}{c}{(5)} &\multicolumn{1}{c}{(6)}\\
\hline
\multicolumn{6}{c}{SMC} \\
\hline
NGC\,121     & 18.26 $\pm$ 0.04 &  3.20 $\pm$ 0.08 & 13.38 $\pm$ 0.50 & 2.39 $\pm$ 0.01 & median\\
NGC\,176     & 20.30 $\pm$ 0.06 &  6.09 $\pm$ 1.78 & 20.60 $\pm$ 4.44 & 2.20 $\pm$ 0.11 & mean  \\
K\,17        & 19.16 $\pm$ 0.08 &  2.87 $\pm$ 0.12 &  7.78 $\pm$ 0.54 & 1.49 $\pm$ 0.01 & mean  \\
NGC\,241+242 & 17.90 $\pm$ 0.17 &  5.90 $\pm$ 3.11 &  4.76 $\pm$ 1.96 & 0.60 $\pm$ 0.07 & mean  \\
NGC\,290     & 17.75 $\pm$ 0.09 &  3.09 $\pm$ 0.39 &  6.07 $\pm$ 0.97 & 1.11 $\pm$ 0.03 & mean  \\
L\,48        & 18.86 $\pm$ 0.10 &  4.27 $\pm$ 0.32 &  8.77 $\pm$ 0.90 & 1.32 $\pm$ 0.02 & mean  \\
K\,34        & 19.23 $\pm$ 0.06 &  2.48 $\pm$ 0.15 &  9.66 $\pm$ 0.82 & 2.39 $\pm$ 0.04 & median\\
NGC\,330     & 16.49 $\pm$ 0.16 &  2.34 $\pm$ 0.08 &  6.77 $\pm$ 0.74 & 1.48 $\pm$ 0.03 & mean  \\
L\,56        & 16.32 $\pm$ 0.35 &  2.16 $\pm$ 0.08 &  2.20 $\pm$ 0.45 & 0.51 $\pm$ 0.01 & median\\
NGC\,339     & 21.09 $\pm$ 0.04 &  2.77 $\pm$ 0.31 & 31.41 $\pm$ 3.54 & 7.23 $\pm$ 0.71 & mean  \\
NGC\,346     & 17.90 $\pm$ 0.11 &  2.53 $\pm$ 0.22 &  9.68 $\pm$ 0.86 & 2.01 $\pm$ 0.03 & median\\
IC\,1611     & 18.98 $\pm$ 0.10 &  2.06 $\pm$ 0.19 &  6.24 $\pm$ 1.01 & 1.49 $\pm$ 0.06 & median\\
IC\,1612     & 18.22 $\pm$ 0.19 &  2.13 $\pm$ 0.11 &  2.12 $\pm$ 0.32 & 0.49 $\pm$ 0.01 & mean  \\
L\,66        & 17.41 $\pm$ 0.13 &  2.76 $\pm$ 0.12 &  3.79 $\pm$ 0.37 & 0.74 $\pm$ 0.01 & mean  \\
NGC\,361     & 20.32 $\pm$ 0.04 &  2.10 $\pm$ 0.08 & 15.72 $\pm$ 0.95 & 4.34 $\pm$ 0.07 & median\\
K\,47        & 19.02 $\pm$ 0.05 & 10.77 $\pm$ 3.37 & 19.35 $\pm$ 4.03 & 1.74 $\pm$ 0.15 & mean  \\
IC\,1624     & 19.64 $\pm$ 0.06 &  2.72 $\pm$ 0.16 &  9.94 $\pm$ 0.84 & 1.97 $\pm$ 0.03 & median\\
NGC\,411     & 19.75 $\pm$ 0.05 &  3.26 $\pm$ 0.14 & 17.64 $\pm$ 1.05 & 3.12 $\pm$ 0.04 & median\\
NGC\,416     & 18.43 $\pm$ 0.02 &  3.11 $\pm$ 0.09 & 12.63 $\pm$ 0.46 & 2.30 $\pm$ 0.01 & median\\
NGC\,419     & 18.25 $\pm$ 0.02 &  2.70 $\pm$ 0.06 & 14.74 $\pm$ 0.47 & 2.93 $\pm$ 0.01 & median\\
NGC\,458     & 18.94 $\pm$ 0.06 &  3.43 $\pm$ 0.10 & 14.46 $\pm$ 0.73 & 2.48 $\pm$ 0.02 & mean  \\
L\,114       & 18.99 $\pm$ 0.29 &  2.06 $\pm$ 0.11 &  3.36 $\pm$ 0.53 & 0.80 $\pm$ 0.02 & mean  \\
\hline
\multicolumn{6}{c}{LMC} \\
\hline
NGC\,1783    & 18.68 $\pm$ 0.03 &  2.72 $\pm$ 0.10 & 21.06 $\pm$ 1.00 & 4.90 $\pm$ 0.06 & median\\
NGC\,1818    & 17.16 $\pm$ 0.04 &  3.03 $\pm$ 0.15 & 10.16 $\pm$ 0.63 & 1.88 $\pm$ 0.01 & median\\
NGC\,1831    & 18.88 $\pm$ 0.02 &  3.42 $\pm$ 0.16 & 22.07 $\pm$ 0.77 & 3.79 $\pm$ 0.02 & median\\
NGC\,1847    & 17.30 $\pm$ 0.05 &  2.09 $\pm$ 0.12 &  4.56 $\pm$ 0.49 & 1.07 $\pm$ 0.01 & mean  \\
NGC\,1856    & 16.95 $\pm$ 0.03 &  2.42 $\pm$ 0.04 &  9.57 $\pm$ 0.27 & 2.04 $\pm$ 0.00 & median\\
NGC\,1866    & 17.27 $\pm$ 0.03 &  2.84 $\pm$ 0.06 & 16.03 $\pm$ 0.59 & 3.09 $\pm$ 0.01 & mean  \\
NGC\,1868    & 17.69 $\pm$ 0.05 &  2.85 $\pm$ 0.03 &  7.47 $\pm$ 0.24 & 1.44 $\pm$ 0.00 & mean  \\
NGC\,1870    & 16.38 $\pm$ 0.06 &  3.15 $\pm$ 0.07 &  4.18 $\pm$ 0.19 & 0.76 $\pm$ 0.00 & median\\
NGC\,1978    & 18.35 $\pm$ 0.03 &  2.82 $\pm$ 0.08 & 19.67 $\pm$ 0.73 & 3.81 $\pm$ 0.02 & mean  \\
NGC\,2004    & 15.45 $\pm$ 0.20 &  2.46 $\pm$ 0.16 &  3.82 $\pm$ 0.61 & 0.81 $\pm$ 0.02 & mean  \\
NGC\,2011    & 16.64 $\pm$ 0.22 &  2.29 $\pm$ 0.09 &  2.98 $\pm$ 0.43 & 0.66 $\pm$ 0.01 & mean  \\
NGC\,2100    & 16.77 $\pm$ 0.07 &  2.46 $\pm$ 0.10 &  6.80 $\pm$ 0.49 & 1.44 $\pm$ 0.01 & mean  \\
NGC\,2121    & 20.59 $\pm$ 0.02 &  3.66 $\pm$ 0.36 & 47.21 $\pm$ 3.85 & 7.78 $\pm$ 0.44 & median\\
NGC\,2157    & 16.89 $\pm$ 0.08 &  3.65 $\pm$ 0.12 & 11.92 $\pm$ 0.69 & 1.97 $\pm$ 0.01 & mean  \\
NGC\,2159*    & 19.08 $\pm$ 0.08 &  2.42 $\pm$ 0.26 & 12.66 $\pm$ 1.83 & 2.70 $\pm$ 0.16 & median\\
NGC\,2164    & 16.71 $\pm$ 0.08 &  3.29 $\pm$ 0.06 &  9.09 $\pm$ 0.40 & 1.60 $\pm$ 0.01 & mean  \\
NGC\,2210    & 16.98 $\pm$ 0.03 &  3.06 $\pm$ 0.05 &  7.65 $\pm$ 0.25 & 1.41 $\pm$ 0.00 & mean  \\
NGC\,2213    & 18.74 $\pm$ 0.10 &  2.22 $\pm$ 0.06 &  6.02 $\pm$ 0.48 & 1.36 $\pm$ 0.01 & mean  \\
NGC\,2214    & 17.74 $\pm$ 0.07 &  2.10 $\pm$ 0.04 &  6.49 $\pm$ 0.36 & 1.52 $\pm$ 0.01 & mean  \\
H\,11        & 19.83 $\pm$ 0.03 &  3.42 $\pm$ 0.12 & 22.37 $\pm$ 0.97 & 3.84 $\pm$ 0.03 & median\\
HS\,314      & 16.26 $\pm$ 0.25 &  2.25 $\pm$ 0.12 &  2.22 $\pm$ 0.41 & 0.50 $\pm$ 0.01 & mean  \\
\hline
\end{tabular}
\begin{list}{Table Notes.}
\item Col. 2: Central surface brightness. Col. 3: Gamma parameter. Col. 4: $a$
parameter. Col. 5: Core radius. Col. 6: Statistical adopted for the surface
brightness. * For NGC\,2159 we used $\mu_{0,B}$.
\end{list}
\end{table*}

\begin{table*}[!h]
\caption[]{Luminosity in the V band and mass estimate calculated using structural parameters.}
\renewcommand{\tabcolsep}{3.15mm}
\renewcommand{\arraystretch}{1.1}
\label{tab:lum}
\begin{tabular}{lccrccccc}
\hline
\hline
\multicolumn{1}{l}{Cluster} &\multicolumn{1}{c}{Age} & \multicolumn{1}{c}{Ref.} &\multicolumn{1}{c}{$r_m$} &\multicolumn{1}{r}{$M/L_V$} &\multicolumn{1}{c}{$\log L_m$} &\multicolumn{1}{c}{$\log L_{\infty}$} &\multicolumn{1}{c}{$\log M_m$} &\multicolumn{1}{c}{$\log M_{\infty}$}\\
&\multicolumn{1}{c}{(Gyr)} &&\multicolumn{1}{c}{(pc)} &\multicolumn{1}{r}{} &\multicolumn{1}{c}{(${\rm L}_{\odot,V}$)} &\multicolumn{1}{c}{(${\rm L}_{\odot,V}$)} &\multicolumn{1}{c}{(${\rm M}_{\odot}$)} &\multicolumn{1}{c}{(${\rm M}_{\odot}$)}\\
\multicolumn{1}{l}{(1)} &\multicolumn{1}{c}{(2)} &\multicolumn{1}{c}{(3)} &\multicolumn{1}{c}{(4)} &\multicolumn{1}{c}{(5)} &\multicolumn{1}{c}{(6)} &\multicolumn{1}{c}{(7)} &\multicolumn{1}{c}{(8)} &\multicolumn{1}{c}{(9)}\\
\hline
\multicolumn{9}{c}{SMC} \\
\hline
NGC\,121     & 11.90 $\pm$ 1.30 & 2   & 12.88 & 2.74 & 5.00 $\pm$ 0.05 & 5.09 $\pm$ 0.05 & 5.44 $\pm$ 0.14 & 5.53 $\pm$ 0.13 \\
NGC\,176     &  0.46 $\pm$ 0.01 & 3   & 7.29  & 0.23 & 4.17 $\pm$ 0.34 & 4.43 $\pm$ 0.27 & 3.53 $\pm$ 0.08 & 3.80 $\pm$ 0.06 \\
K\,17        &  0.30 $\pm$ 0.10 & 9   & 6.80  & 0.28 & 4.23 $\pm$ 0.10 & 4.40 $\pm$ 0.09 & 4.81 $\pm$ 0.38 & 4.98 $\pm$ 0.35 \\
NGC\,241+242 &  0.07 $\pm$ 0.04 & 10  & 1.46  & 0.17 & 3.75 $\pm$ 0.56 & 3.83 $\pm$ 0.50 & 2.98 $\pm$ 0.09 & 3.06 $\pm$ 0.09 \\
NGC\,290     &  0.03 $\pm$ 0.01 & 14  & 4.86  & 0.10 & 4.52 $\pm$ 0.24 & 4.65 $\pm$ 0.21 & 3.52 $\pm$ 0.02 & 3.65 $\pm$ 0.02 \\
L\,48        &  0.15 $\pm$ 0.04 & 3   & 9.48  & 0.20 & 4.19 $\pm$ 0.11 & 4.21 $\pm$ 0.12 & 3.50 $\pm$ 0.02 & 3.51 $\pm$ 0.02 \\
K\,34        &  0.24 $\pm$ 0.12 & 10  & 11.42 & 0.22 & 4.60 $\pm$ 0.19 & 4.90 $\pm$ 0.16 & 3.94 $\pm$ 0.04 & 4.24 $\pm$ 0.03 \\
NGC\,330     &  0.03 $\pm$ 0.01 & 1   & 17.01 & 0.09 & 5.50 $\pm$ 0.17 & 5.76 $\pm$ 0.15 & 4.45 $\pm$ 0.02 & 4.71 $\pm$ 0.01 \\
L\,56        & 0.006 $\pm$ 0.01 & 14  & 12.15 & 0.14 & 4.77 $\pm$ 0.35 & 5.17 $\pm$ 0.31 & 3.92 $\pm$ 0.05 & 4.32 $\pm$ 0.04 \\
NGC\,339     &  6.30 $\pm$ 1.30 & 2   & 19.98 & 1.66 & 4.67 $\pm$ 0.26 & 4.97 $\pm$ 0.20 & 4.89 $\pm$ 0.43 & 5.19 $\pm$ 0.33 \\
NGC\,346     & $\sim$ 0.003     & 17  & 2.92  & 0.02 & 4.65 $\pm$ 0.27 & 5.31 $\pm$ 0.20 & 2.95 $\pm$ 0.01 & 3.61 $\pm$ 0.00 \\
IC\,1611     &  0.11 $\pm$ 0.05 & 10  & 9.23  & 0.16 & 4.46 $\pm$ 1.95 & 5.44 $\pm$ 1.38 & 3.66 $\pm$ 0.31 & 4.65 $\pm$ 0.22 \\
IC\,1612     &  $\sim$ 0.10     & 16  & 3.89  & 0.14 & 3.84 $\pm$ 0.53 & 4.47 $\pm$ 0.40 & 2.98 $\pm$ 0.07 & 3.62 $\pm$ 0.06 \\
L\,66        &  0.15 $\pm$ 0.10 & 11  & 6.32  & 0.08 & 4.42 $\pm$ 0.12 & 4.53 $\pm$ 0.12 & 3.32 $\pm$ 0.01 & 3.44 $\pm$ 0.01 \\
NGC\,361     &  8.10 $\pm$ 1.20 & 2   & 19.13 & 2.03 & 4.70 $\pm$ 0.49 & 5.57 $\pm$ 0.35 & 5.01 $\pm$ 1.00 & 5.87 $\pm$ 0.71 \\
K\,47        & $\sim$ 0.007     & 16  & 4.86  & 0.10 & 4.23 $\pm$ 0.28 & 4.24 $\pm$ 0.25 & 3.23 $\pm$ 0.03 & 3.24 $\pm$ 0.02 \\
IC\,1624     &  0.06 $\pm$ 0.03 & 10  & 9.72  & 0.12 & 4.31 $\pm$ 0.14 & 4.50 $\pm$ 0.12 & 3.39 $\pm$ 0.02 & 3.58 $\pm$ 0.01 \\
NGC\,411     &  0.20 $\pm$ 0.10 & 3   & 15.91 & 0.63 & 4.63 $\pm$ 0.08 & 4.71 $\pm$ 0.07 & 4.43 $\pm$ 0.05 & 4.51 $\pm$ 0.05 \\
NGC\,416     &  6.90 $\pm$ 1.10 & 2   & 14.58 & 1.79 & 4.93 $\pm$ 0.06 & 5.01 $\pm$ 0.05 & 5.18 $\pm$ 0.10 & 5.26 $\pm$ 0.09 \\
NGC\,419     & $\sim$ 0.40      & 16  & 15.18 & 0.40 & 5.22 $\pm$ 0.06 & 5.41 $\pm$ 0.05 & 4.82 $\pm$ 0.02 & 5.02 $\pm$ 0.02 \\
NGC\,458     &  0.05 $\pm$ 0.01 & 3   & 17.01 & 0.25 & 4.76 $\pm$ 0.06 & 4.81 $\pm$ 0.06 & 4.16 $\pm$ 0.02 & 4.21 $\pm$ 0.01 \\
L\,114       &  5.60 $\pm$ 0.50 & 14  & 14.58 & 0.20 & 4.10 $\pm$ 1.12 & 4.90 $\pm$ 0.82 & 3.40 $\pm$ 0.22 & 4.20 $\pm$ 0.16 \\
\hline 
\multicolumn{9}{c}{LMC} \\
\hline 
NGC\,1783    &  1.30 $\pm$ 0.40  & 4  & 20.41 & 0.63 & 5.39 $\pm$ 0.09 & 5.62 $\pm$ 0.07 & 4.74 $\pm$ 0.02 & 4.96 $\pm$ 0.02 \\
NGC\,1818    &  0.02 $\pm$ 0.01  & 3  & 9.72  & 0.08 & 5.24 $\pm$ 0.10 & 5.36 $\pm$ 0.08 & 4.15 $\pm$ 0.01 & 4.26 $\pm$ 0.01 \\
NGC\,1831    &  0.32 $\pm$ 0.12  & 4  & 17.25 & 0.32 & 5.12 $\pm$ 0.07 & 5.21 $\pm$ 0.06 & 4.63 $\pm$ 0.02 & 4.71 $\pm$ 0.02 \\
NGC\,1847    &  0.02 $\pm$ 0.01  & 3  & 4.86  & 0.09 & 4.77 $\pm$ 0.82 & 5.67 $\pm$ 0.59 & 3.72 $\pm$ 0.07 & 4.62 $\pm$ 0.05 \\
NGC\,1856    &  0.12 $\pm$ 0.04  & 4  & 18.10 & 0.18 & 5.54 $\pm$ 0.06 & 5.78 $\pm$ 0.05 & 4.80 $\pm$ 0.01 & 5.04 $\pm$ 0.01 \\
NGC\,1866    &  0.09 $\pm$ 0.01  & 7  & 17.01 & 0.18 & 5.65 $\pm$ 0.05 & 5.80 $\pm$ 0.05 & 4.91 $\pm$ 0.01 & 5.06 $\pm$ 0.01 \\
NGC\,1868    &  0.33 $\pm$ 0.03  & 3  & 15.79 & 0.40 & 4.89 $\pm$ 0.04 & 4.96 $\pm$ 0.04 & 4.49 $\pm$ 0.02 & 4.57 $\pm$ 0.02 \\
NGC\,1870    &  0.72 $\pm$ 0.30  & 19 & 5.83  & 0.16 & 4.79 $\pm$ 0.06 & 4.85 $\pm$ 0.05 & 3.99 $\pm$ 0.01 & 4.06 $\pm$ 0.01 \\
NGC\,1978    &  2.50 $\pm$ 0.50  & 18 & 16.52 & 2.10 & 5.37 $\pm$ 0.07 & 5.56 $\pm$ 0.05 & 5.69 $\pm$ 0.14 & 5.88 $\pm$ 0.11 \\
NGC\,2004    &  0.03 $\pm$ 0.01  & 4  & 5.83  & 0.08 & 5.30 $\pm$ 0.24 & 5.54 $\pm$ 0.22 & 4.20 $\pm$ 0.02 & 4.45 $\pm$ 0.02 \\
NGC\,2011    & $<$ 0.01          & 6  & 12.88 & 0.05 & 4.81 $\pm$ 0.22 & 5.05 $\pm$ 0.20 & 3.50 $\pm$ 0.01 & 3.75 $\pm$ 0.01 \\
NGC\,2100    & 0.032 $\pm$ 0.019 & 4  & 11.66 & 0.07 & 5.29 $\pm$ 0.14 & 5.52 $\pm$ 0.12 & 4.14 $\pm$ 0.01 & 4.36 $\pm$ 0.01 \\
NGC\,2121    &  0.70 $\pm$ 0.20  & 3  & 18.10 & 1.33 & 4.92 $\pm$ 0.17 & 5.11 $\pm$ 0.12 & 5.05 $\pm$ 0.23 & 5.24 $\pm$ 0.16 \\
NGC\,2157    &  0.03 $\pm$ 0.02  & 3  & 14.58 & 0.11 & 5.37 $\pm$ 0.07 & 5.40 $\pm$ 0.07 & 4.41 $\pm$ 0.01 & 4.44 $\pm$ 0.01 \\
NGC\,2159*    &  0.06 $\pm$ 0.03  & 3  & 9.72  & -- & -- & -- & -- & -- \\
NGC\,2164    &  0.05 $\pm$ 0.03  & 3  & 17.98 & 0.13 & 5.31 $\pm$ 0.05 & 5.34 $\pm$ 0.05 & 4.43 $\pm$ 0.01 & 4.46 $\pm$ 0.01 \\
NGC\,2210    & 15.85 $\pm$ 1.26  & 13 & 17.01 & 3.37 & 5.13 $\pm$ 0.04 & 5.17 $\pm$ 0.04 & 5.66 $\pm$ 0.14 & 5.70 $\pm$ 0.12 \\
NGC\,2213    &  1.30 $\pm$ 0.50  & 5  & 15.55 & 0.87 & 4.55 $\pm$ 0.18 & 4.94 $\pm$ 0.14 & 4.49 $\pm$ 0.15 & 4.88 $\pm$ 0.12 \\
NGC\,2214    &  0.04 $\pm$ 0.01  & 3  & 17.01 & 0.11 & 5.08 $\pm$ 0.25 & 5.75 $\pm$ 0.18 & 4.12 $\pm$ 0.03 & 4.79 $\pm$ 0.02 \\
H\,11        & 15.00 $\pm$ 3.00  & 8  & 17.74 & 3.25 & 4.75 $\pm$ 0.06 & 4.84 $\pm$ 0.05 & 5.27 $\pm$ 0.20 & 5.35 $\pm$ 0.18 \\
HS\,314      & $<$ 0.01          & 15 & 9.72  & 0.03 & 4.72 $\pm$ 0.32 & 5.01 $\pm$ 0.28 & 3.20 $\pm$ 0.01 & 3.49 $\pm$ 0.01 \\
\hline
\end{tabular}
\begin{list}{Table Notes.}
\item (1) \citet{DaCosta98}; (2) \citet{Mighell98};
(3) \citet{Hodge83}; (4) \citet{Santos04}; (5) \citet{DaCosta85};
(6) \citet{Goul06}; (7) \citet{Becker83}; (8) \citet{Mighell96};
(9) \citet{HodgeF87}; (10) \citet{EF85}; (11) \citet{Piatti05}; 
(12) \citet{Piatti05}; (13) \citet{Geisler97}; (14) \citet{Ahum02};
(15) \citet{Bica96}; (16) \citet{Chiosi}; (17) \citet{Sabbi07};
(18) \citet{Sagar89}; (19) \citet{AL87}. * A B image was used.
\end{list}
\end{table*}

\subsection{Comparison with previous studies}
\label{sec:com}

\begin{figure}[!h]
  \begin{center}
   \includegraphics[scale=0.46,viewport=0 0 650 730,clip]{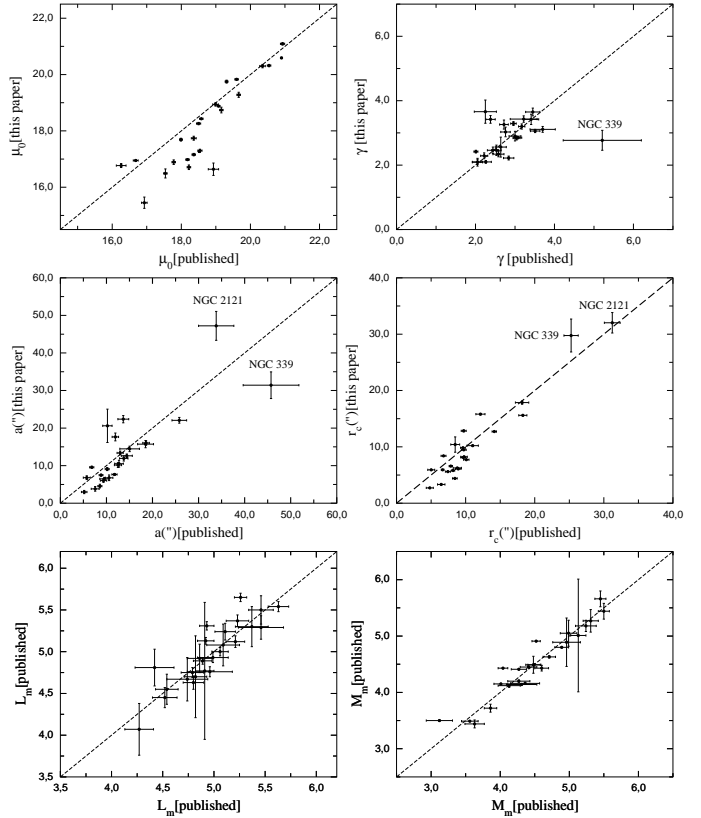}
   \caption[]{\small Measured values of the structural parameters, core radii,
   luminosity and mass compared with those in common with
   \citet{Mack03a,Mack03b}. The dashed line is plotted for reference and
   indicates identity.}
   \label{fig:com}
  \end{center}
\end{figure}

Comparison of the values measured in this study with HST results from
\citet{Mack03a,Mack03b} are plotted in Fig. \ref{fig:com}. Agreement is in
general good. Some deviant points are indicated.

\subsection{Comparison of cluster properties}
\label{sec:pro}

\begin{figure}[!h]
  \begin{center}
   \includegraphics[scale=0.5]{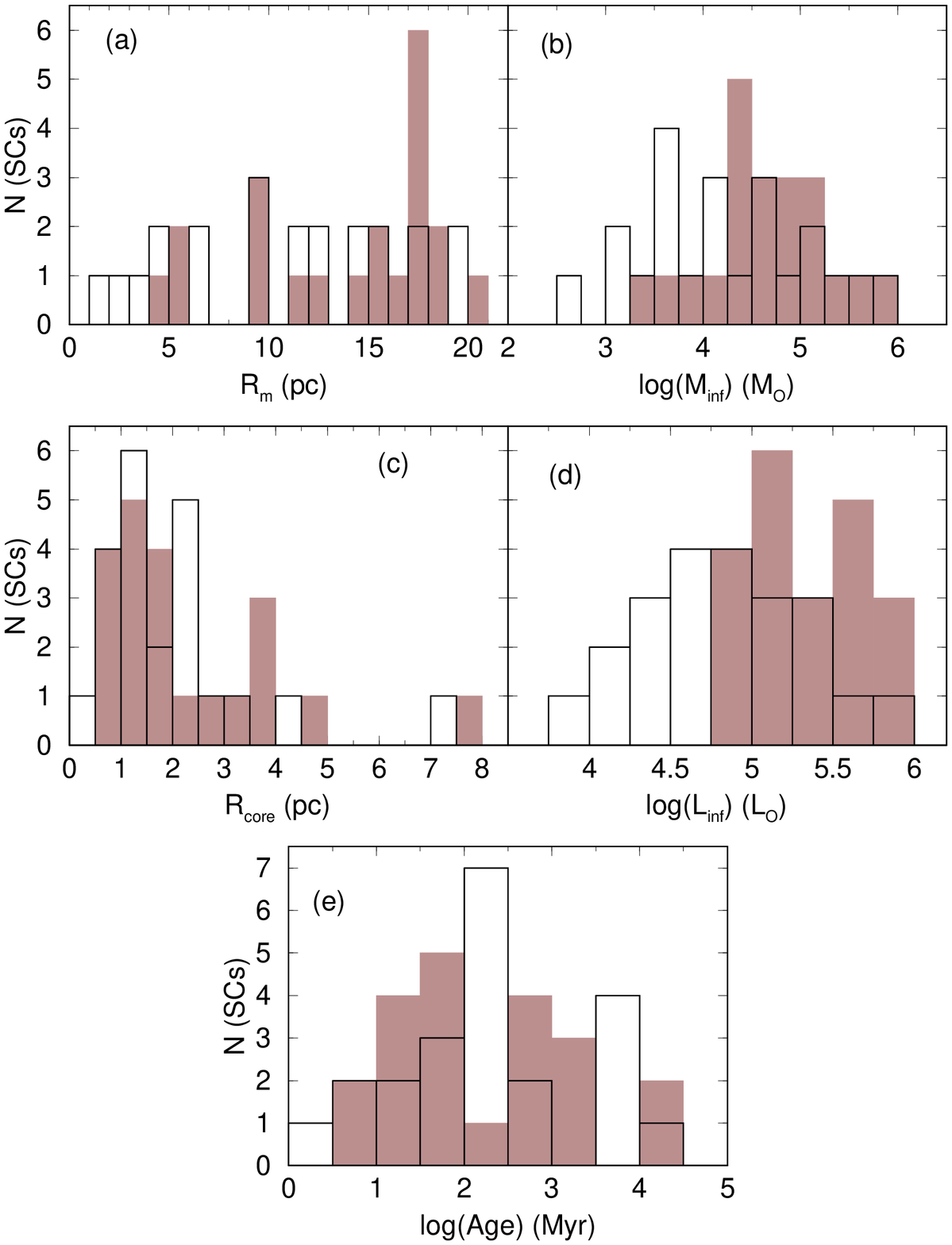}
   \caption[]{Comparison among LMC (shaded histograms) and SMC (empty) star
   clusters.}
   \label{fig:his}
  \end{center}
\end{figure}

We analyse in Fig. \ref{fig:his} statistical properties of the present star
cluster sample. As a caveat, we note that are dealing with small fractions
of the catalogued star clusters in both Clouds, $\approx4.5\%$ in the
SMC (\citealt{BicaSchM95}) and $\approx0.9\%$ in the LMC (\citealt{Bica99}).
At least for the present sample, the LMC clusters are more massive and luminous
than the SMC ones (panels a and b). In terms of radii, both samples present
similar distributions (panels c and d); however, the core radius distribution is
significantly different from that of $r_m$. The age distributions are comparable
between the present LMC and SMC clusters (panel e).

\begin{figure}[!h]
  \begin{center}
   \includegraphics[scale=0.5]{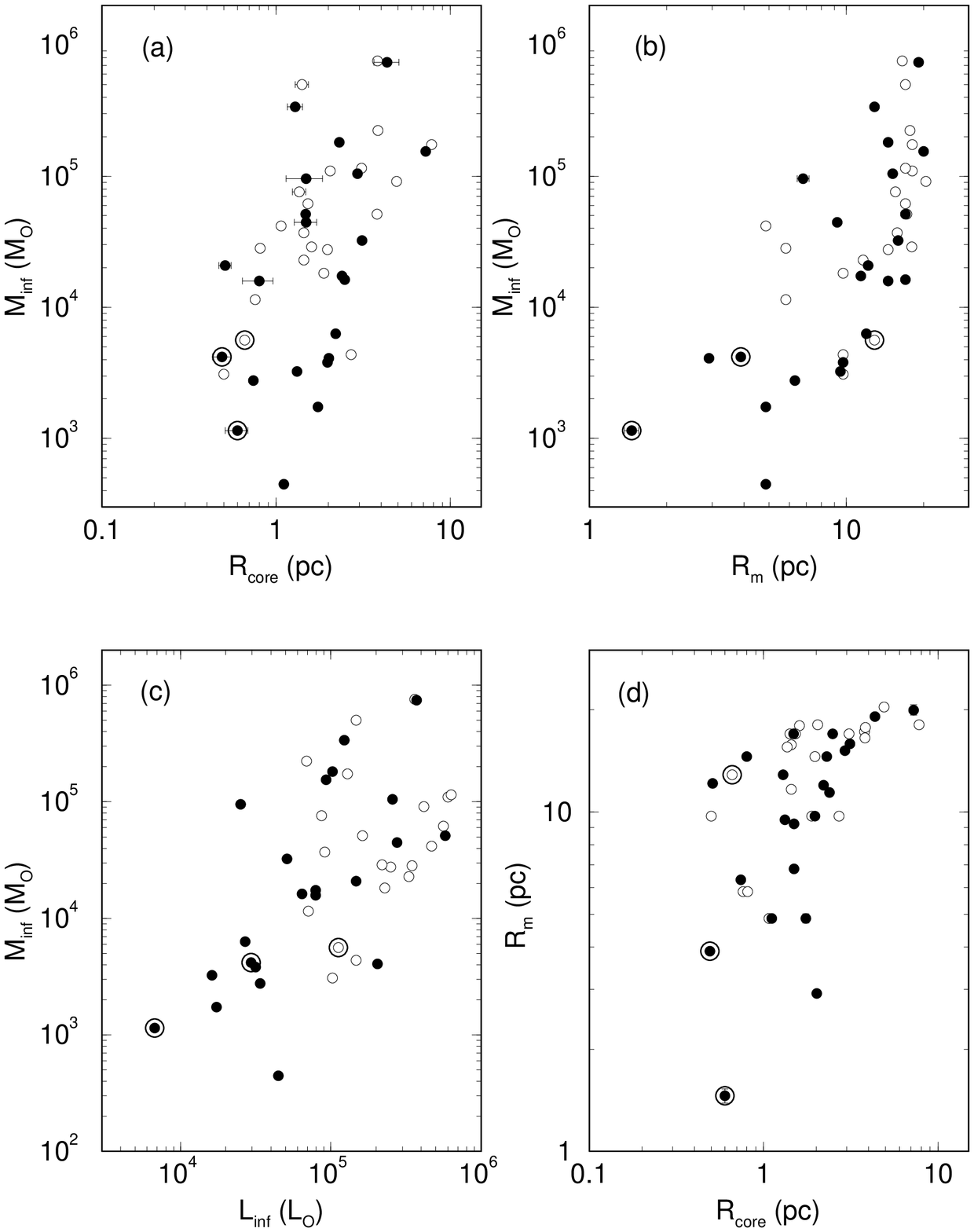}
   \caption[]{Relations among cluster parameters. Open circles are SMC and
   filled circles are LMC clusters. Binary clusters are encircled.}
   \label{fig:par}
  \end{center}
\end{figure}

In Fig. \ref{fig:par} we investigate relations between parameters. The total
mass correlates with the core radius (panel a), similarly to Galactic open
clusters (\citealt{BB2007}); in the present case, the mass reaches about two
orders of magnitude higher than in the Galactic ones, while for the core radius
the factor is about 4. Similar results apply for the maximum radius $r_m$ (panel
b). Despite significant age differences, mass and luminosity correlate (panel
c). Finally, the core and maximum radii correlate similarly to Galactic open
clusters (\citealt{BB07c}) for $r_c\la1$\,pc and $r_m\la10$\,pc (panel d); for
larger radii, a saturation effect appears to occur. A possible interpretation is
that the internal evolution continues after truncation of outer regions.

\begin{figure}[!h]
  \begin{center}
   \includegraphics[scale=0.5]{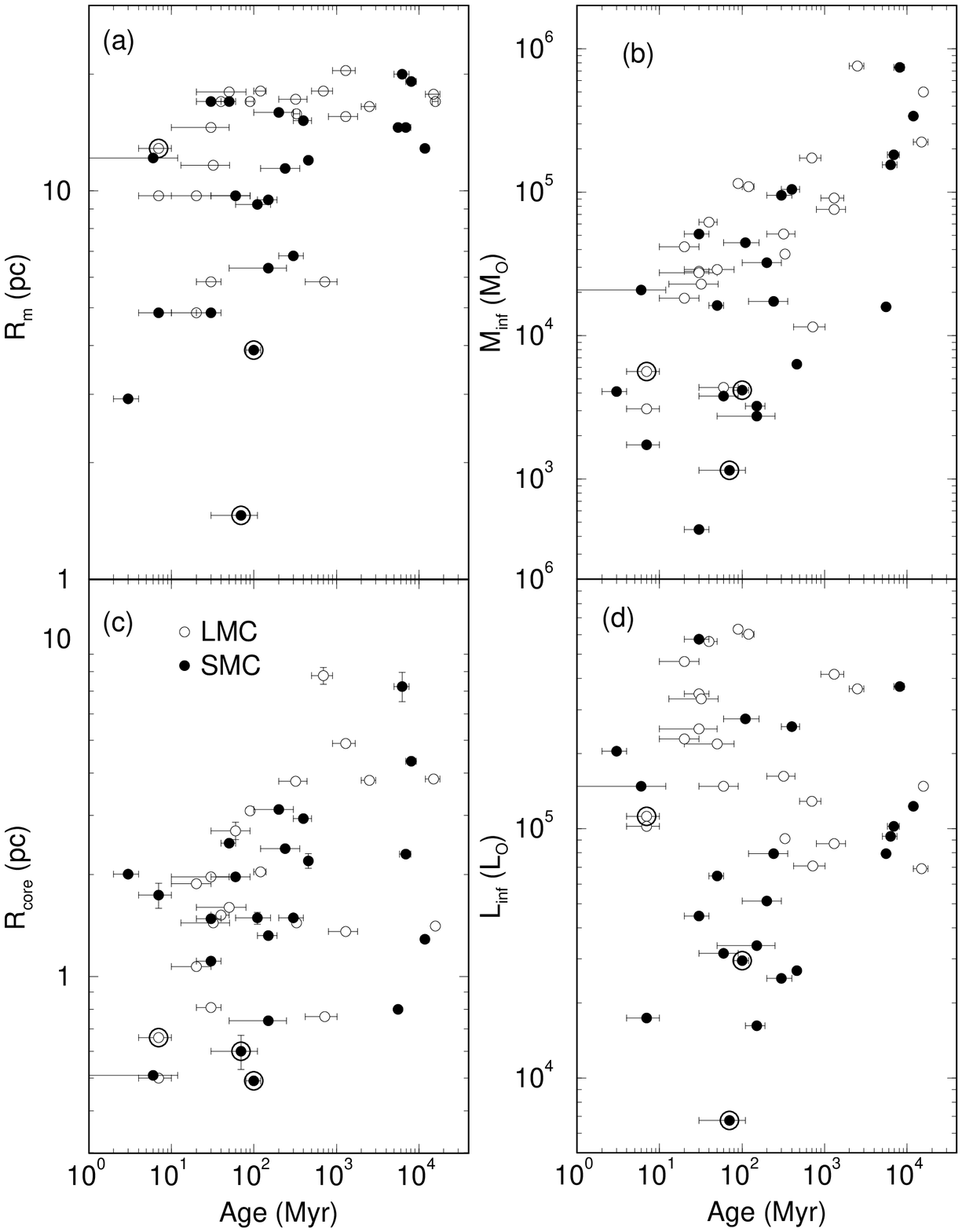}
   \caption[]{Relations of radii, luminosity and mass with cluster age. Symbols as in Fig. \ref{fig:par}.}
   \label{fig:add}
  \end{center}
\end{figure}

Maximum and core radii (Fig. \ref{fig:add} panels a and c) show a
dependence with age, having the maximum radius also evidence of a saturation
effect. Finally, mass has a significant correlation with age while luminosity
does not (panels b and d). The mass-age correlation is probably related to dynamical cluster survival. Low mass clusters disperse into the field long before reaching an old age (\citealt{Goodwin06}).

As compared to \cite{Mack03b} similar behaviours occur in the relations
in common, especially mass vs. age and core radius vs. age.

The binary clusters in Fig. \ref{fig:par} and Fig. \ref{fig:add} tend to be
among the less massive and with smaller radii clusters in the sample. Note that
we measured the main component (Fig. \ref{fig:1}).


\section{Atypical Magellanic clusters}
\label{sec:dis}

\subsection{Extended profile in the very young cluster NGC\,346}
\label{sec:346}

The giant SMC HII region cluster NGC\,346 (Fig. \ref{fig:img}) has a systematic
density excess for r$>$10$''$ (Fig. \ref{fig:1}), likewise R\,136 in 30 Doradus
(\citealt{Mack03a}). This may be attributed to star formation in a dynamically
infant cluster. The excess in NGC\,346 is not a contamination by the
neighbouring intermediate age cluster BS\,90 (\citealt{BicaSchM95,Rochau07}).

\subsection{Binary clusters and merger candidates}
\label{sec:bin}
\begin{figure}[!h]
   \begin{center}
   \fbox{\includegraphics[scale=0.23,viewport=30 30 500 500,clip]{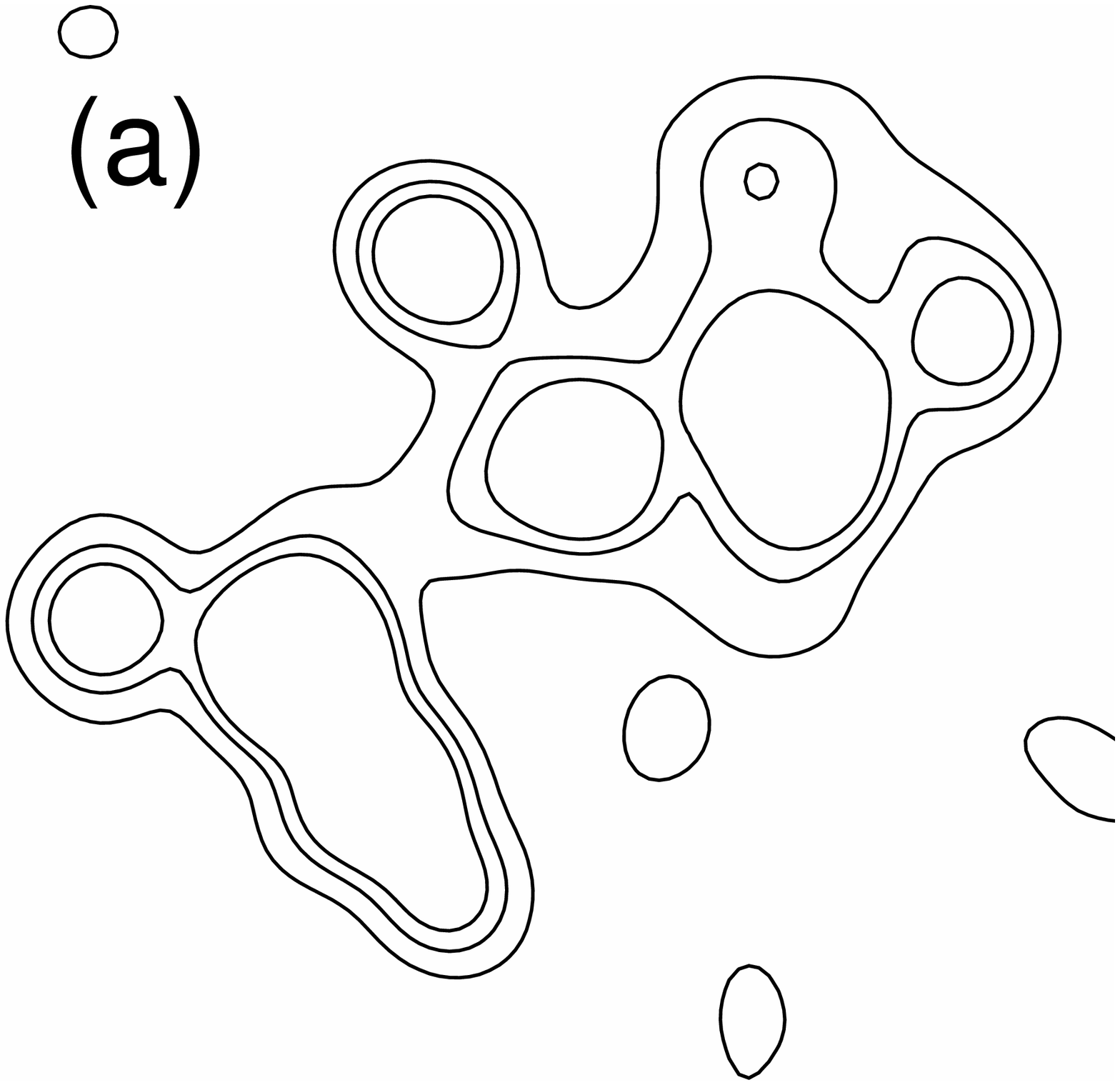}}
   \fbox{\includegraphics[scale=0.23,viewport=30 30 500 500,clip]{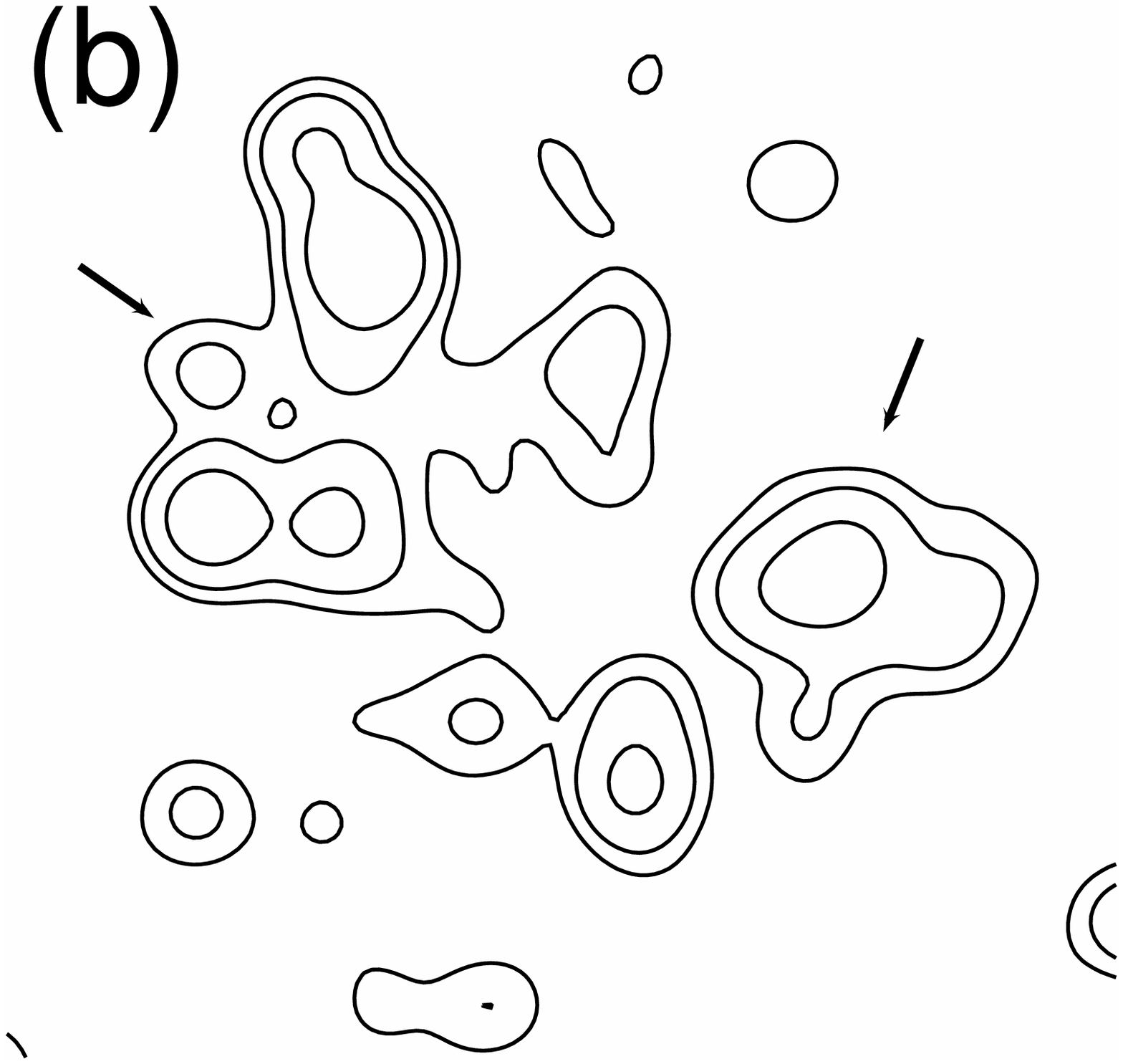}}
   \fbox{\includegraphics[scale=0.23,viewport=30 30 500 500,clip]{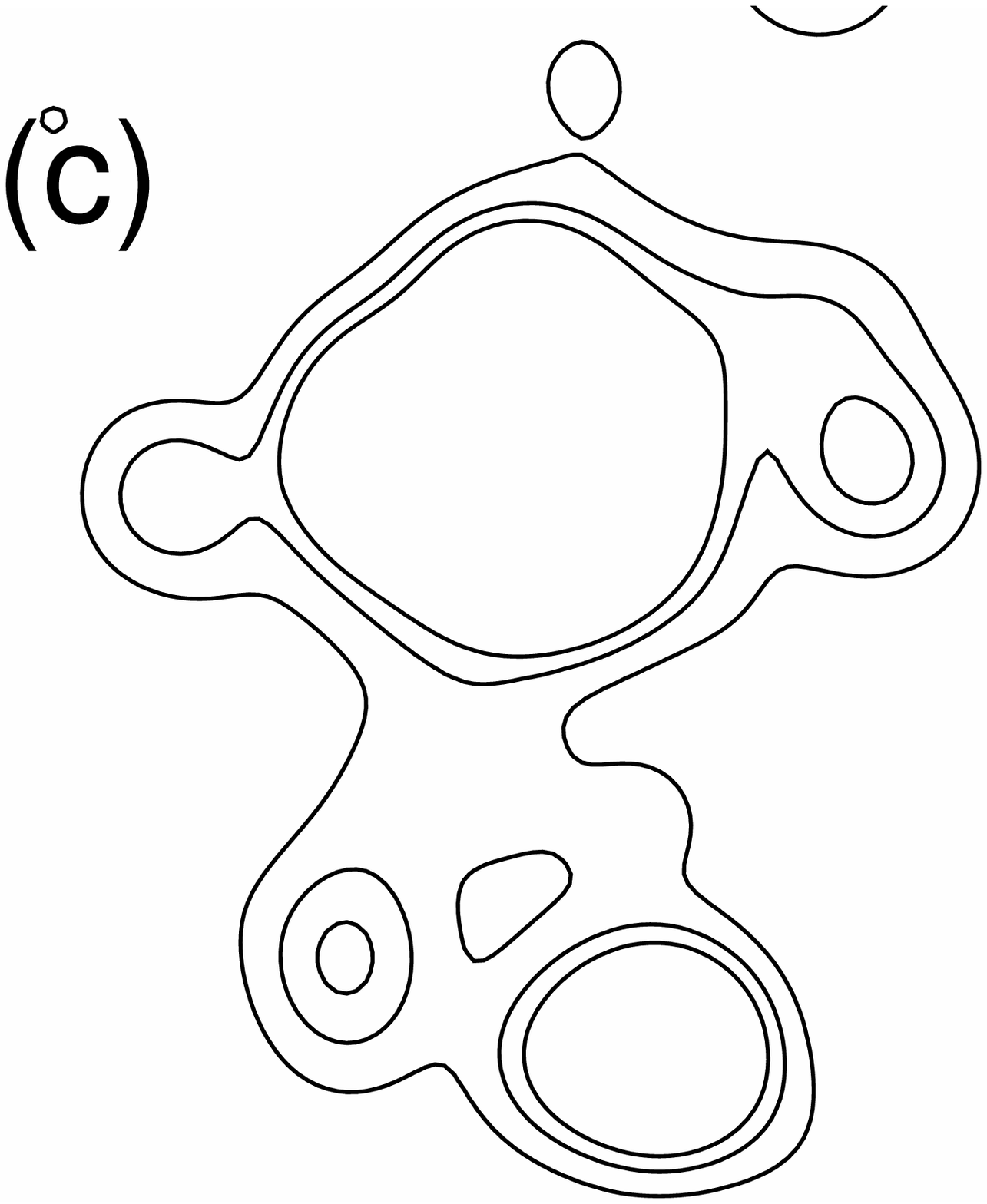}}
   \fbox{\includegraphics[scale=0.23,viewport=30 30 500 500,clip]{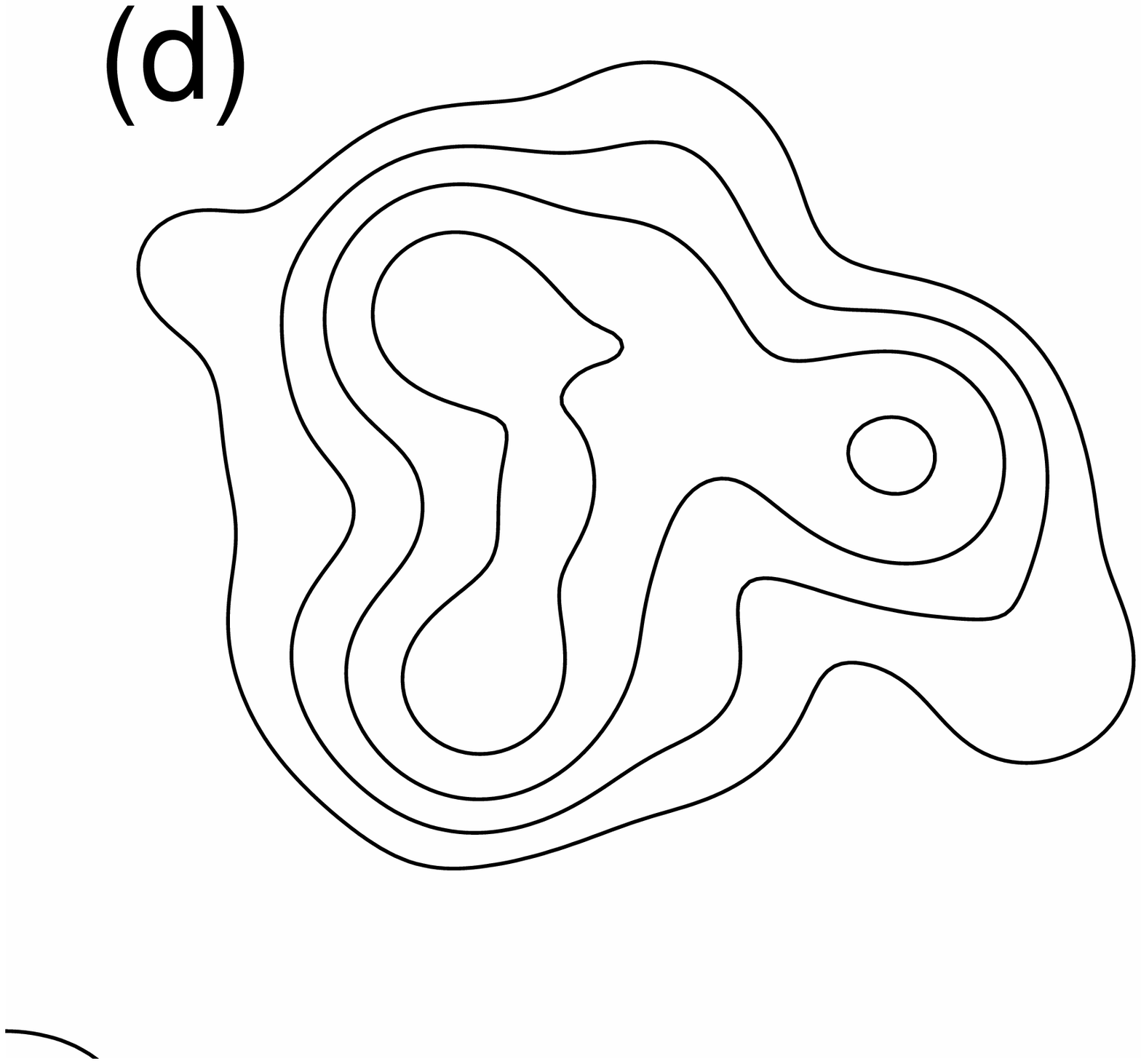}}
   \fbox{\includegraphics[scale=0.23,viewport=30 30 500 500,clip]{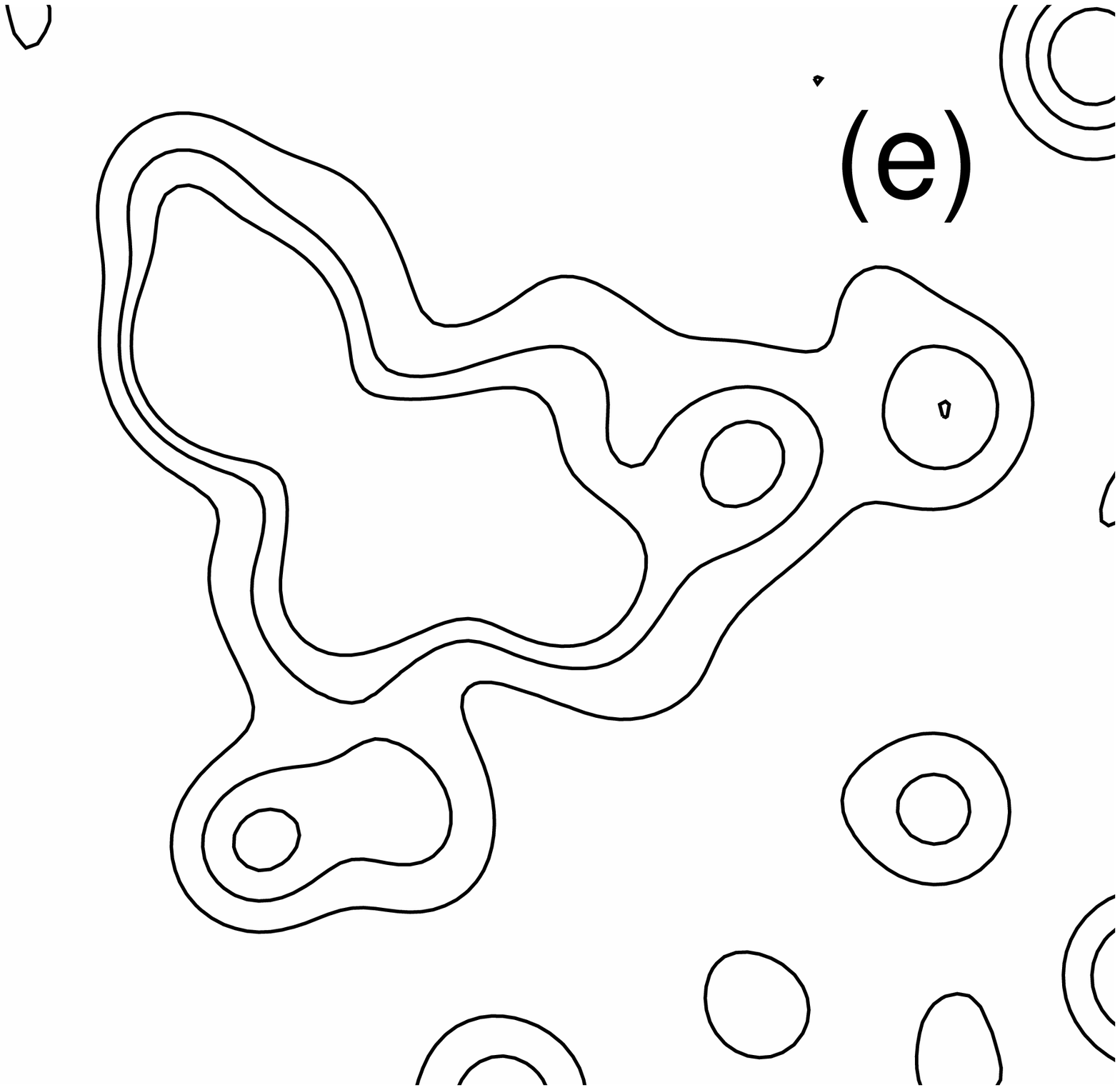}}
   \fbox{\includegraphics[scale=0.23,viewport=30 30 500 500,clip]{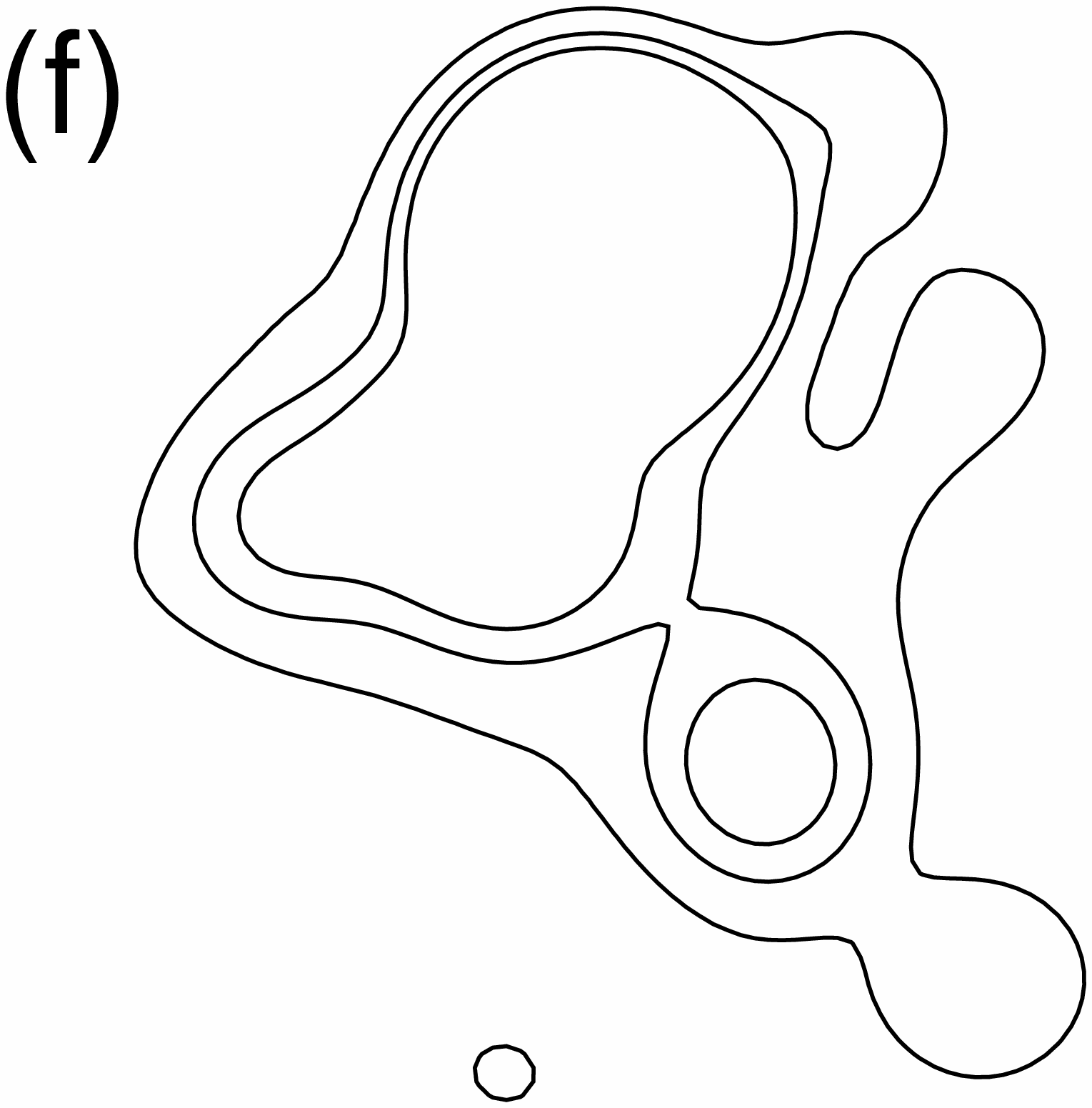}}
   \fbox{\includegraphics[scale=0.23,viewport=30 30 500 500,clip]{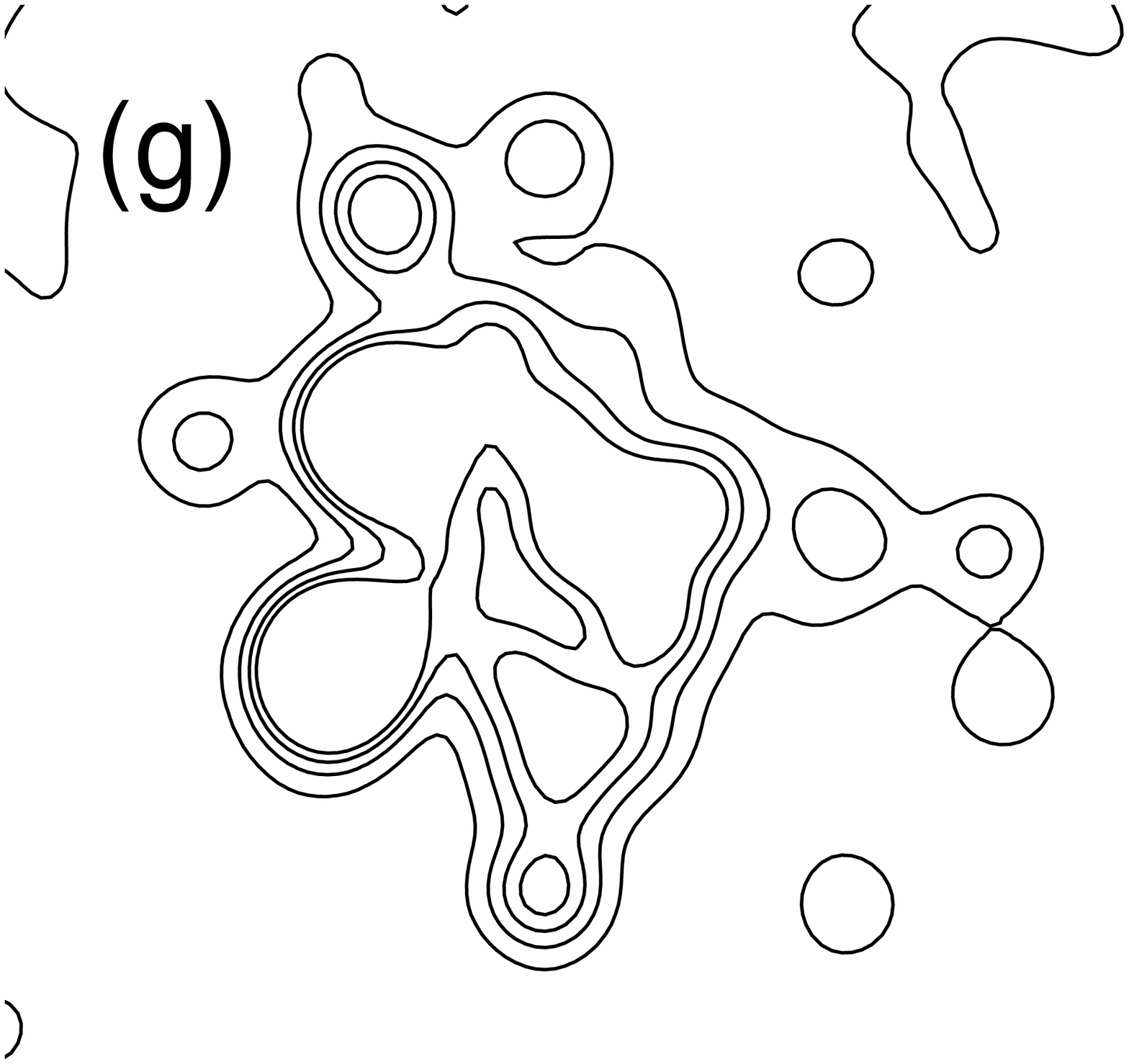}}
   \caption[]{Intensity isophotal maps of (a) NGC\,241+242, (b) IC\,1612 components are marked,
   (c) NGC\,2011, (d) NGC\,376, (e) K\,50, (f) K\,54 and (g) NGC\,1810.}
   \label{fig:iso}
   \end{center}
\end{figure}

We show in Fig. \ref{fig:iso} isophotal maps for the binary clusters and merger
candidates in the present sample. We applied Gaussian filters to smooth the
images using the {\it gauss} routine of the {\sc iraf/images} package.

As it became clear in the analysis of Fig. \ref{fig:1}, seven clusters could not
be satisfactorily fitted with EFF profiles due to the presence of large-scale
structures along the SBPs. NGC\,241+242 and IC\,1612\,+\,H86$-$186 are binary
clusters in the SMC (\citealt{deOliv00a}), while NGC\,2011\,+\,BRHT14b is a
binary cluster in the LMC (\citealt{Bathal91,Dieball,Goul06}). The double bump
representing the cluster members can be seen in the profiles (Fig. \ref{fig:1}).
Isophotal maps of NGC\,241+242, IC\,1612 and NGC\,2011 support this in Fig.
\ref{fig:iso}.

A double peak is also observed in the SBPs of NGC\,376, K\,50 and K\,54, but
these clusters do not have a detected companion, together with NGC\,1810 which
has an unusual profile (Fig. \ref{fig:1}). EFF profiles do not describe them.
Their isophotal maps are complex. NGC\,376, and especially K\,50 and K\,54,
present apparently triangular outer isophotes, while NGC\,376, K\,50 and
NGC\,1810 have isophotal gaps that extend towards the cluster central parts
(Fig. \ref{fig:img} and Fig. \ref{fig:1}). In the numerical simulations of
cluster encounters by \citet{deOliv00b}, similar structures can be seen. We
conclude that unusual cluster profiles and isophotal distributions may be
related to star cluster mergers.

\subsection{Extended profiles}
\label{sec:ext}

The SMC clusters NGC\,176, NGC\,290, L\,66, K\,47, IC\,1624, and the LMC
clusters NGC\,1870, NGC\,2159 and HS\,314, in particular, have extensions beyond
the EFF fitted profiles (Fig. \ref{fig:1}). We are dealing with Blue Magellanic
Clusters, with ages as a rule younger than 500 Myr. \citet{EFF87} found similar
extensions for comparable ages. Such clusters do not seem to be tidally
truncated. Expansion due to mass loss or violent relaxation in the early
cluster may contribute. The present work (Sect. \ref{sec:bin}) suggests that
cluster interactions and eventual mergers may also contribute for clusters to
spill over the Roche limit.

Table \ref{tab:dis} summarizes the main conclusions about structures observed
in the profiles of the star clusters of this study, including finer details.
Note that objects in common with \citet{Mack03a,Mack03b} have similar fine
structures in the profile (e.g. NGC\,176, Fig. \ref{fig:1}). It is interesting
that 35 of the clusters appear to have at least one significant fine structure
($\sim$ 2 pc of typical size) in the observed light profiles. The seeing (Sect. \ref{sec:obs}) implies a resolution better than 0.5 pc. We call attention that the fine structure occurs in a smaller spatial scale as compared to the
larger variations ($\sim$ 5 pc) observed in the profile of the four clusters
with evidence of mergers (Sect. \ref{sec:bin}).

\begin{table}[!h]
\caption[]{Structures in the profiles.}
\label{tab:dis}
\renewcommand{\tabcolsep}{1.8mm}
\renewcommand{\arraystretch}{1.0}
\begin{tabular}{lllc}
\hline
\hline
\multicolumn{1}{l}{Cluster} &\multicolumn{1}{c}{Inner profile} &\multicolumn{1}{c}{External profile} &\multicolumn{1}{c}{Diagnostic}\\
\multicolumn{1}{l}{(1)} &\multicolumn{1}{l}{(2)} &\multicolumn{1}{l}{(3)} &\multicolumn{1}{c}{(4)}\\
\hline
\multicolumn{4}{c}{SMC} \\
\hline
NGC\,121     & NP   & -- & -- \\
NGC\,176     & RD+  & EE & -- \\
K\,17        & RD   & EE & -- \\
NGC\,241+242 &  --  & -- & binary \\
NGC\,290     & RD+  & EE & -- \\
L\,48        & RD   & EE & -- \\
K\,34        & NP   & EE & -- \\
NGC\,330     & RD   & -- & -- \\
L\,56        & RD   & EE & -- \\
NGC\,339     & RD   & -- & -- \\
NGC\,346     & RD   & EE & -- \\
IC\,1611     & RD   & -- & -- \\
IC\,1612     &  --  & -- & binary \\
L\,66        & RD   & EE & -- \\
NGC\,361     & RD   & -- & -- \\
K\,47        & NP   & EE & -- \\
NGC\,376     & RD++ & EE\,\,(flattening) & merger \\
K\,50        & RD++ & EE\,\,(flattening) & merger \\
IC\,1624     & RD   & EE & -- \\
K\,54        & RD++ & -- & merger \\
NGC\,411     & NP   & -- & -- \\
NGC\,416     & NP   & -- & -- \\
NGC\,419     & PD   & -- & -- \\
NGC\,458     & RD+  & -- & -- \\
L\,114       & RD+  & -- & -- \\
\hline
\multicolumn{4}{c}{LMC} \\
\hline
NGC\,1783    & RD   & -- & -- \\
NGC\,1810    & RD++ & EE\,\,(flattening) & merger \\
NGC\,1818    & RD   & -- & -- \\
NGC\,1831    & NP   & -- & -- \\
NGC\,1847    & NP   & EE & -- \\
NGC\,1856    & NP   & -- & -- \\
NGC\,1866    & RD   & -- & -- \\
NGC\,1868    & NP   & -- & -- \\
NGC\,1870    & NP   & EE & -- \\
NGC\,1978    & RD   & -- & -- \\
NGC\,2004    & RD   & -- & -- \\
NGC\,2011    & RD+  & EE & binary \\
NGC\,2100    & RD   & -- & -- \\
NGC\,2121    & NP   & -- & -- \\
NGC\,2157    & RD   & EE & -- \\
NGC\,2159    & RD+  & -- & -- \\
NGC\,2164    & RD   & -- & -- \\
NGC\,2210    & PD   & -- & -- \\
NGC\,2213    & RD   & -- & -- \\
NGC\,2214    & RD   & -- & -- \\
H\,11        & NP   & -- & -- \\
HS\,314      & RD+  & EE & -- \\
\hline
\end{tabular}
\begin{list}{Table Notes.}
\item Col. 2: Caracterization of inner profile. Col. 3: Caracterization of
external profile. Col. 4: Final diagnostic.
NP: normal profile \ RD: radially disturbed \ EE: external excess \ PD: profile
deviation. Plus sign suggests intensity.
\end{list}
\end{table}

\section{Concluding remarks}
\label{sec:con}

Surface brightness profiles in the V band were derived for 25 SMC and 22 LMC
star clusters, including blue and red ones.

Cluster centers were determined using the mirror autocorrelation 
(\citealt{Djorg88}) method with mean error smaller than 0.5$''$.

The observed profiles were fitted with the EFF model. The structural parameters
obtained from the fits were used to determine luminosities and masses of the
star clusters. For those in common with the HST study of
\citet{Mack03a,Mack03b} the agreement is good. For 23 objects the analysis is
carried out for the first time.

It is important to characterize single clusters, binary and candidates to
mergers in the Clouds, clusters which are ideal laboratories for dynamical
studies.

In some cases, the profiles present important deviations from EFF profiles. We
also use isophotal maps to constrain candidates to cluster interactions. The
binary clusters NGC\,241+242, IC\,1612, and NGC\,2011 have a double profile. The
clusters NGC\,376, K\,50, K\,54 and NGC\,1810 do not have detected companions
and present as well significant deviations from EFF profiles with bumps and dips
in a $\sim$ 5 pc scale.

We conclude that important deviations from the body of EFF profiles might be
used as a tool to detect cluster mergers.

\end{document}